\documentclass{iopart}
\usepackage{iopams}
\usepackage{graphicx}
\usepackage{cite}

\eqnobysec 

\begin{document}

\review[Origin of the p-nuclei]{Constraining the astrophysical origin of the p-nuclei through nuclear physics and meteoritic data}

\author{T Rauscher$^{1,2,3}$, N Dauphas$^4$, I Dillmann$^{5,6}$, C Fr\"ohlich$^7$, Zs F\"ul\"op$^2$ and Gy Gy\"urky$^2$}

\address{$^1$ Department of Physics, University of Basel, 4056 Basel, Switzerland}

\address{$^2$ MTA Atomki, 4001 Debrecen, POB 51, Hungary}

\address{$^3$ Centre for Astrophysics Research, University of Hertfordshire, Hatfield AL10 9AB, United Kingdom}

\address{$^4$ Origins Laboratory, Department of the Geophysical Sciences and Enrico Fermi Institute, The University of Chicago, Chicago, IL 60637, USA}

\address{$^5$ GSI Helmholtzzentrum f\"ur Schwerionenforschung GmbH, Darmstadt, Germany }

\address{$^6$ II. Physikalisches Institut, Justus-Liebig-Universit\"at, Gie\ss en, Germany}

\address{$^7$ Department of Physics, North Carolina State University, Raleigh, NC 27695, USA}

\ead{Thomas.Rauscher@unibas.ch}

\begin{abstract}
A small number of naturally occurring, proton-rich nuclides (the p-nuclei) cannot be made in the s- and r-process. Their origin is not well understood. Massive
stars can produce p-nuclei through photodisintegration of pre-existing intermediate and heavy nuclei. This
so-called $\gamma$-process requires high stellar plasma temperatures and occurs mainly in explosive O/Ne
burning during a core-collapse supernova. Although the $\gamma$-process in massive stars
has been successful in producing a large range of p-nuclei, significant deficiences remain. An increasing number of processes and sites has been studied in recent years in search of viable alternatives replacing or supplementing the massive star models. A large number of unstable nuclei, however, with only
theoretically predicted reaction rates are included in the reaction network and thus the nuclear input may also bear considerable uncertainties. The current status of astrophysical models, nuclear input, and observational constraints is reviewed. After an overview of currently discussed models, the focus is on the possibility to better constrain those models through different means. Meteoritic data not only provide the actual isotopic abundances of the p-nuclei but can also put constraints on the possible contribution of proton-rich nucleosynthesis. The main part of the review focusses on the nuclear uncertainties involved in the determination of the astrophysical reaction rates required for the extended reaction networks used in nucleosynthesis studies. Experimental approaches are discussed together with their necessary connection to theory, which is especially pronounced for reactions with intermediate and heavy nuclei in explosive nuclear burning, even close to stability.
\end{abstract}

\submitto{\RPP}
\maketitle

\tableofcontents

\vspace{1.5cm}

\section{Introduction}

The origin of the intermediate and heavy elements beyond iron has been a long-standing, important question in astronomy and astrophysics. The neutron capture s- and r-processes synthesize the bulk of those nuclei. While low-mass, asymptotic giant branch (AGB, $M\lesssim 8$ $M_\odot$) and massive stars ($M \gtrsim 8$ $M_\odot$) were found to contribute to the s-process, the site of the r-process still remains unknown. Moreover, a number of naturally occurring, proton-rich isotopes (the p-nuclei) cannot be made in either the s- or the r-process. Although their natural abundances are tiny compared to isotopes produced in neutron-capture nucleosynthesis, their production is even more problematic. The long-time favored process, photodisintegration of material in the O/Ne-shell of a massive star during its final core-collapse supernova explosion, fails to produce the required amounts of p-nuclei in several mass ranges. Several alternative sites have been proposed but so far no conclusive evidence has been found to favor one or the other. Further important uncertainties stem from the reaction rates used in the modeling of the thermonuclear burning. Investigations in astrophysical and nuclear models, together with various "observational" information (obtained from stellar spectra, meteoritic specimens, and nuclear experiments) comprise the pieces which have to be put together to solve the puzzle of the origin of the p-nuclei. It is an excellent example for the multifaceted, interdisciplinary approaches required to understand nucleosynthesis.

This review attempts to provide a general overview of the conditions required to produce p-nuclei and a summary of the commonly discussed production processes and sites. It then focusses on the possibilities to better constrain astrophysical models through measurements, from meteoritic isotopic abundances to nuclear experiments. The derivation of astrophysical reaction rates for intermediate and heavy nuclei from experiments necessitates the use of theoretical models of nuclear reactions, due to the nuclear properties (such as binding energies and cross sections) demanding extreme thermonuclear conditions to allow the synthesis of proton-rich nuclides.

We start with a brief historic overview and definition of the p-nuclei in \sref{sec:intro}, followed by an overview of the suggested astrophysical processes and sites (\sref{sec:sites}), also outlining the problems encountered. \Sref{sec:meteorites} presents the information which can be extracted from the analysis of meteoritic material, from the actual solar p-abundances (\sref{sec:cosmicabuns}) to constraints from extinct radionuclides (\sref{sec:extinct}) and isotopic anomalies (\sref{sec:anomalies}). The discussion of the relevant nuclear physics starts with basic definitions in \sref{sec:definitions}. Important reactions, the main nuclear uncertainties and the ways in which nuclear experiments can help are examined in sections \ref{sec:importantreactions}, \ref{sec:sensi}, and \ref{sec:expgeneral}. Experimental approaches are reviewed in \sref{sec:methods}, more specifically photodisintegration reactions and their limitations (\sref{sec:gamma}), charged-particle induced reactions (\sref{sec:charged}), elastic scattering (\sref{sec:alphascatt}), and neutron-induced reactions (\sref{sec:neutron}).

\section{The case of the missing nuclides}
\label{sec:intro}

\begin{figure}
\includegraphics[width=\textwidth]{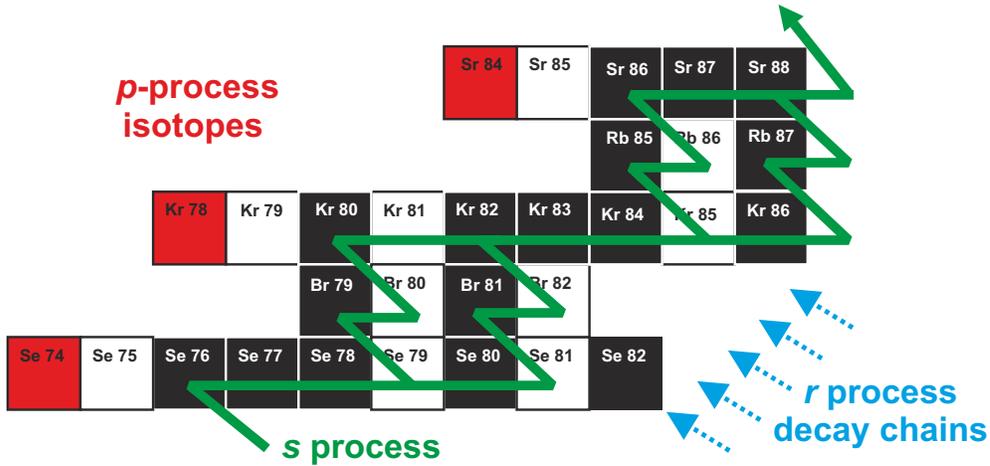}
\caption{The p-isotopes are shielded from r-process decay chains by stable isotopes and are bypassed in the s-process reaction flow.}
\label{fig:flow}
\end{figure}

In the first detailed analysis of solar abundances published by \cite{su56}, it was already indicated that at least two types of processes may be required to produce the abundance distribution above iron, one leading to neutron-rich isotopes and a different one for neutron-deficient nuclides. Only one year later \cite{b2fh} (B$^2$FH) and \cite{agw} made detailed studies on suitable processes and their constraints, based on the data by \cite{su56} and additional astronomical observations and nuclear data. It turned out that two types of neutron-capture processes were required to explain the abundance patterns of intermediate and heavy nuclei, the so-called s- and r-processes \cite{bsmey94,wallerstein,ctt91}. They also realized that
a number of proton-rich isotopes can never be synthesized through sequences of only neutron captures and $\beta^-$ decays (\fref{fig:flow}) and required the postulation of a third process. It was termed \textit{p-process} because it was initially thought to proceed via proton captures at high temperature, perhaps even reaching (partial) (p,$\gamma$)-($\gamma$,p) equilibrium. This nucleosynthesis process was tentatively placed in the H-rich envelope of type II supernovae by B$^2$FH but it was later realized that the required temperatures are not attained there \cite{autru,arngorp}. This also shed doubts on the feasibility to use proton captures for producing \textit{all} of the nuclides missing from the s- and r-process production.

\begin{figure}
\includegraphics[width=\textwidth]{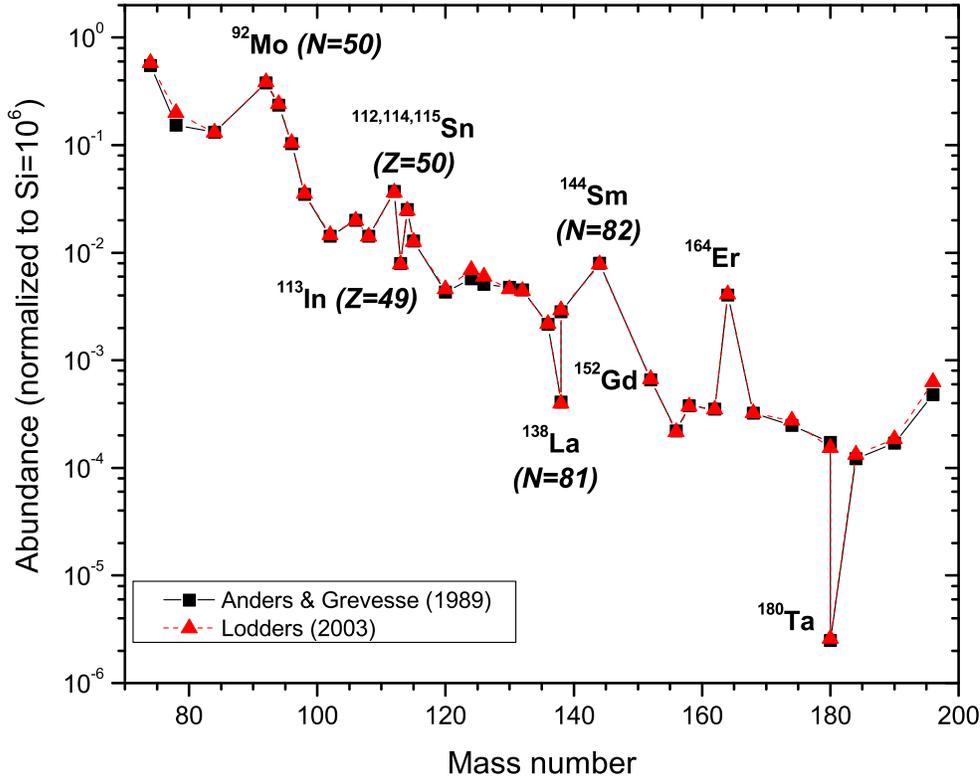}
\caption{Comparison of the solar abundances for the p-nuclei \cite{AnG89,Lod03}; the connecting lines are drawn to guide the eye.}
\label{fig:comp}
\end{figure}

It is somewhat confusing that in the literature the name "p-process" is sometimes used for a proton capture process in the spirit of B$^2$FH but also sometimes taken as a token subsuming whatever production mechanism(s) is/are found to be
responsible for the p-nuclides. For an easier distinction of the production processes, here we prefer to adopt the
modern nomenclature focussing on naming the nuclides in question
the \textit{p-nuclides} (they were called ``excluded isotopes'' by \cite{agw}) and using different names to specify the processes possibly involved.

Historically there were 35 p-nuclides identified (\fref{fig:comp} and \tref{tab:solar}), with $^{74}$Se being the lightest and $^{196}$Hg the heaviest. It is to be noted, however, that this assignment depends on the state-of-the-art of the s-process models (just like the "observed" r-abundances depend on them) and also on estimates of r-process contributions (e.g., to $^{113}$In and $^{115}$Sn \cite{dill30,nemeth}). Almost all p-isotopes are even-even nuclei, with exception of $^{113}$In ($Z$=49), $^{115}$Sn, $^{138}$La, and $^{180}$Ta$^\mathrm{m}$. The isotopic abundances (\tref{tab:solar}) are $1-2$ orders of magnitude lower than for the respective r- and s-nuclei in the same mass region, with exception of $^{92,94}$Mo and $^{96,98}$Ru.

The two neutron-magic p-isotopes $^{92}$Mo (neutron number $N=50$) and $^{144}$Sm ($N=82$), and the proton-magic (charge number $Z=50$) Sn-isotopes $^{112,114}$Sn exhibit larger abundances than the neighboring p-nuclei (\fref {fig:comp}). The abundance of $^{164}$Er also stands out and already B$^2$FH realized that it may contain considerable contributions from the s-process. It was indeed found that there are large s-process contributions to $^{164}$Er, $^{152}$Gd, and $^{180}$Ta \cite{arl99}, thus possibly removing them from the list of p-isotopes. If also the abundances of $^{113}$In and $^{115}$Sn can be explained by modifications of the s-process and/or contributions from the r-process \cite{nemeth}, this would leave only 30 p-isotopes to be explained by other processes.

The nuclei $^{138}$La and $^{180m}$Ta do not fit well the local trend and have much lower abundance than their neighbors. This indicates a further process at work, also because the standard photodisintegration process cannot synthesize them in the required quantity (see \sref{sec:sites}).

\Tref{tab:solar} lists also the isotopic composition as given in \cite{tice09}. The quoted uncertainties are on the abundance relative to other isotopes of the same element. They introduce additional uncertainties in reaction measurements using natural samples (see \sref{sec:expgeneral}). In general the composition uncertainties are below 5\%. It should be noted that there have been recent measurements with higher precision which had not been included in \cite{tice09}, e.g., \cite{2009EPSL.277..334R,brandon} for Os. The modern composition uncertainties for their p-isotopes are below 1\%.

A special obstacle not found in the investigation of the s- and r-process is the fact that there are no elements dominated by p-isotopes. Therefore, our knowledge of p-abundances is limited to solar system abundances derived from meteoritic material (\sref{sec:cosmicabuns}) and terrestrial isotopic compositions. Before astronomical observations of isotopic abundances at the required discrimination level become feasible (if ever), it is impossible to determine p-abundances in stars of different metallicities and thus to obtain the galactic chemical evolution (GCE) picture directly. On the other hand, depending on the actual p-production mechanism this may also be problematic for determining early s-process contributions. If the p-nuclides (or some of them) turn out to be primary (i.e., independent of metallicity) or have a different dependence of production on metallicity than the s-process (perhaps by initially originating from r-nuclei), they may give a larger contribution to elemental abundances in old stars than the s-process. Observations of Ge and Mo in metal-poor stars may cast further light on the origin of the p-nuclei.

\begin{table}
\caption{\label{tab:solar} Contribution of p-isotopes to the isotopic composition of elements \cite{tice09} and solar p-abundances (relative to Si= 10$^6$) from Anders and Grevesse \cite{AnG89} and Lodders \cite{Lod03}. }
\begin{indented}
\item[]\begin{tabular}{@{}lcccc}
\br
Isotope & p-isotope contribution & Solar abund.  & Solar abund.  & Change \\
				& (\%) \cite{tice09} & (2003) \cite{Lod03} & (1989) \cite{AnG89} & (\%) \\
\mr
$^{74}$Se & 0.89 (4) & 5.80$\times$10$^{-1}$ & 5.50$\times$10$^{-1}$ & 5.45 \\
$^{78}$Kr & 0.355 (3) & 2.00$\times$10$^{-1}$ & 1.53$\times$10$^{-1}$ & 30.72 \\
$^{84}$Sr & 0.56 (1) & 1.31$\times$10$^{-1}$ & 1.32$\times$10$^{-1}$ & -0.61 \\
$^{92}$Mo & 14.53 (30) & 3.86$\times$10$^{-1}$ & 3.78$\times$10$^{-1}$ & 2.12 \\
$^{94}$Mo & 9.15 (9) & 2.41$\times$10$^{-1}$ & 2.36$\times$10$^{-1}$ & 2.12 \\
$^{96}$Ru & 5.54 (14) & 1.05$\times$10$^{-1}$ & 1.03$\times$10$^{-1}$ & 2.23 \\
$^{98}$Ru & 1.87 (3) & 3.55$\times$10$^{-2}$ & 3.50$\times$10$^{-2}$ & 1.43 \\
$^{102}$Pd & 1.02 (1) & 1.46$\times$10$^{-2}$ & 1.42$\times$10$^{-2}$ & 2.82 \\
$^{106}$Cd & 1.25 (6) & 1.98$\times$10$^{-2}$ & 2.01$\times$10$^{-2}$ & -1.49 \\
$^{108}$Cd & 0.89 (3) & 1.41$\times$10$^{-2}$ & 1.43$\times$10$^{-2}$ & -1.40 \\
$^{113}$In & 4.29 (5) & 7.80$\times$10$^{-3}$ & 7.90$\times$10$^{-3}$ & -1.27 \\
$^{112}$Sn & 0.97 (1) & 3.63$\times$10$^{-2}$ & 3.72$\times$10$^{-2}$ & -2.55 \\
$^{114}$Sn & 0.66 (1) & 2.46$\times$10$^{-2}$ & 2.52$\times$10$^{-2}$ & -2.38 \\
$^{115}$Sn & 0.34 (1) & 1.27$\times$10$^{-2}$ & 1.29$\times$10$^{-2}$ & -1.94 \\
$^{120}$Te & 0.09 (1) & 4.60$\times$10$^{-3}$ & 4.30$\times$10$^{-3}$ & 6.98  \\
$^{124}$Xe & 0.0952 (3) & 6.94$\times$10$^{-3}$ $^\mathrm{a}$ & 5.71$\times$10$^{-3}$ & 21.54$^\mathrm{a}$ \\
$^{126}$Xe & 0.0890 (2) & 6.02$\times$10$^{-3}$ $^\mathrm{a}$ & 5.09$\times$10$^{-3}$ & 18.27$^\mathrm{a}$ \\
$^{130}$Ba & 0.106 (1) & 4.60$\times$10$^{-3}$ & 4.76$\times$10$^{-3}$ & -3.36 \\
$^{132}$Ba & 0.101 (1) & 4.40$\times$10$^{-3}$ & 4.53$\times$10$^{-3}$ & -2.87 \\
$^{138}$La & 0.08881 (71) & 3.97$\times$10$^{-4}$ & 4.09$\times$10$^{-4}$ & -2.93 \\
$^{136}$Ce & 0.185 (2) & 2.17$\times$10$^{-3}$ & 2.16$\times$10$^{-3}$ & 0.46 \\
$^{138}$Ce & 0.251 (2) & 2.93$\times$10$^{-3}$ & 2.84$\times$10$^{-3}$ & 3.17 \\
$^{144}$Sm & 3.07 (7) & 7.81$\times$10$^{-3}$ & 8.00$\times$10$^{-3}$ & -2.38 \\
$^{152}$Gd & 0.20 (1) & 6.70$\times$10$^{-4}$ & 6.60$\times$10$^{-4}$ & 1.52 \\
$^{156}$Dy & 0.056 (3) & 2.16$\times$10$^{-4}$ & 2.21$\times$10$^{-4}$ & -2.26 \\
$^{158}$Dy & 0.095 (3) & 3.71$\times$10$^{-4}$ & 3.78$\times$10$^{-4}$ & -1.85 \\
$^{162}$Er & 0.139 (5) & 3.50$\times$10$^{-4}$ & 3.51$\times$10$^{-4}$ & -0.28 \\
$^{164}$Er & 1.601 (3) & 4.11$\times$10$^{-3}$ & 4.04$\times$10$^{-3}$ & 1.73 \\
$^{168}$Yb & 0.123 (3) & 3.23$\times$10$^{-4}$ & 3.22$\times$10$^{-4}$ & 0.31 \\
$^{174}$Hf & 0.16 (1) & 2.75$\times$10$^{-4}$ & 2.49$\times$10$^{-4}$ & 10.44 \\
$^{180}$Ta$\rm ^m$ & 0.01201 (32) & 2.58$\times$10$^{-6}$ & 2.48$\times$10$^{-6}$ & 4.03 \\
$^{180}$W  & 0.12 (1) & 1.53$\times$10$^{-4}$ & 1.73$\times$10$^{-4}$ & -11.56 \\
$^{184}$Os & 0.02 (1) & 1.33$\times$10$^{-4}$ & 1.22$\times$10$^{-4}$ & 9.02 \\
$^{190}$Pt & 0.012 (2) & 1.85$\times$10$^{-4}$ & 1.70$\times$10$^{-4}$ & 8.82 \\
$^{196}$Hg & 0.15 (1) & 6.30$\times$10$^{-4}$ & 4.80$\times$10$^{-4}$ & 31.25 \\
\br
\end{tabular}
\item[] $^{\rm a}$ Abundances by \cite{Rei02}: $^{124}$Xe: 6.57$\times$10$^{-3}$ ($+5.63$\%); $^{126}$Xe: 5.85$\times$10$^{-3}$ ($+2.91$\%).
\end{indented}
\end{table}

\section{Processes and sites possibly contributing to the production of p-nuclei}
\label{sec:sites}

\subsection{General considerations}
There are several possibilities to get to the proton-rich side. Sequences of proton captures may reach a p-isotope from elements with lower charge number. They are suppressed by the Coulomb barriers, however, and it is not possible to arbitrarily compensate for that by just requiring higher plasma temperatures. At high temperature ($\gamma$,p) reactions become faster than proton captures and prevent the build-up of proton-rich nuclides. Only a very proton-rich environment allows to have fast proton-captures, see  \eref{eq:caprate}. Photodisintegrations are an alternative way to make p-nuclei, either by directly producing them through destruction of their neutron-richer neighbor isotopes through sequences of ($\gamma$,n) reactions (these are the predominant photodisintegration processes for most stable nuclei), or by flows from
heavier, unstable nuclides via ($\gamma$,p) or ($\gamma$,$\alpha$) reactions and subsequent $\beta$-decays.

There are three ingredients influencing the resulting p-abundances and these three are differently combined in the various sites proposed as the birthplace of the p-nuclides. The first one is, obviously, the temperature variation as a function of time, defining the timescale of the process and the peak temperature. This already points to explosive conditions which accommodate both the necessary temperatures for photodisintegrations (or proton captures on highly charged nuclei) and a short timescale. The latter is required because it has to be avoided that too much material is transformed in order to achieve the tiny solar p-abundances (assuming that these are typical). The second parameter is the proton density. While photodisintegrations and $\beta$ decays are not sensitive to the proton abundance, proton captures are. With a high number of protons available, proton captures can prevail over ($\gamma$,p) reactions even at high temperature. Last but not least, the seed abundances, i.e., the number and composition of nuclei on which the photodisintegrations or proton captures act initially, are also highly important but not well constrained. In most suggested production mechanisms (see below), the final p-abundances depend sensitively on these seeds and therefore are secondary. Thus, they depend
on some s- and/or r-process nuclides being already present in the material, either because the star inherited those abundances from its proto-stellar cloud or because some additional production occured within the site before the onset of p-nucleus production.

So far it seems to be impossible to reproduce the solar abundances of the remaining 30 p-isotopes by one single process. In our current understanding several (independently operating) processes seem to contribute to the p-abundances. These processes can be realized in different sites, e.g., the $\gamma$-process discussed below will occur in any sufficiently hot plasma. It was first discovered in simulations of massive star explosions but also appears in type Ia supernovae.

\begin{figure}
\begin{center}
\includegraphics[width=8cm]{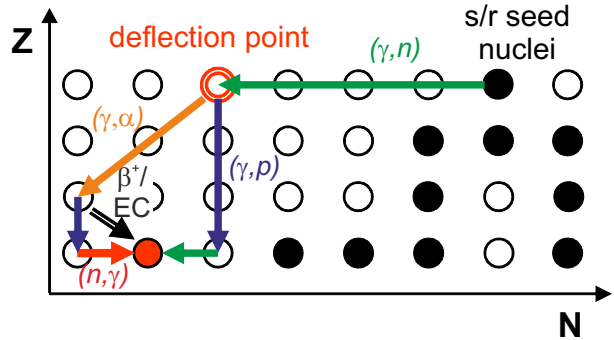}
\caption{Reaction flow in the $\gamma$-process.}
\label{fig:gamma}
\end{center}
\end{figure}

\subsection{Shells of exploding massive stars}
\label{sec:massive}
For a long time, the favored process for production of p-nuclei
has been the \textit{$\gamma$-process} occurring during explosive O/Ne-shell burning in massive stars \cite{woohow,arn,rayet90,rayet95,rhhw02,haya08}. It was realized early
that the abundances of most p-nuclei are inversely correlated with their photodisintegration rates \cite{agw,woohow}, pointing to an important
contribution of photodisintegration. At temperatures of $2\leq T \leq 3.5$ GK, pre-existing seed nuclei in the p-nuclear mass range can be partially photodisintegrated, starting with sequences of ($\gamma$,n) reactions and creating proton-rich isotopes. Several mass units away from stability, the ($\gamma$,n) reactions compete with $\beta$-decays but also with ($\gamma$,p) and/or ($\gamma$,$\alpha$) (see \fref{fig:gamma} and \sref{sec:importantreactions}).

Massive stars provide the required conditions of transforming s- and r-process material already present
in the proto-stellar cloud or produced in-situ in the weak s-process, during the He- and C-burning phase. The $\gamma$-process occurs naturally in simulations of massive stars and does not require any artifical fine-tuning. During the final core-collapse supernova (ccSN) a shockwave ejects and heats the outer layers of the star, in just the right amount needed to produce p-nuclei through photodisintegration. It is crucial that a range of temperatures be present as nuclei in the lower mass part of the p-nuclides require higher temperatures for photodisintegration ($2.5-3.5$ GK) whereas the ones at higher mass are photodissociated more easily and should not be exposed to too high temperature ($T<2.5$ GK), as otherwise all heavy p-nuclei would be destroyed. Stars with higher mass may reach $\gamma$-process
temperatures already pre-explosively \cite{rhhw02} although some or all of those pre-explosive p-nuclei may be destroyed again in the explosion.

\begin{figure}
\includegraphics[angle=-90,width=\textwidth]{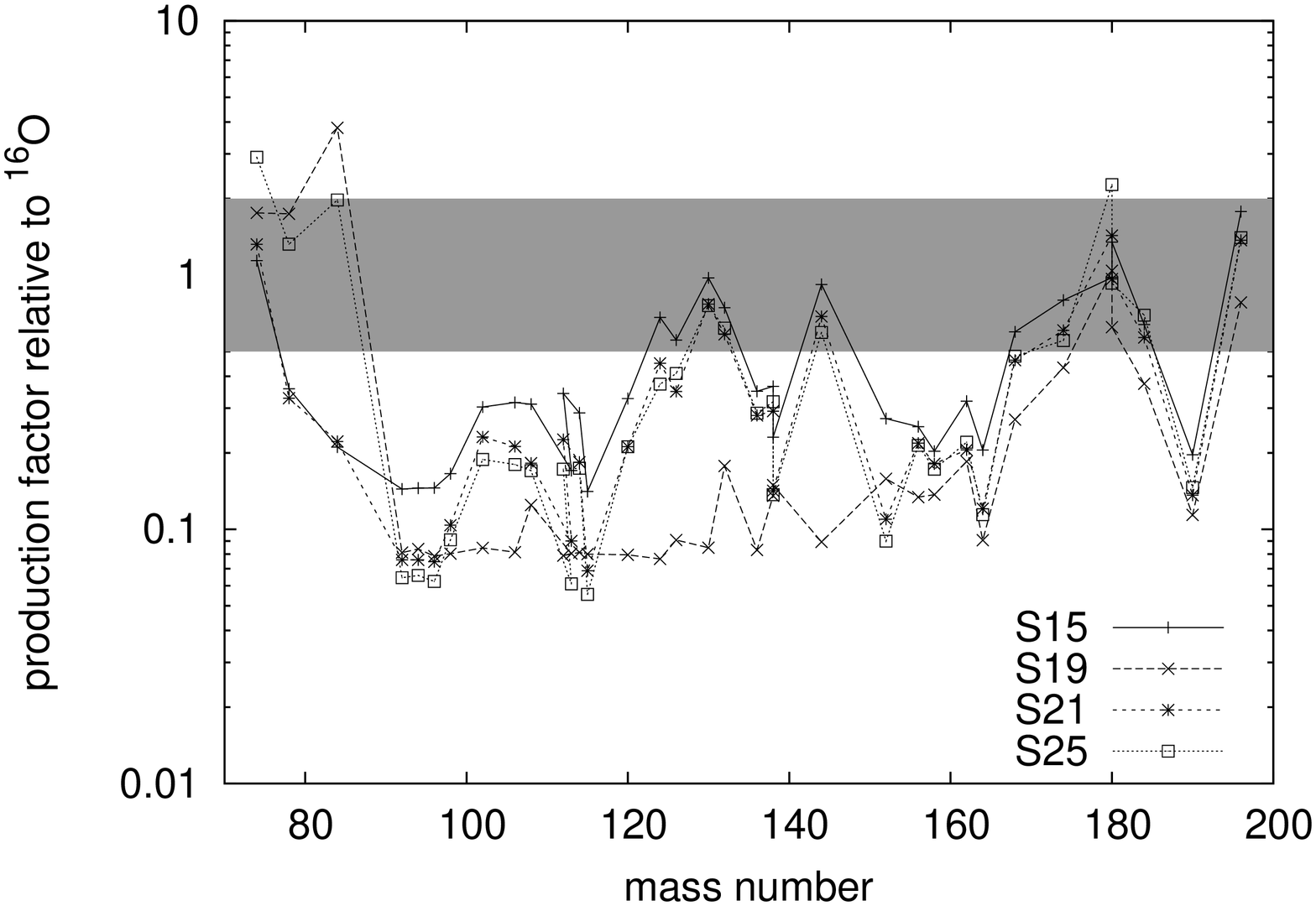}
\includegraphics[angle=-90,width=\textwidth]{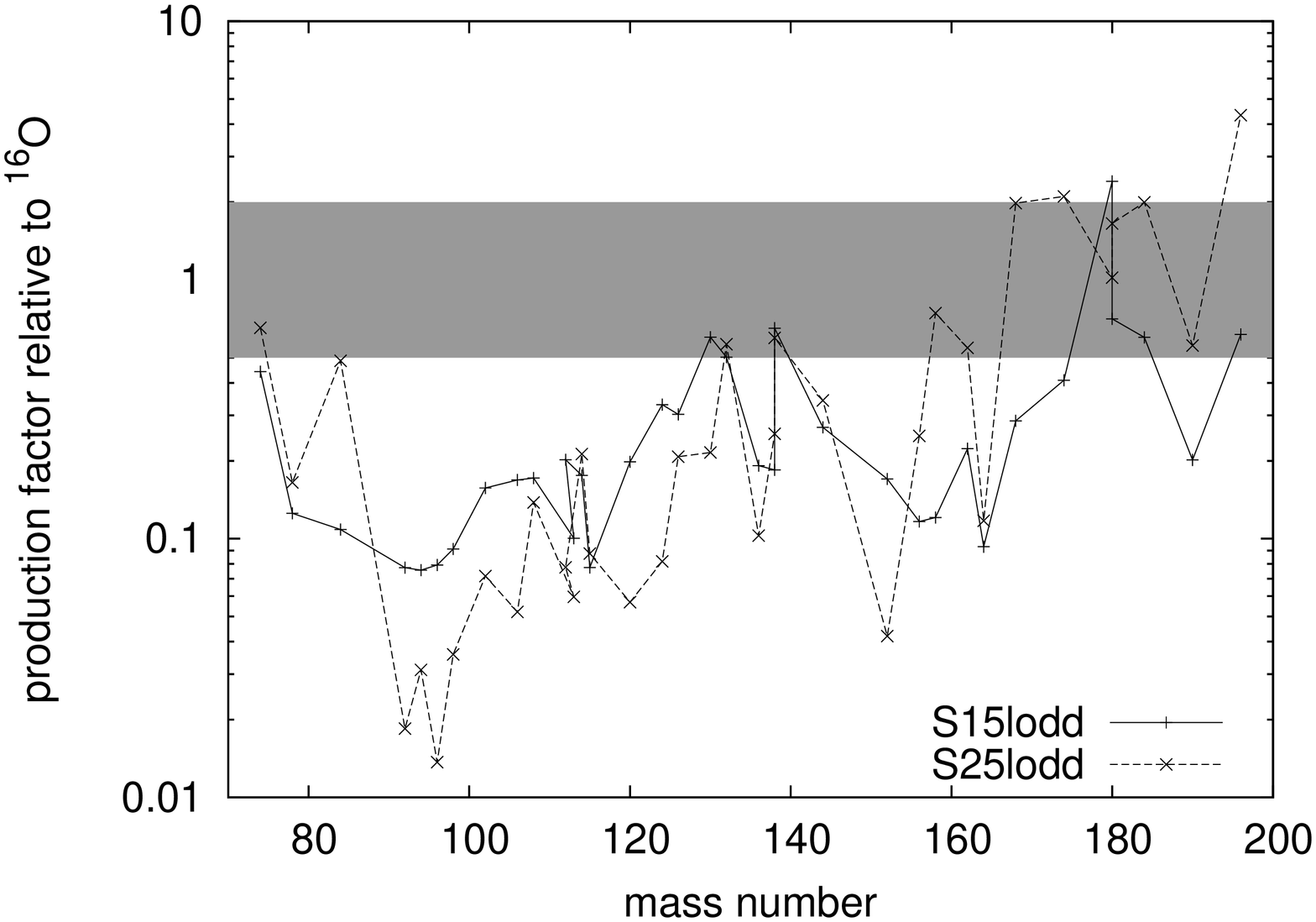}
\caption{\label{fig:raumodels}Production factors $F_i$ relative to $^{16}$O of p-nuclei for massive star models taken from \cite{rhhw02} with initial solar metallicity, for progenitors with 15 (S15, S15lodd), 19 (S19), 21 (S21), and 25 (S25, S25lodd) solar masses. The initial metallicity is based on \cite{AnG89} (top) and \cite{Lod03} (bottom). The shaded area gives a factor of 2 around the $^{16}$O production factor as acceptable range of co-production. The lines are drawn to guide the eye.}
\end{figure}

All p-nuclei are secondary in this production mechanism, i.e., depending on the initial metallicity of the star. \Fref{fig:raumodels} shows the production factors relative to $^{16}$O for stellar models with initially solar composition. These models were the first ones to self-consistently follow the $\gamma$-processes through the pre-supernova stages and the supernova explosion. The production factors $F_i$ relative to $^{16}$O are defined as
\begin{equation}
F_i=\frac{f_i}{f_\mathrm{^{16}O}}=\frac{Y_i^\mathrm{final}/Y_i^\mathrm{initial}}{Y_\mathrm{^{16}O}^\mathrm{final}/Y_\mathrm{^{16}O}^\mathrm{initial}} \quad,
\end{equation}
with the initial and final abundances $Y_i^\mathrm{initial}$, $Y_i^\mathrm{final}$, respectively. The nuclide $^{16}$O is the main "metal" produced in massive stars and its production factor is often used as fiducial point to define a band of acceptable agreement in production.

As seen in the upper part of \fref{fig:raumodels}, the p-nucleus production not only depends on the initial metallicity but also on details of the stellar evolution which are also determined by the stellar mass. Similar trends, however, can be found in all models. The p-nuclei in the mass ranges $124\leq A \leq 150$ and $168\leq A \leq 200$, are produced in solar abundance ratios within about a factor of 2 relative to $^{16}$O. Below $A < 124$ and between $150 \leq A \leq 165$ the p-isotopes are severely underproduced. The S19 model shows a special behavior due to the partial merging of convective shells. In general, the total production of the proton-rich isotopes increases for higher entropy in the oxygen shell, i.e., with increasing mass, but also depends on details of stellar structure and the composition of the star at the time of core collapse. To consider the total contribution of massive stars to the Galactic budget of p-nuclei, the individual yields have to be averaged over the stellar mass distribution, giving more weight to stars with less mass. Since the $\gamma$-process yields of stars with different mass can vary wildly, a fine grid of masses has to be used.

The results shown in the upper part of \fref{fig:raumodels} used the solar abundances of \cite{AnG89}. Recent new abundance determinations have brought considerable changes especially in the relative abundances of light nuclei (see \sref{sec:cosmicabuns}). The lower part of \fref{fig:raumodels} shows models calculated with abundances by \cite{Lod03} (with amendments as given in \cite{Lod10}). The differences in the final p-nuclei production compared to the older solar abundances is not due to different solar intermediate and heavy element abundances, as can also be verified by inspection of \tref{tab:solar}. Rather, the different light element abundances (including $^{16}$O) and their impact on the hydrostatic burning phases (in particular helium burning \cite{turhegeraustin10}) lead to a different pre-supernova structure of the star, affecting the $\gamma$-process nucleosynthesis later on. Obviously, the normalization to the $^{16}$O production factor is also affected (\sref{sec:cosmicabuns}).

The very rare $^{138}$La cannot be produced in a $\gamma$-process but it was suggested to be formed through neutrino reactions on stable nuclei (the \textit{$\nu$-process}) \cite{neutrino,hekol05}. This was shown to be feasible with neutrinos emitted by the nascent neutron star emerging from the core collapse of the massive star, thus producing this isotope in the same site as the other p-nuclides but in a different process. This $\nu$-process is included in the results shown in \fref{fig:raumodels}. The equally rare $^{180m}$Ta probably also received a large contribution from the $\nu$-process. The prediction of its yield from massive stars, however, suffers from the problem to accurately predict the final isomeric state to ground state (g.s.) ratio after freeze-out of nuclear reactions (see, e.g., \cite{rhhw02,belic,mokago}). It entails not only the population of these states through neutrino- and $\gamma$-induced reactions but also to follow internal $\gamma$-transitions throughout the nucleosynthesis phase.

The longstanding shortcomings in the production of the light p-nuclei with $A<124$ and also those in the mass range $150\leq A \leq165$ have triggered a number of investigations in astrophysics and nuclear physics aimed at resolving these deficiences. It has already been realized early on that it is unfeasible to produce the most abundant p-isotopes, $^{92,94}$Mo and $^{96,98}$Ru, by photodisintegration in exploding massive stars due to lack of seed nuclei in their mass region \cite{woohow,howmey,fullmey}. Therefore a different production environment has to be found. The underproduction at higher masses, on the other hand, may still be cured by improved astrophysical reaction rates (see \sref{sec:sensi}).

\subsection{White Dwarf explosions}
\label{sec:WD}
A crucial problem in obtaining solar p-production in the $\gamma$-process in massive stars is the seed distribution which is strongly constrained.
Thermonuclear explosions of strongly s-process enriched matter provide an alternative. If the appropriate temperature range is covered in such explosions,
a $\gamma$-process ensues but with a different seed distribution compared to the one found in massive stars.
Such an environment is provided in the thermonuclear explosion of a mass-accreting White Dwarf, mainly composed of C and O \cite{howmey}.
Exploratory, parameterized calculations for the canonical type Ia supernova (SNIa) model -- the explosion of a White Dwarf (WD) after it has accreted enough mass from a companion star to reach the Chandrasekhar limit -- also found an underproduction of light p-nuclei, even when
assuming a seed enrichment of factors $10^3-10^4$ in s-process nuclei from an AGB companion \cite{arngorp}. Recent studies, based on the carbon deflagration model of \cite{nom}, found similar problems \cite{kusa05,kusa11}. In contrast, post-processing of high-resolution 2D models considering two types of explosions, deflagration and deflagration-detonation, find that they can co-produce all p-nuclei (with the exception of $^{113}$In, $^{115}$Sn, $^{138}$La, $^{152}$Gd, and $^{180}$Ta) when using strong enhancements in the assumed s-process seeds \cite{travWD}. It was concluded in \cite{kusa11,travWD} that a high-resolution treatment of the outer zones of the type Ia supernova is crucial to accurately follow the production of p-nuclides.

Another alternative is a subclass of type Ia
supernovae which is supposed to be caused by the disruption of a sub-Chandrasekhar WD due to a thermonuclear
runaway in a He-rich accretion layer \cite{gorsubchandra}. High neutron fluxes are built up in the early phase
of the explosion and a weak r-process ensues. Once the temperature exceeds $T_9\approx2$, photodisintegrations take over and move the
nucleosynthesis to the proton-rich side where two processes act: the $\gamma$-process, as described above, and additionally
rapid proton captures on proton-rich unstable nuclei at $3<T_9 \leq 3.5$. The latter is somewhat similar to the rp-process but at much lower proton densities and
thus closer to stability. The proton captures are in equilibrium with ($\gamma$,p) reactions but nuclei with low capture Q-value cannot be bridged efficiently within the short timescale of the explosion. A large number of neutrons, however, is released in the reactions $^{18}$O($\alpha$,n)$^{21}$Ne, $^{22}$Ne($\alpha$,n)$^{25}$Mg, and $^{26}$Mg($\alpha$,n)$^{29}$Si during the detonations \cite{arngorp}. The waiting points with low (p,$\gamma$) Q-value thus can be
bypassed by (n,p) reactions \cite{arngorp,gorsubchandra}. This so-called \textit{pn-process} can efficiently
produce the light p-nuclides from Se to Ru but it overproduces them in relation to the heavier ones. Again, a strong increase
in the photodisintegration seed abundances would be required to produce all p-nuclei at solar relative abundances. (It has to be
noted that here the nuclei produced in the pn-process would be primary whereas the others are secondary.) It was concluded, nevertheless, that subChandrasekhar He-detonation models are not an efficient site for p-nucleus synthesis \cite{gorsubchandra,gorsub3D}.

Both WD scenarios
(canonical and sub-Chandrasekhar mass type Ia supernovae) suffer from the fact that they are difficult to simulate and
self-consistent hydrodynamical models including accretion, pre-explosive burning, explosion, and explosive burning in turbulent
layers are missing. A complete model would also allow to study the actual amount of seed enhancement (if any) in these
environments, through strong s-processing either in the companion star or during accretion. Another open question is the actual p-nucleus contribution of type Ia supernovae during the chemical evolution of the Galaxy. It is still under debate what fraction of type Ia supernova events is actually comprised of the single-degenerate type (with a companion star) and how much double-degenerate events (collision of two WDs) contribute.

\subsection{Thermonuclear burning on the surface of neutron stars}
\label{sec:rp}

Explosive H- and He-burning on the surface of a mass-accreting neutron star can explain a type of bursts observed in galactic point-like X-ray sources on timescales of a few to a few tens of seconds \cite{wall,taam1,taam2,schatz,wooxray}.
Such a \textit{rp-process} involves sequences of proton captures and $\beta$-decays along the proton dripline \cite{wall,schatz,wooxray}.
How far the burning can move up the nuclear chart depends on details of the hydrodynamics (convection) and the amount of accreted
protons. There is a definitive endpoint, however, when the rp-process path runs into the region of Te $\alpha$-emitters \cite{endpoint}.
Therefore, if the rp-process actually runs that far, only p-nuclides with $A<110$ can be reached through decays of very proton-rich
progenitors. This would conveniently allow to only account for the underproduction of the light p-nuclides, which would be primary.

Currently it is unclear whether the produced nuclides can be ejected into the interstellar medium or whether
they are trapped in the gravitational field and eventually only modify the surface composition of the neutron star. The fact that p-nuclides are only produced at the bottom of the burning zone in each burst and have to be quickly moved outwards by strong convection complicates the ejection of significant amounts \cite{rpeject}.

\subsection{Neutrino-wind and accretion disk outflows}
\label{sec:neutrino}
Conditions suited for the synthesis of nuclides in the range of the p-nuclei may also be established by strong neutrino flows acting on hot matter, moving out from ccSN explosions or from accretion disks around compact objects. They would give rise to a primary production of p-nuclides.

For example, very proton-rich conditions were found in the innermost ejected layers of a ccSN due to the interaction of
$\nu_\mathrm{e}$ with the ejected, hot, and dense matter at early times. The hot matter freezes out from nuclear statistical equilibrium
and sequences of proton captures and $\beta$-decays ensue, similar as in the late phase of the rp-process (\sref{sec:rp}) and in the pn-process (\sref{sec:WD}). The nucleosynthesis
timescale is given by the explosion and subsequent freeze-out and is shorter than in the rp-process. The flow towards heavier
elements would be hindered by nuclides with low proton capture Q-value and long $\beta$-decay halflives. Similar as in the pn-process,
(n,p) reactions accelerate the upward flow. The neutrons stem from the reaction $\bar{\nu}_\mathrm{e}+\mathrm{p}\rightarrow
\mathrm{n}+\mathrm{e}^+$. Although at early times the $\bar{\nu}_\mathrm{e}$ flux is small, the sheer number of free protons
guarantees a constant neutron supply. Thus, this process was termed the \textit{$\nu$p-process} by \cite{frohlich}. It was confirmed by the calculations of \cite{pruet,wanajo06}. Like the rp- and pn-processes it may contribute to the light p-nuclides via decays of very proton-rich species, although it is not known up to which mass. The details depend sensitively on the explosion mechanism, the neutrino emission from the proto-neutron star, and the hydrodynamics governing the ejecta motion. Nuclear uncertainties are discussed in \sref{sec:importantreactions} and in \cite{wanajo11,frohrauOmeg,frohnupnuc}.

Accretion disks around compact objects have also been discussed as a viable site for the production of p-nuclei. Such disks can be formed from fallback of material in ccSN or in neutron star mergers. The production strongly depends on the assumed accretion rate which determines the neutrino trapping inside the disk. Fully consistent hydrodynamic simulations have not been performed yet and the models have to make assumptions on accretion rates, disk properties, and especially the amount of ejected material in wind outflows. A production of p-nuclei comparable to the one in the O/Ne-shell $\gamma$-process has been found in parameterized, hot ccSN accretion disks, including the underproduction problem for light p-nuclei \cite{fuji03disk}. Conversely, it was shown that only light p-nuclei up to $^{94}$Mo can be synthesized in parameterized black hole accretion disks under slightly neutron-rich QSE (quasi-statistical equilibrium) conditions at high accretion rates \cite{suroutflow} but that a full $\nu$p-process can ensue at lower accretion rates \cite{kizivat}. The production of the lightest p-nuclei under QSE conditions was previously also discussed in the directly ejected matter in a ccSN \cite{hoff96,mey98,wanajo11}. For a detailed discussion of QSE conditions, see, e.g., \cite{mey98}.

Recently, magnetically driven jets ejected from rapidly rotating, collapsing massive stars have been studied regarding their nucleosynthesis. Obviously, ejection of the produced nuclei is guaranteed in this case. Again, axisymmetric jet simulations show light p-nucleus production up to $^{92}$Mo through QSE proton captures but it was also suggested that $^{113}$In, $^{115}$Sn, and $^{138}$La can be synthesized through fission in very neutron-rich zones of the jet \cite{fuji07jet}. The conclusions regarding the fission products are highly uncertain, however, because they sensitively depend on unknown fission properties, such as fission barriers and fragment distributions. Conversely, full 3D magneto-hydrodynamic simulations of such jets found very neutron-rich conditions, closer to neutron star merger conditions than to the ones found in ccSN neutrino-wind outflows, precluding the production of p-nuclei \cite{winteler}.

\subsection{Galactic chemical evolution}

The yields of different sources add up throughout the history of the Galaxy. Therefore, the evolution of chemical enrichment has to be followed in order to fully understand the distribution of p-nuclides in the Galaxy. This is hampered by several problems. Not only have the frequency, spatial distribution, and yields of the different sources to be known but also their dependence on metallicity (if the p-nucleus production is secondary) and how the products are mixed in the interstellar medium. Clearly, this poses great challenges both for GCE models and for the simulations of each site, including the determination of the ejection of the produced nuclei. For ccSN, it has been shown that the deficiencies in the $\gamma$-process essentially remain also when integrating over a range of stellar masses \cite{rayet95,heger07}. Although this was derived using stars of solar metallicity, it is not expected to be different when including stars at lower metallicity, as the p-production in the $\gamma$-process is secondary and scales with the amount of seed nuclei present in the star.

As pointed out above, in the case of the p-nuclides there are not enough observables to constrain well GCE models or even single production sites, as the isotopic abundances of p-nuclei cannot be separately determined in stellar spectra. This underlines the importance of analyzing meteoritic material as explained in the following \sref{sec:meteorites}. Combining the isotopic information, e.g., of extinct radioactivities, with GCE allows to put severe constraints on the possible production processes. For example, \sref{sec:GCEextinct} shows that processes making $^{92}$Mo but not $^{92}$Nb cannot have contributed much to the composition of the presolar cloud. This would rule out a significant contribution of p-nuclides at and above Mo from very proton-rich environments.

\section{Meteoritic constraints on p-isotope abundances and nucleosynthesis}
\label{sec:meteorites}

Unlike large planetary bodies, which have had their compositions modified by core/mantle segregation and silicate mantle differentiation, meteorites provide a minimally altered record of the composition of the dust present in the solar protoplanetary disk. The study of meteorites has helped define the cosmic abundance of the nuclides, and has revealed the presence of presolar grains and extinct radionuclides in the early solar system. In the past decade, significant progress in mass spectrometric techniques have put new constraints on p-nucleosynthesis, in particular on the roles of rp- and $\nu$p-processes to the production of light p-nuclides ${\rm ^{92}Mo}$, ${\rm ^{94}Mo}$, ${\rm ^{96}Ru}$, and ${\rm ^{98}Ru}$.

\subsection{Cosmic abundances of p-nuclides}
\label{sec:cosmicabuns}

Solar spectroscopy provides critical constraints on the cosmic abundance of key elements such as H, He, C, N and O \cite{2009ARAA..47..481A}. However, remote techniques do not allow to establish the abundance of p-isotopes. A class of meteorites known as CI chondrites (CI stands for carbonaceous chondrite of Ivuna-type) have been shown to contain most elements in proportions that are nearly identical to those measured in the solar spectrum \cite{AnG89,Lod03,2005mcp..book...41P}. For this reason, CI chondrites have been extensively studied to establish the relative abundances and isotopic compositions of heavy elements that cannot be directly measured in the solar photosphere. The virtue of this approach is that CI chondrite specimens (e.g., the Orgueil meteorite that fell in France in 1864 and weighted 14 kg total) can be measured in the laboratory by mass spectrometry, providing highly precise and accurate data. The only elements for which CI-chondrites do not match the present solar composition are Li that is burned in the Sun, and volatile elements (e.g., H, C, N, noble gases) that did not fully condense in solids and were removed from the protoplanetary disk when nebular gas was dissipated. Thus, the relative proportions and isotopic ratios of p-nuclides are well known from meteorite measurements (\tref{tab:solar}).

Silicon is most commonly used to normalize meteoritic to solar photosphere abundances. For the purpose of comparing meteoritic p-isotope abundances with nucleosynthetic model predictions, it is more useful to normalize the data to $^{16}$O. The ratios of p-isotopes to $^{16}$O are still uncertain because of uncertainties in the abundance of O in the solar photosphere, which  was drastically revised downward \cite{2001ApJ...556L..63A,2009ARAA..47..481A}. A difficulty persists, however, as helioseismology requires a larger abundance of O (and other metals) to account for the inferred sound speed at depth, as well as other observables \cite{2008PhR...457..217B}. It is not known at present what is the cause for this discrepancy but the abundances inferred by one of these methods (helioseismology versus solar spectroscopy) must be incorrect. The solar O abundance was revised from 8.93 dex in 1989 \cite{AnG89} to 8.69 dex in 2009 \cite{2009ARAA..47..481A}, i.e., a factor of 1.74. This is significant with respect to p-nucleosynthesis as a factor of two mismatch in predicted to measured abundances of p-nuclides relative to $^{16}$O is often taken as cutoff between success and failure (see \sref{sec:massive}).

In p-isotope abundances $Y_\mathrm{p}$, there is an overall decrease with increasing atomic mass (\fref{fig:comp}), reflecting the decreasing abundance of seed s- and r-process nuclides at higher masses \cite{haya08}. This trend can be fit by an empirical formula $Y_\mathrm{p}/Y_{^{16}\mathrm{O}}=4.86\times10^{8}\times A^{-8.65}$. A second empirical relationship is also found between pairs of p-isotopes separated by two atomic mass units, such as $^{156}$Dy and $^{158}$Dy. For such pairs, the ratio of their abundances increases with atomic mass following approximately, $Y_{A+2}/Y_{A}=0.019\times A^{-1.409}$. An empirical relationship of this kind had been used to estimate the relative abundance of the short-lived p-isotope $^{146}$Sm to the stable $^{144}$Sm \cite{1972Natur.237..447A}.

\subsection{Clues on the synthesis of Mo and Ru p-isotopes from extinct nuclides $^{92}$Nb and $^{146}$Sm}
\label{sec:extinct}

Meteorites contain extinct radionuclides with short half-lives (i.e., relative to the age of the solar system) that were present when the solar system was formed but have now decayed below detection level \cite{2011AREPS..39..351D}. An example of extinct radionuclide is $^{26}$Al ($t_{1/2}=0.7$ Myr), which was present in sufficient quantities when planetesimals were accreted ($^{26}$Al/$^{27}$Al$\sim5\times 10^{-5}$) to induce melting and core segregation. Although extinct nuclides have since long completely decayed, their past presence can be inferred from measurement of isotopic variations in their decay products. Some phases formed with high parent-to-daughter ratios, inducing variations in the daughter nuclide by decay of the parent nuclide. These isotopic variations are most always small and discoveries of new extinct radionuclides depend on developments in mass spectrometry to measure isotopic ratios precisely, as well as sample selection to find phases with high parent-to-daughter ratios.  Two documented extinct radionuclides in meteorites origin from a process also making p-nuclei: $^{92}$Nb and $^{146}$Sm. Other short-lived radionuclides of such origin may have been present in the solar protoplanetary disk when meteorites were formed but have not been found yet.

\begin{figure}
\includegraphics[width=0.7\textwidth]{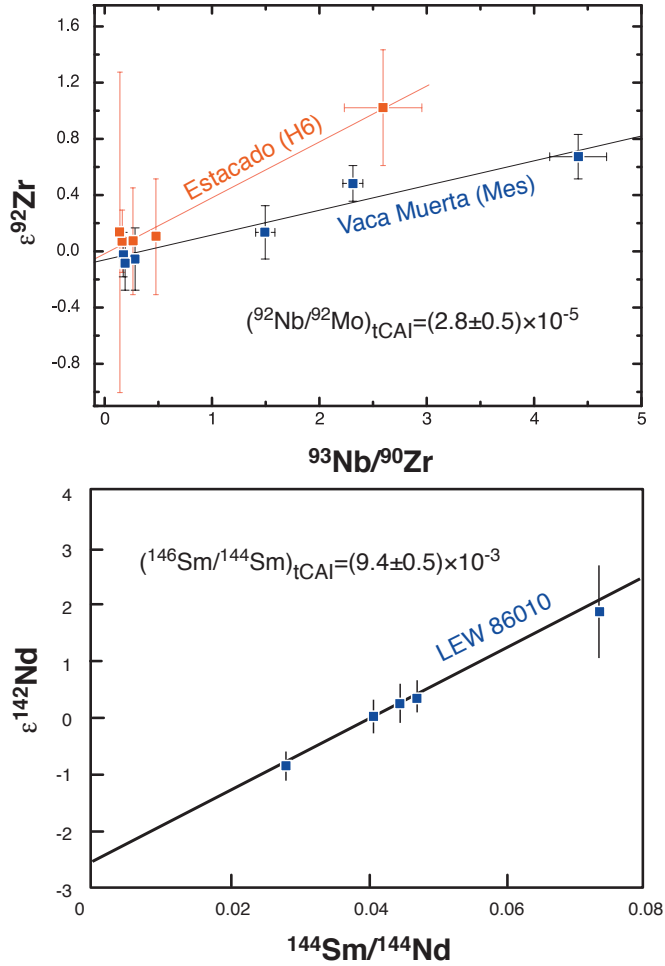}
\caption{Meteoritic evidence for the presence of $^{92}$Nb \cite{schoenbaechler} and $^{146}$Sm \cite{1992GeCoA..56.1673L} in the early solar system (also see \cite{1992GeCoA..56..797P,harper,kino12,2010EPSL.291..172B,1989ApJ...344L..81P}). In both diagrams, the variations in the daughter isotope ($^{92}$Zr or $^{142}$Nd) correlate with the parent-to-daughter ratio (Nb/Zr or Sm/Nd), demonstrating the presence of short-lived nuclides $^{92}$Nb and $^{146}$Sm in meteorites (Estacado is an ordinary chondrite, Vaca Muerta is a mesosiderite, and LEW 86010 is an angrite). The slopes of these correlations give the initial abundances of the extinct radionuclides, see (\ref{eq:SmNd}). The $\epsilon$ notation is explained in \fref{fig:144Sm_142Nd}.
\label{fig:Nb92_Sm146}}
\end{figure}

\subsubsection{Samarium-146 ($t_{1/2}$=68 My)}
It was first suggested by \cite{1972Natur.237..447A} that this nuclide might be present in the solar system and that it could be a useful nuclear cosmochronometer. Early efforts to estimate the solar system initial $^{146}$Sm/$^{144}$Sm ratio yielded uncertain results \cite{1977EPSL..35..273L,1984EPSL..67..137J}. The initial abundance of $^{146}{\rm Sm}$ in meteorites was first solidly established by measuring $^{142}$Nd/$^{144}$Nd (isotope ratio of the decay product of $^{146}$Sm to a stable isotope of Nd) and $^{147}$Sm/$^{144}$Nd (ratio of stable isotopes of Sm and Nd) in achondrite meteorites \cite{1989ApJ...344L..81P,1992GeCoA..56.1673L,1992GeCoA..56..797P}. Many studies followed, confirming the $^{146}$Sm/$^{144}$Sm value, albeit with improved precision and accuracy \cite{2010EPSL.291..172B,kino12}. To estimate the initial abundance of $^{146}$Sm, one has to establish what is known as an extinct radionuclide isochron. If $^{146}$Sm was present when meteorites were formed, one would expect to find a correlation between the ratios $^{142}$Nd/$^{144}$Nd and $^{147}$Sm/$^{144}$Nd measured by mass spectrometry,
\begin{equation}
\left(\frac{^{142}\mathrm{Nd}}{^{144}\mathrm{Nd}} \right)_\mathrm{present}=\left(\frac{^{142}\mathrm{Nd}}{^{144}\mathrm{Nd}} \right)_0+\left(\frac{^{146}\mathrm{Sm}}{^{144}\mathrm{Sm}} \right)_0\times \left(\frac{^{147}\mathrm{Sm}}{^{144}\mathrm{Nd}} \right)_\mathrm{present} \quad,
\label{eq:SmNd}
\end{equation}
where 0 subscripts correspond to the values at the time of formation of the meteorite investigated. The slope of this correlation gives the initial $^{146}$Sm/$^{144}$Sm ratio (\fref{fig:Nb92_Sm146}). All samples formed at the same time from the same reservoir will plot on the same correlation line, which is called an isochron for this reason. While most meteorites formed early, there may have been some delay between collapse of the molecular cloud core that made the Sun and formation of the meteorites investigated. For this reason, one often has to correct initial abundances inferred from meteorite measurements to account for this decay time. Extinct radionuclides with very short half-lives (e.g., $^{36}$Cl, $t_{1/2}$=0.30 Myr) are very sensitive to this correction. Time zero is often taken to be the time of condensation from nebular gas of refractory solids known as Calcium-Aluminum-rich inclusions (CAIs), which must correspond closely to the formation of the solar system \cite{2010Icar..208..455C}. The most important development with $^{146}$Sm in the past several years with respect to p-nucleosynthesis is a drastic revision of its half-life from 104 Myr to 68 Myr \cite{kino12}. Using this new half-life and the most up-to-date meteorite measurements, the initial $^{146}$Sm/$^{144}$Sm ratio at CAI formation is estimated to be $(9.4\pm 0.5)\times 10^{-3}$. Note that both $^{144}$Sm and $^{146}$Sm are p-nuclides.

\subsubsection{Niobium-92 ($t_{1/2}$=34.7 Myr)}
This nuclide, which decays into $^{92}$Zr, was first detected by \cite{harper} who measured the Zr isotopic composition of rutile (TiO$_2$) extracted from a large mass of the Toluca iron meteorite. The correction to the time of CAI formation was uncertain, yet they were able to estimate an initial $^{92}$Nb/$^{93}$Nb ratio of $\sim1\times 10^{-5}$. This result was questioned by several studies that reported higher initial $^{92}$Nb/$^{93}$Nb ratios of $\sim 10^{-3}$ \cite{2000EPSL.184...75S,2000Sci...289.1538M,2000ApJ...536L..49Y}. The later studies might have suffered from unresolved analytical artifacts and the most reliable estimate of the initial $^{92}$Nb/$^{93}$Nb ratio is $(1.6\pm0.3)\times 10^{-5}$ \cite{schoenbaechler}. It is customary in meteoritic studies to normalize the abundance of an extinct radionuclide to the abundance of a stable isotope of the parent nuclide, i.e., $^{93}$Nb (a pure s-process nuclide) for $^{92}$Nb. For the purpose of examining p-nucleosynthesis and comparing meteoritic abundances with predictions from GCE, it is more useful to normalize $^{92}$Nb to a neighbor p-nuclide such as $^{92}$Mo \cite{harper}. The early solar system initial $^{92}$Nb/$^{92}$Mo ratio is thus estimated to be $(2.8\pm 0.5)\times 10^{-5}$.

\subsubsection{Technetium-97 ($t_{1/2}$=4.21 Myr) and Technetium-98 ($t_{1/2}$=4.2 Myr)}
These two nuclides have the same origin as p-nuclides and may have been present at the birth of the solar system but they have not been detected yet (i.e., only upper-limits could be derived). This stems from several difficulties; their expected abundances in meteorites are low, little fractionation is expected between parent and daughter nuclides (i.e., Tc/Mo and Tc/Ru ratios), and no stable isotope of the parent nuclide exists (one has to rely on a proxy element such as Re). Available Mo and Ru isotopic analyses yield the following constraints on $^{97}$Tc and $^{98}$Tc abundances at CAI formation: $^{97}$Tc/$^{92}$Mo$<3\times 10^{-6}$ (or $^{97}$Tc/$^{98}$Ru$<4\times 10^{-4}$) \cite{2002ApJ...565..640D} and $^{98}$Tc/$^{98}$Ru$<2\times 10^{-5}$ \cite{Becker200343}. Note that $^{92}$Mo and $^{98}$Ru are both pure p-nuclides.

\begin{table}
\caption{Extinct p-radionuclides in the early solar system; updated from \cite{dauphas, 2011AREPS..39..351D}.\label{table:extrad}}
\footnotesize\rm
\begin{tabular}{lcccc}
\br
Ratio & $t_{1/2}$ (Myr) & ${\rm R_{meteorite}}$ & ${\rm R_{production}}$ & ${\rm R_{ISM}}$ 4.5 Ga (model) \\
\mr
$^{97}{\rm Tc}/^{98}{\rm Ru}$ & 4.21 & $<4\times 10^{-4}$ & $(4.4\pm 1.3)\times 10^{-2}$ & $(1.1\pm 0.3)\times 10^{-4}$ \\
$^{98}{\rm Tc}/^{98}{\rm Ru}$ & 4.2 & $<8\times 10^{-5}$ & $(7.3\pm 2.5)\times 10^{-3}$ & $(1.8\pm 0.6)\times 10^{-5}$ \\
$^{92}{\rm Nb}/^{92}{\rm Mo}$ & 34.7 & $(2.8\pm 0.5)\times 10^{-5}$ & $(1.5\pm 0.6)\times 10^{-3}$ & $(3.0\pm 1.3)\times 10^{-5}$ \\
$^{146}{\rm Sm}/^{144}{\rm Sm}$ & 68 & $(9.4\pm 0.5)\times 10^{-3}$ & $(1.8\pm 0.6)\times 10^{-1}$ & $(7.1\pm 2.4)\times 10^{-3}$ \\
\br
\end{tabular}
\end{table}

\subsubsection{Galactic chemical evolution}
\label{sec:GCEextinct}
The abundances of extinct p-radionuclides in the early solar system can be compared with predictions from models of the chemical evolution of the Galaxy. Meteoriticists have often used simple closed-box GCE or uniform production models, which predict that the ratio of an extinct radionuclide in the interstellar medium $R_\mathrm{ISM}$ (e.g., $^{146}$Sm/$^{144}$Sm) should be related to the production ratio $R_\mathrm{production}$ through $R_\mathrm{ISM}=R_\mathrm{production}\tau /T_\mathrm{G}$, where $\tau$ is the mean-life of the nuclide ($t_{1/2}/ln 2$) and $T_\mathrm{G}$ is the time elapsed between Milky Way formation and solar system birth. The closed box model however fails to reproduce first order astronomical observables such as the metallicity distribution of G-dwarfs \cite{1997nceg.book.P,2004AA...418..989N}. Open-box models involving growth of the Galaxy by infall of low-metallicity gas are more realistic. In such models, the abundance of a short-lived p-nuclide in the average interstellar medium (ISM) at the birth of the solar system becomes
\begin{equation}
R_\mathrm{ISM}/R_\mathrm{production}=(k+2)\tau /T_\mathrm{G} \quad,
\label{eq:gce}
\end{equation}
where $k$ is a constant \cite{1988MNRAS.234....1C,dauphas,2009GeCoA..73.4922H}. Early work estimated its value to be between 1 and 3 \cite{1988MNRAS.234....1C}. A value of $k=1.7\pm 0.4$ was obtained using a non-linear infall GCE model constrained by recent astronomical observations \cite{dauphas}. The time elapsed between the formation of the Galaxy and the formation of the solar system can be estimated, using the same infall GCE model and the U/Th ratio, to be $14.5-4.5\simeq 10$ Gyr \cite{2005Natur.435.1203D}. Therefore, the only unknown in the above equation is the production ratio, which can be estimated using nucleosynthesis calculations. However, a complication remains to compare the predicted abundances and the measured ones as (\ref{eq:gce}) only gives the predicted abundance averaged over all ISM reservoirs while the solar system must have formed from material that was partially isolated from fresh nucleosynthetic inputs. The earliest models used a free-decay interval to account for this isolation period, so the ratio in the early solar system would be the ratio in the ISM decreased by some free decay, $R_\mathrm{ESS}=R_\mathrm{ISM}e^{-(\Delta _\mathrm{free-decay}/\tau )}$. More realistically, isolation from fresh nucleosynthetic inputs was not complete. To address this issue, \cite{1983ApJ...268..381C} devised a three-phase ISM mixing model between (1) dense molecular clouds from which stellar systems form, (2) large H-I clouds, and (3) smaller H-I clouds that can be evaporated by supernova shocks. Using the same parameters as those used by \cite{1983ApJ...268..381C}, the expected ratio in the molecular cloud core from which the Sun was born is,
\begin{equation}
R_\mathrm{ESS}=R_\mathrm{ISM}/[1+1.5T_\mathrm{mix}/\tau+0.4(T_\mathrm{mix}/\tau)^2],
\label{eq:ism3}
\end{equation}
where $\tau$ is the nuclide mean-life and $T_\mathrm{mix}$ is the 3-phase ISM mixing timescale. Realistic values for the mixing timescale are probably on the order of 10-100 Myr. This is the only free parameter in the model and there are two extinct p-radionuclides to uniquely constrain its value.

\begin{figure}
\includegraphics[width=\textwidth]{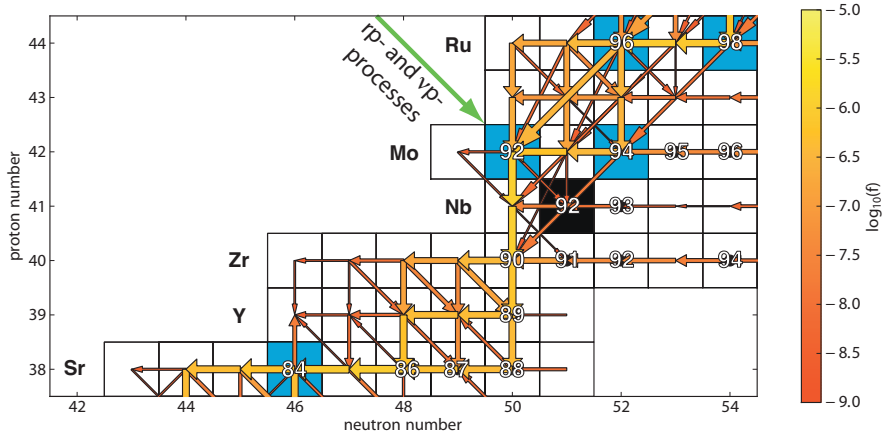}
\caption{Reaction flows in the $\gamma$-process producing $^{92}$Mo and the extinct radionuclide $^{92}$Nb. Size and shading of the arrows show the magnitude of the reaction flows $f$ on a logarithmic scale, nominal p-nuclides are shown as filled squares. The nuclide $^{92}$Nb can be produced by the $\gamma$-process but it cannot be produced by the rp- and $\nu$p-processes (or any process involving a decay of proton-rich nuclei contributing to $^{92}$Mo) as it is shielded from contributions by these processes by the stable $^{92}$Mo. The presence of $^{92}$Nb in meteorites indicates that proton-rich processes did not contribute much to the nucleosynthesis of Mo and Ru p-isotopes \cite{dauphas}.
\label{fig:Nb92_shielding}}
\end{figure}

It was shown by \cite{dauphas} how $^{92}$Nb can put important constraints on the role of the rp- and $\nu$p-processes in the nucleosynthesis of Mo and Ru p-isotopes. The calculations given below are updated with more recent data (\tref{table:extrad}). As discussed in \sref{sec:sites}, the site and exact nuclear pathway for the nucleosynthesis of light p-nuclides is still a matter of debate. The production ratios of extinct p-nuclides in the $\gamma$-process are $^{97}$Tc/$^{98}$Ru=$(4.4\pm 1.3)\times 10^{-2}$, $^{98}$Tc/$^{98}$Ru=$(7.3\pm 2.5)\times 10^{-3}$, $^{92}{\rm Nb}/^{92}{\rm Mo}=(1.5\pm 0.6)\times 10^{-3}$, and $^{146}{\rm Sm}/^{144}{\rm Sm}=(1.8\pm 0.6)\times 10^{-1}$ \cite{rhhw02}. The corresponding ratios in the average ISM 4.5 Gyr ago, see \eref{eq:gce}, are $^{97}$Tc/$^{98}$Ru=$(1.1\pm 0.4)\times 10^{-4}$, $^{98}$Tc/$^{98}$Ru=$(1.8\pm 0.7)\times 10^{-5}$, $^{92}{\rm Nb}/^{92}{\rm Mo}=(3.0\pm 1.3)\times 10^{-5}$, and $^{146}{\rm Sm}/^{144}{\rm Sm}=(7.1\pm 2.5)\times 10^{-3}$. The upper limits on $^{97}$Tc and $^{98}$Tc from meteorite measurements, $^{97}{\rm Tc}/^{98}{\rm Ru}<4\times 10^{-4}$ and $^{98}{\rm Tc}/^{98}{\rm Ru}<8\times 10^{-5}$, are consistent with the inferred ISM ratio from galactic chemical evolution and $\gamma$-process modeling. The predicted values for $^{92}$Nb and $^{146}$Sm are also in excellent agreement with meteorite data; in meteorites $^{92}{\rm Nb}/^{92}{\rm Mo}=(2.8\pm 0.5)\times 10^{-5}$ while we predict in the ISM $^{92}{\rm Nb}/^{92}{\rm Mo}=(3.0\pm 1.3)\times 10^{-5}$; in meteorites $^{146}{\rm Sm}/^{144}{\rm Sm}=(9.4\pm 0.5)\times 10^{-3}$ while we predict in the ISM $^{146}{\rm Sm}/^{144}{\rm Sm}=(7.1\pm 2.5)\times 10^{-3}$ (\tref{table:extrad}). We are comparing here the abundances measured in meteorites with those in the average ISM 4.5 Gyr ago predicted from GCE modeling. Partial isolation of ISM material from fresh nucleosynthetic inputs can be taken into account using \eref{eq:ism3}. We find that the 3-phase ISM mixing timescale must be small, less than about 20 to 30 Myr. Otherwise the predicted abundances would not match the measured values in meteorites. For example, adopting $T_\mathrm{mix}=30$ Myr would decrease the expected ratios in the early solar system to $^{92}{\rm Nb}/^{92}{\rm Mo}=1.5\times 10^{-5}$ and $^{146}{\rm Sm}/^{144}{\rm Sm}=4.7\times 10^{-3}$; factors of 2 lower than meteorite values. To summarize, $^{92}{\rm Nb}/^{92}{\rm Mo}$ and $^{146}{\rm Sm}/^{144}{\rm Sm}$ production ratios from the $\gamma$-process can reproduce extremely well the measured abundances of these extinct radionuclides in the early solar system. However, it is well documented that state-of-the-art $\gamma$-process calculations underproduce $^{92}$Mo, $^{94}$Mo, $^{96}$Ru and $^{98}$Ru (\sref{sec:massive}), isotopes that are in similar abundance to s-process isotopes of the same elements. What if other processes, such as the rp-process or the $\nu$p-process, had produced the missing p-isotopes of Mo and Ru? This implies that $\sim 10$\% of $^{92}$Mo would be produced by the $\gamma$-process while $\sim 90$\% would be produced by the rp- or the $\nu$p-process. However, $^{92}$Nb cannot be produced in these processes because it is shielded from a contribution by proton-rich progenitors during freeze-out by the stable $^{92}$Mo (\fref{fig:Nb92_shielding}). If 90\% of $^{92}$Mo had been made by  processes that cannot produce $^{92}$Nb, this would have decreased the effective ${\rm ^{92}Nb/^{92}Mo}$ production ratio by a factor of $\sim10$. The predicted ${\rm ^{92}Nb/^{92}Mo}$ ratio in the early solar system would also be lower than the ratio measured in meteorites by a factor of 10. A significant contribution to Mo-Ru by any process that does not make $^{92}$Nb can therefore be excluded \cite{dauphas}. Independently, another study concluded that a significant contribution to Mo p-isotopes from a $\nu$p-process was unlikely, based on the inferred $^{94}{\rm Mo}/^{92}{\rm Mo}$ isotope ratio in this process \cite{2009ApJ...690L.135F}.

The calculation outlined above can also be done the other way around, by using the $^{92}$Nb/$^{92}$Mo ratio in meteorites to calculate its production ratio.  The ${\rm ^{146}Sm/^{144}Sm}$ ratio indicates that the 3-phase ISM mixing timescale must be small; for the purpose of simplicity we assume that it is zero. Taking a non-zero value would result in a higher inferred ${\rm ^{92}Nb/^{92}Mo}$ production ratio, therefore our estimate corresponds to a conservative upper limit. We find the ${\rm ^{92}Nb/^{92}Mo}$ production ratio to be $0.0015^{+0.0012}_{-0.0009}$ (or higher). Models should take this value as a fundamental constraint on p-nucleosynthesis in the Mo-Ru mass region. For instance, it remains to be tested whether the recently reported $\gamma$-process predictions for SNIa \cite{travWD}, which do not experience a Mo-Ru underproduction problem, can reproduce the abundances of $^{92}$Nb and $^{146}$Sm in meteorites.

\subsection{p-Isotope anomalies in meteorites}
\label{sec:anomalies}

When the solar system was formed, temperatures in the inner part of protosolar nebula were sufficiently high to induce vaporization of the dust present. However, some presolar grains survived heating and these grains can be retrieved from primitive meteorites. They are found by measuring their isotopic compositions, which are non-solar, indicating that the grains condensed in the outflows of stars that lived before the solar system was formed. Several comprehensive reviews have been published on this topic \cite{1993Metic..28..490A,2004ARAA..42...39C,2005ChEG...65...93L}. A variety of phases from various types of stars have been documented. Six types of grains have a supernova origin; nanodiamonds, silicon carbide (SiC) of type X, low-density graphite, silicon nitride (Si$_3$N$_4$), a small number of presolar corundum (Al$_2$O$_3$) grains, and nanospinels. In the case of nanospinels, it is still unknown whether the grains condensed in the outflows of ccSN or SNIa as these grains are characterized by large excesses in the neutron-rich isotope $^{54}$Cr, which can be produced by both kinds of stars \cite{2010ApJ...720.1577D,2011GeCoA..75..629Q}. Because heavy elements are present at low concentrations and presolar grains are small (i.e., up to a few tens of micrometers but most often much less than that), it is analytically challenging to measure their isotopic compositions. However, the development of the technique of Resonant Ionization Mass Spectrometry (RIMS) has allowed cosmochemists to measure the isotopic composition of trace heavy elements in single presolar grains \cite{Levine200936}. The advantages of this technique over Secondary Ionization Mass Spectrometry (SIMS) for this type of measurements are that it selectively ionizes the element of interest using tunable lasers, so isobaric interferences are almost non-existent, and it has high yield, meaning that a significant fraction of the atoms in the sample make it to the detector. The Sr, Zr, Mo, and Ba isotopic compositions of presolar X-type SiC grains of supernova origin were measured by \cite{2000LPI....31.1917P}. Notably, they reported large excesses of $^{95}$Mo and $^{97}$Mo relative to other Mo isotopes. It was shown that these signatures could be explained by an episode of neutron-burst that took place in a He-shell during passage of the supernova shock, which produced $^{95}$Y, $^{95}$Zr, and $^{97}$Zr radioactive progenitors that rapidly decayed into $^{95}$Mo and $^{97}$Mo \cite{2000ApJ...540L..49M}. This is expected to occur in a very localized region of the ccSN and the reason why the grains record such a signature is not understood but it must reflect a selection bias in dust formation/preservation.

To this day, no $\gamma$-process signature has been documented in heavy trace elements in presolar grains of supernova origin. However, the types and numbers of grains that have been studied by RIMS are limited and further work is required to document isotope signatures in grains of supernova origin. The next generation of RIMS instruments may have the capability to tackle this question in a more systematic manner \cite{2011LPI....42.1995S}.

\begin{figure}
\includegraphics[width=0.8\textwidth]{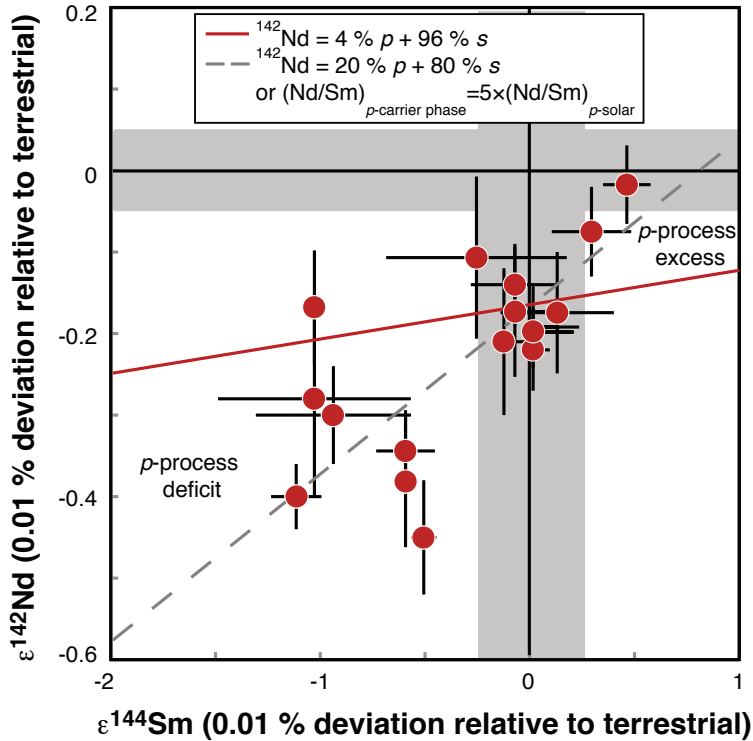}
\caption{Correlation in bulk chondrites between $^{142}$Nd and $^{144}$Sm isotopic variations \cite{Gannoun22042011}. The correlation probably corresponds to mixing between a solar component and a presolar end-member enriched in p-isotopes. The slope of the correlation cannot be explained by assuming a solar mixture of 4\% p and 96\% s for $^{142}{\rm Nd}$ in the presolar end-member. Instead, the grains that carry those anomalies may have a fractionated Nd/Sm p-isotope contribution ratio resulting from chemical fractionation of those two elements upon condensation in a circumstellar environment.
\label{fig:144Sm_142Nd}}
\end{figure}

Isotopic anomalies for heavy elements can also be present in macroscopic objects, in some cases reaching the scale of bulk planets \cite{2002ApJ...565..640D}. The isotopic anomalies are much more subdued than those documented in presolar grains but these anomalies can be measured with other instruments (TIMS: Thermal Ionization Mass Spectrometers, and MC-ICPMS: Multi-Collector Inductively Coupled Plasma Mass Spectrometers) at greater precision, reaching a few parts per million on isotopic ratios (i.e., $\sim$0.001\% for TIMS or MC-ICPMS versus $\sim$1\% for SIMS or RIMS). Isotopic variations (1 $\epsilon$ deficit, where $\epsilon =0.01$\%) in the isotopic abundance of the p-isotope $^{144}$Sm in bulk meteorites of carbonaceous type are documented \cite{2006Sci...314..806A}. Similar results were found using a more aggressive sample digestion technique known as flux fusion, ensuring that the $^{144}$Sm isotopic variations did not result from incomplete digestion of the samples \cite{2007Sci...316.1175C}. Variations in $^{144}$Sm were also resolved by \cite{Gannoun22042011} and a correlation with isotopic variations in $^{142}$Nd was found (\fref{fig:144Sm_142Nd}). This is significant because isotopic variations in $^{142}$Nd in planetary materials have been ascribed to decay of $^{146}$Sm, with important implications on planetary differentiation processes in the early solar system \cite{2005Sci...309..576B,2007Natur.450..525D,2008Natur.452..336C}.

It was calculated that in order to explain the $^{142}$Nd-$^{144}$Sm correlation by variations in the $\gamma$-process component, a 20\% $\gamma$-process contribution to $^{142}$Nd would be needed \cite{Gannoun22042011}. This is at odds with the predominant s-process nature of this nuclide. Such a high contribution can most likely be ruled out based on several lines of evidence. Firstly, the s-process in this mass region is well understood and reproduces well the cosmic abundance of $^{142}$Nd \cite{arl99}. If anything, the s-process produces too much of this isotope rather than too little. Secondly, a 20\% $\gamma$-process contribution would plot far off the empirical trend of abundance versus mass for p-isotopes. The empirical relationship gives $^{142}{\rm (Nd}-p)/^{16}{\rm O}=0.00012$, while a 20\% $\gamma$-process contribution would mean $^{142}{\rm (Nd}-p)/^{16}{\rm O}=0.003$, i.e., a factor of $\sim 30$ higher than predicted. Thirdly, the $^{142}$Nd produced in the $\gamma$-process comes from the $\alpha$-decay of $^{146}$Sm. The $^{146}$Sm/$^{144}$Sm ratio in the $\gamma$-process depends strongly on the relative strength of the reactions $^{148}$Gd($\gamma$,n)$^{147}$Gd and $^{148}$Gd($\gamma$,$\alpha$)$^{144}$Sm but is independent of the seed nuclei. Nuclear physics experiments (see \sref{sec:sensi}) rule out a ratio which would allow to contribute significantly to $^{142}$Nd.

If confirmed, the cause for the correlation between $^{142}$Nd and $^{144}$Sm remains to be explained. A likely interpretation is that the presolar phase controlling the $\gamma$-process Sm-Nd isotopic anomalies in planetary materials does not contain these elements in solar proportions. Indeed, chemical fractionation of Sm and Nd during grain condensation can produce a correlation with a steeper slope than expected (see, e.g., \cite{2004EPSL.226..465D} for a similar discussion in the context of correlated Mo-Ru isotope anomalies). Isotope anomalies also have been reported for the p-isotopes $^{184}$Os \cite{2009EPSL.277..334R} and $^{180}$W \cite{2010MPSA..73.5116S} but further work is needed before these variations can be understood.

\section{Reaction rates and reaction mechanisms}
\label{sec:definitions}

\subsection{Introduction}
The previous sections discussed the astrophysical sites and observational constraints for the production of p-nuclei. Important for modelling nucleosynthesis is a reliable foundation in the nuclear physics required to predict astrophysical reaction rates. It is important to note that there are fundamental differences between experimental and theoretical studies of reactions on intermediate and heavy nuclei, as appearing in p-nucleosynthesis, and reactions on lighter nuclei up to Si. Different nuclear properties are important at low mass than at high mass and new challenges arise, due to the higher nuclear level density (NLD) and the higher Coulomb barriers encountered in heavier nuclei. Experimental and theoretical approaches well suited for reactions on lighter nuclei are not directly applicable to heavy nuclei involved in explosive nucleosynthesis. Because of the high Coulomb barriers and the higher temperatures, giving rise to pronounced stellar effects not encountered in light nuclei to such an extent, experimental investigations are not able to completely determine the astrophysical reaction rate in most cases and have to be supplemented by theory. In the remainder of this review, we discuss the challenges arising and the methods currently available to address them in the quest for providing reliable and accurate reaction rates to study the origin of the p-nuclei.

\subsection{Definition of the stellar reaction rate}
\label{sec:ratedefs}
The astrophysical reaction rate $r^*$ for an interaction between two particles or nuclei in a stellar environment is obtained by folding the Maxwell-Boltzmann energy distribution $\Phi$, describing the thermal center-of-mass (c.m.) motion of the interacting nuclei in a plasma of temperature $T$, with a quantity $\sigma^*$ which is related to the probability that the reaction occurs, and by multiplying the result with the number densities $n_a$, $n_A$, i.e., the number of interacting particles in a unit volume,
\begin{equation}
\label{eq:rate}
r^*=\frac{n_a n_A}{1+\delta_{aA}} \int_0^\infty \sigma^*(E) \Phi(E,T)\;dE = \frac{n_a n_A}{1+\delta_{aA}} R^* \quad.
\end{equation}

The stellar reactivity (or rate per particle pair) is denoted by $R^*$. To avoid double counting of pairs, the Kronecker symbol $\delta_{aA}$ is introduced. It is unity when the nuclei $a$ and $A$ are the same and zero otherwise. The asterisk superscript indicates \textit{stellar} quantities, i.e., including the effect of thermal population of excited nuclear states in a stellar plasma. Depending on temperature and nuclear level structure, a fraction of nuclei is present in an excited state in the plasma, instead of being in the ground state (g.s.). This has to be considered when calculating the interactions and rates
by using the \textit{stellar cross section} $\sigma^*$ \cite{raureview,sensipaper}
\begin{equation}
\sigma^*(E,T)=\frac{\sigma^\mathrm{eff}(E)}{G_0(T)}=\frac{1}{G_0(T)} \sum_i \sum_j \frac{2J_i+1}{2J_0+1} W_i \sigma^{i \rightarrow j}(E-E_i)
\quad,
\label{eq:effcs}
\end{equation}
which involves a weighted sum over transitions from all initial excited states $i$, with spin $J_i$ and energy $E_i$, up to the interaction energy $E$, leading to all accessible final states $j$.
As usual, cross sections for individual transitions $\sigma^{i \rightarrow j}$ are zero for negative energies. The quantity
\begin{equation}
G_0=\sum_i {\frac{\left( 2J_i + 1 \right) \exp \left(-\frac{E_i}{kT}\right)}{2J_0+1}}
\end{equation}
is nothing else than the partition function of the target nucleus normalized to the ground state spin, and $\sigma^\mathrm{eff}$ is usually called effective cross section \cite{holmes}.
The weights
\begin{equation}
\label{eq:weights}
W_i(E)=\frac{E-E_i}{E}=1-\frac{E_i}{E}
\end{equation}
of the contributions of excited states to the effective cross section depend linearly on the excitation energy because Maxwell-Boltzmann energy-distributed projectiles act on each excited state \cite{raureview}. The relevant energies $E$ are given by the energy range contributing most to the integral in equation (\ref{eq:rate}), see \sref{sec:energywindows}.

Although not discussed in further detail here, it is worth mentioning that also weak interactions and decays are affected by the thermal population of excited states. This is important because it changes the decay lifetimes of nuclei such as, e.g., $^{92}$Nb and $^{180}$Ta, while temperatures are still high during nucleosynthesis.

It is straightforward to show that the stellar reactivities defined with the effective cross section -- connecting all initial states to all final states -- also obey reciprocity, just as the individual transitions $\sigma^{i \rightarrow j}$ do \cite{raureview,fowler,holmes}.  The reciprocity relations for a reaction $A(a,b)B$ and its reverse reaction $B(b,a)A$ are \cite{raureview,holmes,ilibook}
\begin{equation}
\label{eq:revrate}
\frac{R^*_{Bb}}{R^*_{Aa}}=\frac{(2J_0^A+1) (2J_a+1)}{(2J_0^B+1) (2J_b+1)} \frac{G^A_0(T)}{G^B_0(T)} \left( \frac{m_{Aa}}{m_{Bb}}\right) ^{3/2}e^{-Q_{Aa}/(kT)}
\end{equation}
when $a$, $b$ are particles, and
\begin{equation}
\frac{R_{B\gamma}^*}{R^*_{Aa}}=\frac{(2J_0^A+1) (2J_a+1)}{(2J_0^B+1)} \frac{G^A_0(T)}{G^B_0(T)}
\left( \frac{m_{Aa}kT}{2\pi \hbar^2}\right)^{3/2} e^{-Q_{Aa}/(kT)} \label{eq:revphoto}
\end{equation}
when $b$ is a photon. The normalized partition functions $G^A_0$ and $G^B_0$ of the nuclei $A$ and $B$, respectively, are defined as before.
The photodisintegration reactivity
\begin{equation}
\label{eq:photoreac}
R_{B\gamma}^*=\int_0^\infty \sigma_\gamma^*(E) \Phi_\mathrm{Planck}(E,T)\;dE
\end{equation}
includes a stellar photodisintegration cross section $\sigma_\gamma^*(E)$ defined in complete analogy to the stellar cross section $\sigma^*$ in equation (\ref{eq:effcs}).
In relating the capture rate of $A(a,\gamma)B$ to the photodisintegration rate $\lambda_{B\gamma}^*=n_B R_{B\gamma}^*$, however, it has to be assumed that the denominator $\exp(E/(kT))-1$ of the Planck distribution $\Phi_\mathrm{Planck}$ for photons appearing in equation (\ref{eq:photoreac}) can be replaced by the one from the Maxwell-Boltzmann distribution $\exp(E/(kT))$. The validity of this approximation has been investigated independently several times \cite{wardfowler,sm144lett,ilibook,raureview,mathewsreci}.
The contributions to the integral in (\ref{eq:photoreac}) have to be negligible at the low energies where $\Phi$ and $\Phi_\mathrm{Planck}$ differ considerably.
This is ensured by either a sufficiently large and positive $Q_{Aa}$, which causes the integration over the Planck distribution to start not at zero energy but rather at a sufficiently large threshold energy, or by vanishing effective cross sections at low energy due to, e.g., a Coulomb barrier. It turns out that the change in the denominator is a good approximation for the calculation of the rate integrals, especially for the temperatures and reactions encountered in p-nucleus production.

The photodisintegration rate of a nucleus only depends on plasma temperature $T$ whereas rates of reactions with particles in the entrance channel depend on $T$ and the number density $n_a$ of the projectile. The number density $n_a$ scales with the plasma matter density $\rho$ and the abundance of the projectile $Y_a$. A variation of plasma density, on the other hand, will not affect photodisintegration of a nucleus. Therefore the ratio of temperature and density sets the ratio of photodisintegration to capture rates (for a fixed Q-value)
\begin{equation}
\label{eq:caprate}
\frac{r^*_{B\gamma}}{r^*_{Aa}} \propto \frac{Y_B}{Y_A} \frac{T^{3/2}e^{-Q_\mathrm{Aa}/(kT)}}{N_A \rho Y_a}\quad.
\end{equation}
This is the reason why at conditions with large $\rho Y_a$ (e.g., high proton densities), capture reactions can balance photodisintegrations even at high $T$, leading to equilibrium values for the abundances $Y_A$, $Y_B$.

It has to be emphasized that \eref{eq:revrate} and \eref{eq:revphoto} only hold when using the stellar and effective cross sections $\sigma^*$ and $\sigma^\mathrm{eff}$, respectively, in the calculation of the reactivity, \textit{not} with the usual laboratory cross sections
\begin{equation}
\label{eq:labcs}
\sigma_i=\sum_j \sigma^{i \rightarrow j},
\end{equation}
where $i=0$ if the target is in the g.s. In reactions involving intermediate and heavy nuclei at high plasma temperatures, transitions on excited states of the target will be significant in most cases (see \sref{sec:expgeneral}).

\subsection{Reaction mechanisms}

\subsubsection{Relevant energy windows}
\label{sec:energywindows}

In order to decide which type of reactions have to be considered in the prediction of the stellar cross sections, the relative interaction energies $E$ appearing in the astrophysical plasma have to be known.
These energies and the location of the maximum of the integrand can be found in \cite{energywindows}. They have to be derived by inspection of the actual integrand in (\ref{eq:rate}). The simple, frequently used formula for estimating the Gamow window (see, e.g., \cite{clayton,cauldrons}) from the charges of projectile and target is only applicable when the energy dependence of the cross section is fully given by the entrance channel. This has been found inadequate for many reactions except those involving light nuclei \cite{energywindows,ilibook,newton} and thus is not applicable for the reactions appearing in the production of the p-nuclei.

The effective temperature range for the formation of p-nuclei is $1.5\leq T\leq 3.5$ when considering both a $\gamma$-process and a proton-rich environment, such as required for the $\nu$p-process. This translates into relative interaction energies of about $1.5-3.5$ MeV for protons (at the high end of the temperature range and for the light p-nuclei) in the $\gamma$-process. The relevant energies shift more strongly with charge for $\alpha$-particles (at the low end of the temperature range and for the heavier p-nuclei) and energies are in the ranges of, e.g., $6-9$, $7-10$, and $9-11$ MeV, respectively, at 2 GK and for charges $Z\approx 62$, $Z\approx 74$, and $Z\approx 82$, respectively. Proton captures are only of limited importance in the $\nu$p-process because a (p,$\gamma$)-($\gamma$,p) equilibrium is upheld most of the time \cite{frohlich,raufrohOmeg,frohnupnuc}. Non-equilibrium proton captures at the low end of the temperature range yield interaction energies of about $1-2$ MeV. Crucial for $\nu$p-processing is the acting of (n,p) reactions at all temperatures. This results in neutron energies up to 1 MeV in (n,p). Of further importance in both $\gamma$-process and proton-rich nucleosynthesis are (n,$\gamma$) reactions and their inverses, also for the whole temperature range. This implies neutron energies of up to 1400 keV for neutron captures.

\subsubsection{Compound reactions}
\label{sec:compound}
The compound formation and excitation energy $E_\mathrm{form}=E+Q$ depends on the interaction energy $E$ and the reaction Q-value $Q$.
The nuclear level density (NLD) integrated over the compound excitation energies obtained with the relevant energy ranges from \sref{sec:energywindows} determines which reaction mechanism will dominate. With only few levels within the energy range appearing in the integration of the reactivity, individual resonances have to be considered. With a high NLD, individual resonances are not resolved anymore and the sum over individual resonances can be replaced by averaged resonances which should reproduce the integrated properties (widths) of all resonances at all energies \cite{desc,raureview}. This is the statistical Hauser-Feshbach model of compound nuclear reactions. Its astrophysical application has been discussed in many recent publications \cite{raureview,holmes,nonsmoker,RTK96,sanantonio} and we do not want to repeat all the details here.

Averaged properties can be predicted with higher accuracy than individual resonances in most cases. It is important to note, however, that there is a difference in the application of the statistical model to the calculation of cross sections and to reactivities. Since the calculation of the reactivity involves an additional averaging over the Maxwell-Boltzmann distribution, it is more ``forgiving'' when fluctuations in the cross sections are not reproduced as long as the average value is correct across the relevant energy range. Therefore, the Hauser-Feshbach model can be applied at lower NLD at compound formation energy for the calculation of the rate than for the calculation of the cross section. Put differently, even when fluctuations due to resonances are seen in the experimental cross sections but not in the Hauser-Feshbach predictions, the model may still yield a reliable rate provided it gives the same Maxwell-Boltzmann average. See \sref{sec:expgeneral} for further details.

Combining the above with the typical interaction energies quoted in \sref{sec:energywindows}, it is easy to see that the statistical model is applicable for all reactions involved in p-nucleosynthesis (see sections \ref{sec:sites} and \ref{sec:importantreactions} for the important reactions). Problems may arise for (p,$\gamma$) and ($\gamma$,p) on nuclei with magic proton number or very proton-rich nuclei. The proton separation energy decreases for proton-richer isotopes and this shifts the compound formation energy to regions of lower NLD. Both, the $\gamma$- and the $\nu$p-process, however, do not involve nuclei close to the driplines where it is known that the statistical model cannot be applied anymore \cite{RTK96,schatz,rembges}. Moreover, there is a (p,$\gamma$)-($\gamma$,p) equilibrium in the $\nu$p-process, diminishing the importance of individual reactions \cite{raufrohOmeg,frohnupnuc}. This leaves proton reactions at charge $Z=28$ and $Z=50$, the latter only being important in the $\gamma$-process, as possible cases where the statistical model may not be fully applicable at all temperatures and individual resonances may have to be considered.

Reactions involving $\alpha$-particles should not be problematic for the statistical model even though they are important in a region of $\alpha$-emitters with negative ($\alpha$,$\gamma$) Q-values (see \sref{sec:importantreactions}), due to their much higher astrophysically relevant interaction energies.

\subsubsection{Direct reactions}
\label{sec:direct}
It is well known that direct reactions are important at interaction energies above several tens of MeV because of the reduced compound formation probability \cite{desc,sat83,gh92}. In light nuclei with widely spaced energy levels, direct reactions can give important contributions to the cross sections between resonances. In the p-nucleus mass range, however, direct reactions are not expected to contribute to the astrophysical reaction rates due to the generally higher NLD at the compound formation energy $E_\mathrm{form}$, as discussed in the previous \sref{sec:compound}.

It has been realized recently, however, that low-energy direct inelastic scattering has to be included in the analysis of ($\alpha$,$\gamma$) laboratory cross sections \cite{raucoulexproc,raucoulexlett}. With high Coulomb barriers, Coulomb excitation \cite{alder} can become non-negligible at low interaction energies, modifying the experimental yield. Data is scarce close to the astrophysically relevant energy region (see \sref{sec:charged}) but an overprediction of $\alpha$-induced cross sections relative to the experimental values was observed for several cases while standard predictions worked well for others. Many attempts to consistently describe the data with modified global $\alpha$+nucleus optical potentials have failed. Accounting for Coulomb excitation at low energies can explain the deviations at least partially and will pave the way for an improved global understanding of reaction rates involving $\alpha$-particles. \Sref{sec:sensi} gives an example of how strong the Coulomb excitation effect can be. Nuclear excitation is negligible at the low energies relevant for astrophysics as the Coulomb scattering takes place far outside the nucleus.

Obviously, there is no Coulomb excitation for ($\gamma$,$\alpha$) rates which are needed for the $\gamma$-process. Therefore, an optical potential describing the $\alpha$-emission has to be used. Due to detailed balance considerations, this has to be similar to one describing compound formation in the absence of Coulomb excitation. Nevertheless, even when using charged-particle induced reactions, direct reactions not leading to a compound formation can be considered in the Hauser-Feshbach model quite generally by simply renormalizing the $\alpha$-transmission coefficients $T_\ell$ in the entrance channel for each partial wave $\ell$ \cite{raucoulexproc,raucoulexlett}
\begin{equation}
\label{eq:coulextrans}
T'_\ell=f_\ell T_\ell = \left( \frac{T_\ell}{T_\ell+T_\ell ^\mathrm{direct}} \right) T_\ell ,
\end{equation}
where the transmission coefficient $T_\ell ^\mathrm{direct}$ into the direct channel can be derived from the direct cross section $\sigma_\ell ^\mathrm{direct}$. Here, $\sigma_\ell ^\mathrm{direct}$ is the Coulomb excitation cross section calculated in a fully quantum mechanical approach as, e.g., shown in \cite{alder}. In this approach, the transmission coefficients $T_\ell$ for compound formation are computed using an optical potential \textit{not} including the direct reactions, or more specifically the Coulomb excitation, in its imaginary part. This optical potential thus only describes the absorption into the compound channel and not into all inelastic channels. Only with such a potential the stellar reactivity of the ($\gamma$,$\alpha$) reaction can be computed by applying \eref{eq:revphoto}.

\section{Nuclear aspects of the p-nucleus production}
\subsection{Important reactions}
\label{sec:importantreactions}

Although various astrophysical sites were presented in \sref{sec:sites}, the actual types of participating reactions are limited. In environments producing the light p-nuclei under QSE conditions, individual proton captures in the vicinity of these p-nuclei only play a role in the brief freeze-out phase. While the reactions in a given mass region are in equilibrium, the abundances are determined by the known nuclear mass differences \cite{mey98,raureview,sanantonio}. A similar situation is encountered for the proton captures in the $\nu$p-process which are in (p,$\gamma$)-($\gamma$,p) equilibrium most of the time \cite{wanajo11,raufrohOmeg,frohnupnuc}. Therefore a variation of the proton capture rates only has limited impact but may locally redistribute isotopic abundances \cite{raufrohOmeg,frohrauOmeg,frohnupnuc}. Not in equilibrium, however, are the (n,p) reactions required to overcome waiting points with low proton-capture Q-value. In the mass region relevant for the production of light p-nuclei, they occur at the $N=Z$ line, starting at $^{56}$Ni and $2-3$ mass units towards stability \cite{raufrohOmeg,frohnupnuc}. They crucially determine the onset of the $\nu$p-process as well as how quickly matter can be processed towards higher masses.

\begin{figure}
\includegraphics[angle=-90,width=\textwidth]{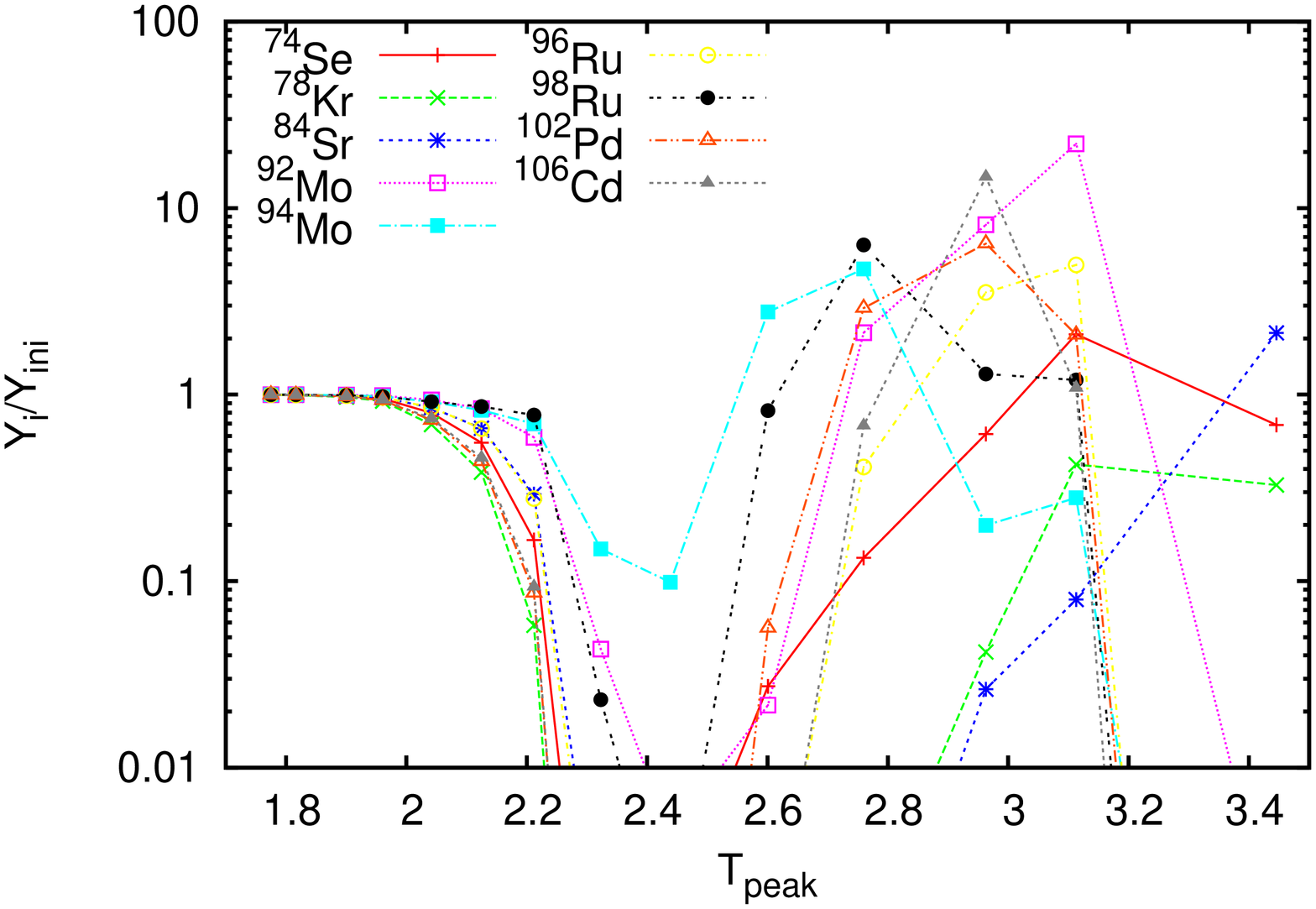}
\includegraphics[angle=-90,width=\textwidth]{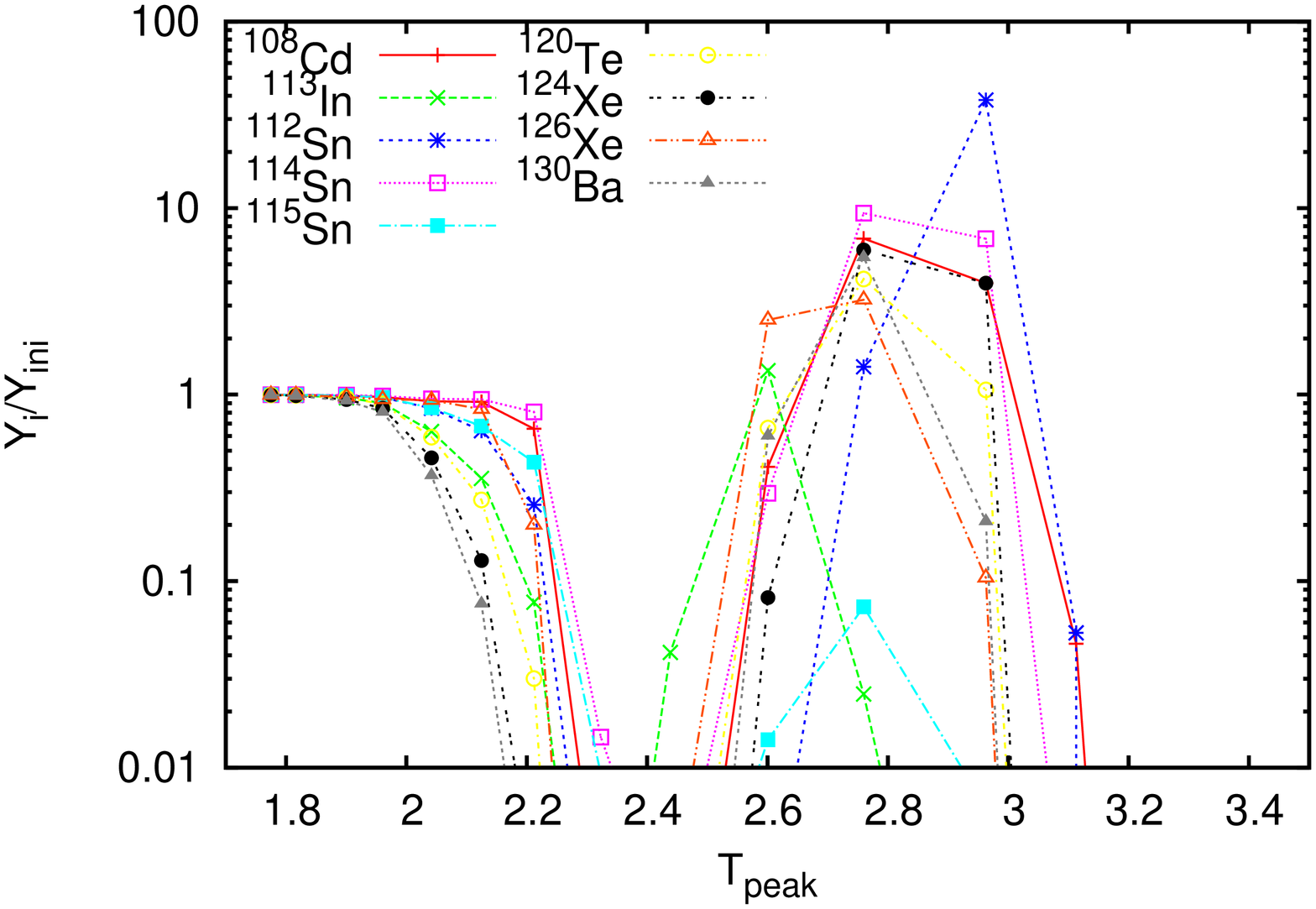}
\caption{\label{fig:peak1}Production factors $Y_i/Y_\mathrm{ini}$ relating the final abundance $Y_i$ to the initial abundance $Y_\mathrm{ini}$ as function of peak temperature attained in a burning zone. Shown are the production factors for p-nuclides with mass numbers $74\leq A\leq 106$ (top) and $108\leq A\leq 130$ (bottom). Initial solar abundances were used, the trajectories were similar to the ones from \cite{rayet95} but reaction rates were taken from \cite{reaclib}.}
\end{figure}

\begin{figure}
\includegraphics[angle=-90,width=\textwidth]{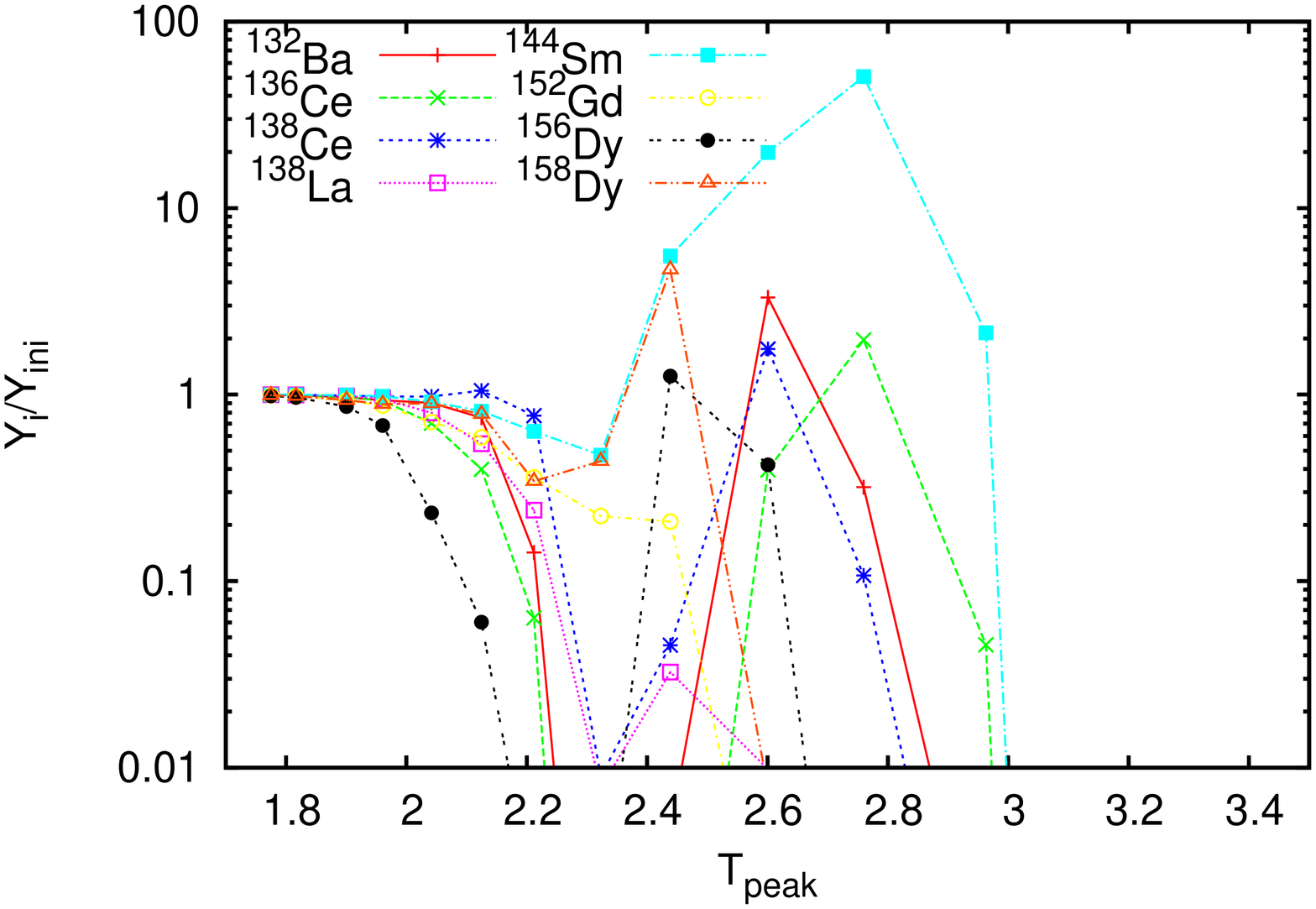}
\includegraphics[angle=-90,width=\textwidth]{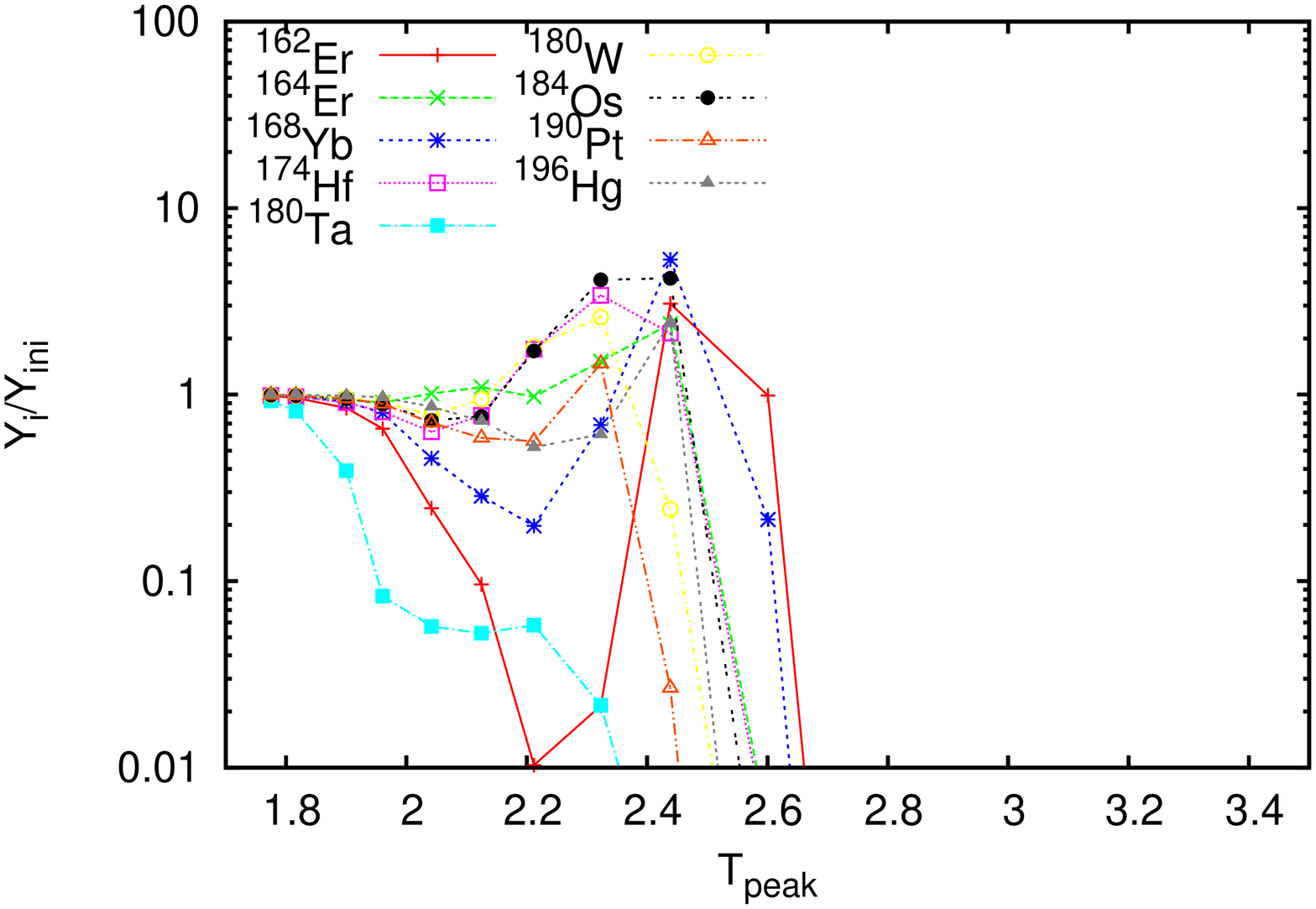}
\caption{\label{fig:peak2}Same as \fref{fig:peak1} but for p-nuclides with mass numbers $132\leq A\leq 158$ (top) and $162\leq A\leq 196$ (bottom). The isotopes $^{138}$La and $^{180}$Ta are underproduced because no $\nu$-process was included here and the population of the $^{180m}$Ta isomer was not followed separately. For $^{148}$Gd($\gamma$,$\alpha$)$^{144}$Sm the rate from \cite{nonsmoker} was used.}
\end{figure}

\begin{figure}
\includegraphics[angle=-90,width=\textwidth]{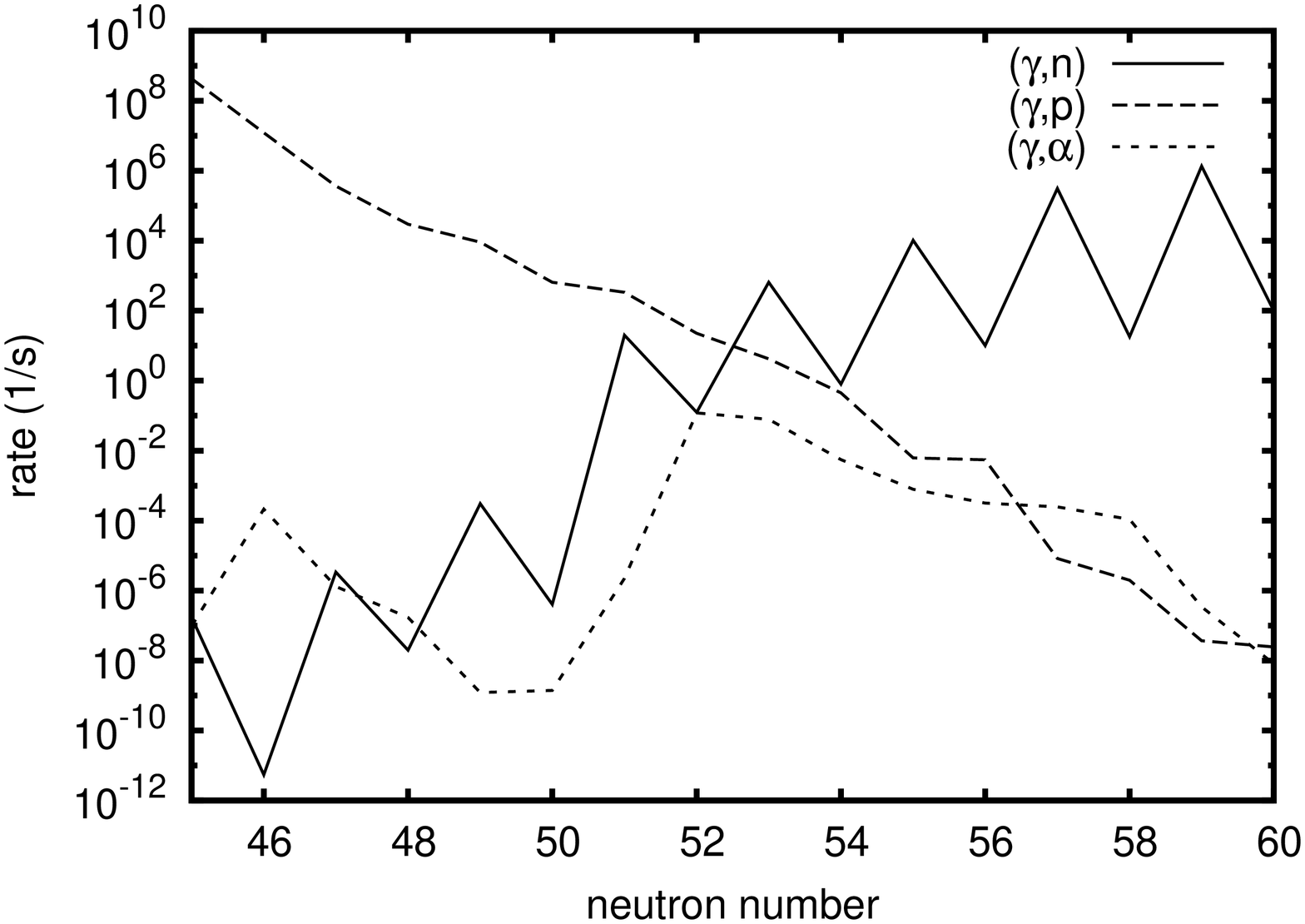}
\caption{\label{fig:ratecomp}Comparison of photodisintegration rates for Mo isotopes at 2.5 GK. The ($\gamma$,n) rate per nucleus is shown as solid line, the ($\gamma$,p) rate per nucleus as dashed line, and the ($\gamma$,$\alpha$) rate per nucleus as short-dashed line.}

\end{figure}

The schematic reaction path for the $\gamma$-process has been shown in \fref{fig:flow}.
The nuclei and reactions involved in the $\gamma$-processes in various hot environments do not largely differ. The permitted temperature range is rather tightly constrained by the fact that photodisintegrations must be possible but not so strong as to completely destroy the p-nuclei or their radioactive progenitors. The main difference between the sites suggested in \sref{sec:sites} is the amount and distribution of seed nuclei to be photodisintegrated. This will change the flow in a given reaction sequence accordingly but it should be noted that abundance ratios of nuclei originating from the same seed are not affected by this, such as the $^{146}$Sm/$^{144}$Sm production ratio discussed in sections \ref{sec:GCEextinct} and \ref{sec:anomalies}. Another difference arises from the temperature evolution which is not necessarily the same in, say, ccSN and SNIa. Since different types of reactions exhibit a different temperature dependence, deflections and branchings in the $\gamma$-process may shift with temperature. The time spent at a certain temperature is weighted by the temperature evolution and thus also how rates compete at a given nucleus. For instance, the ($\gamma$,n)/($\gamma$,$\alpha$) branching at $^{148}$Gd is temperature sensitive. A higher temperature favors ($\gamma$,n) with respect to ($\gamma$,$\alpha$) and thus increases $^{146}$Sm production \cite{woohow,woohowlett}. The competition between ($\gamma$,n), ($\gamma$,p), and ($\gamma$,$\alpha$) at different temperatures has been studied in detail by \cite{branchings,rauNIC2010}.

\begin{figure}
\includegraphics[width=\textwidth]{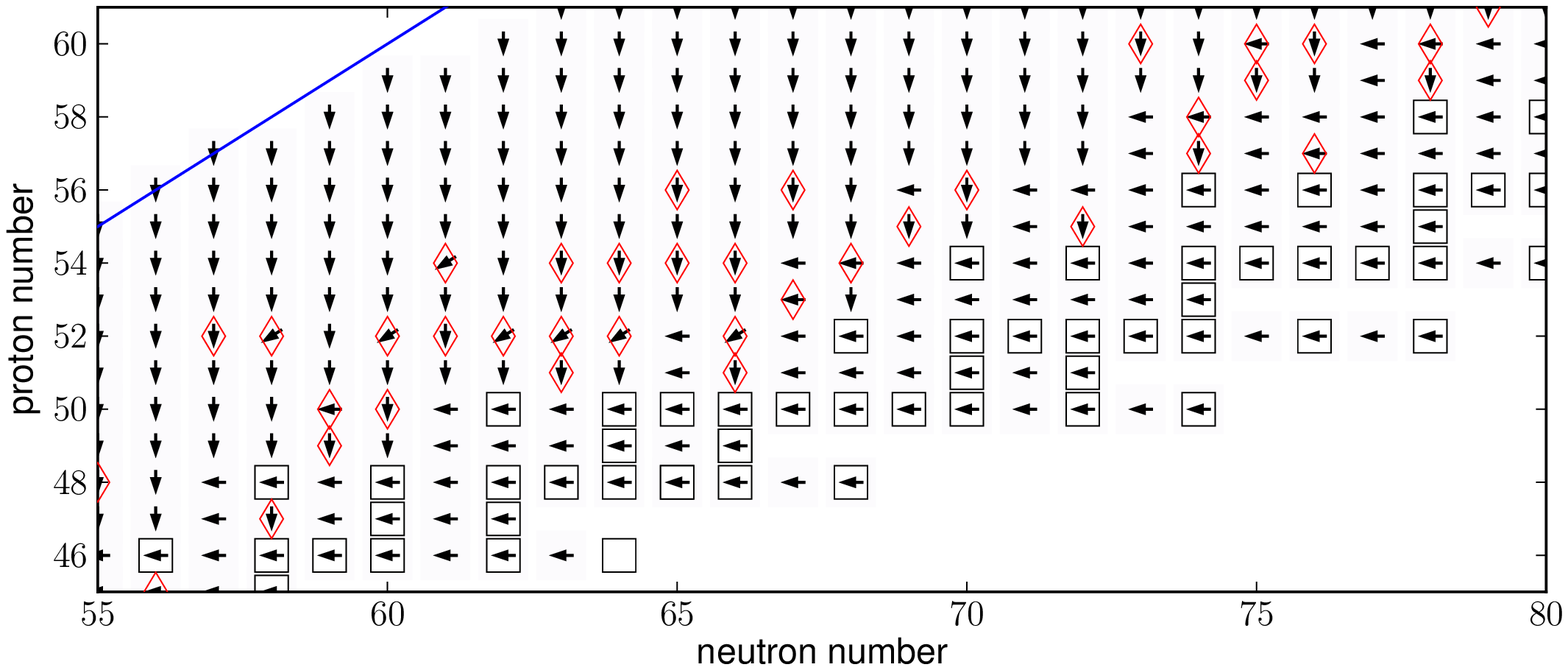}
\includegraphics[width=\textwidth]{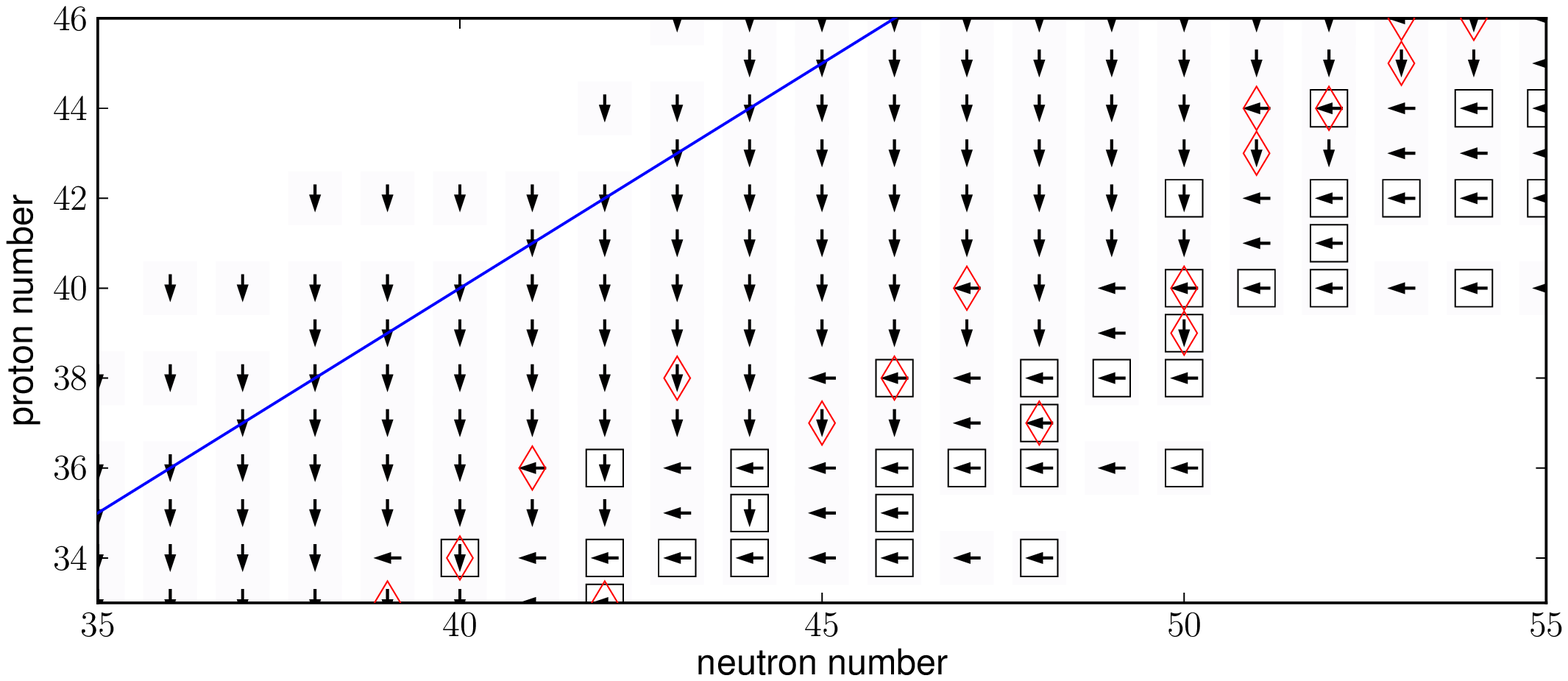}
\caption{Reactivity field plots for $33 \leq Z \leq 46$ (bottom) and $45\leq Z\leq 61$ (top) at 3 GK. The arrows give the dominant destruction reaction for each nucleus. Competition points are marked by red diamonds.  The first competition point in a sequence of ($\gamma$,n) reactions from stability is mainly important. The $N=Z$ line is shown by the straight blue line.}
\label{fig:deflectlow}
\end{figure}

\begin{figure}
\includegraphics[width=\textwidth]{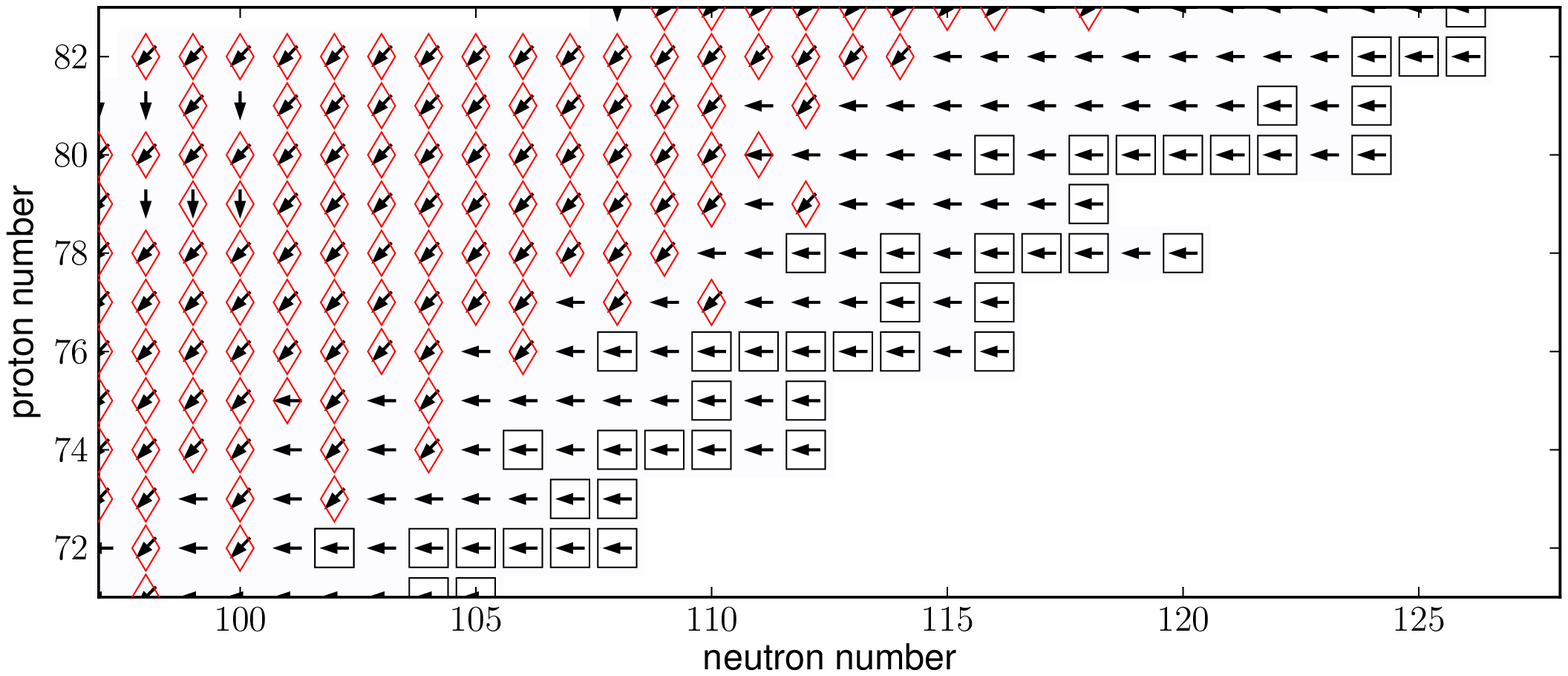}
\includegraphics[width=\textwidth]{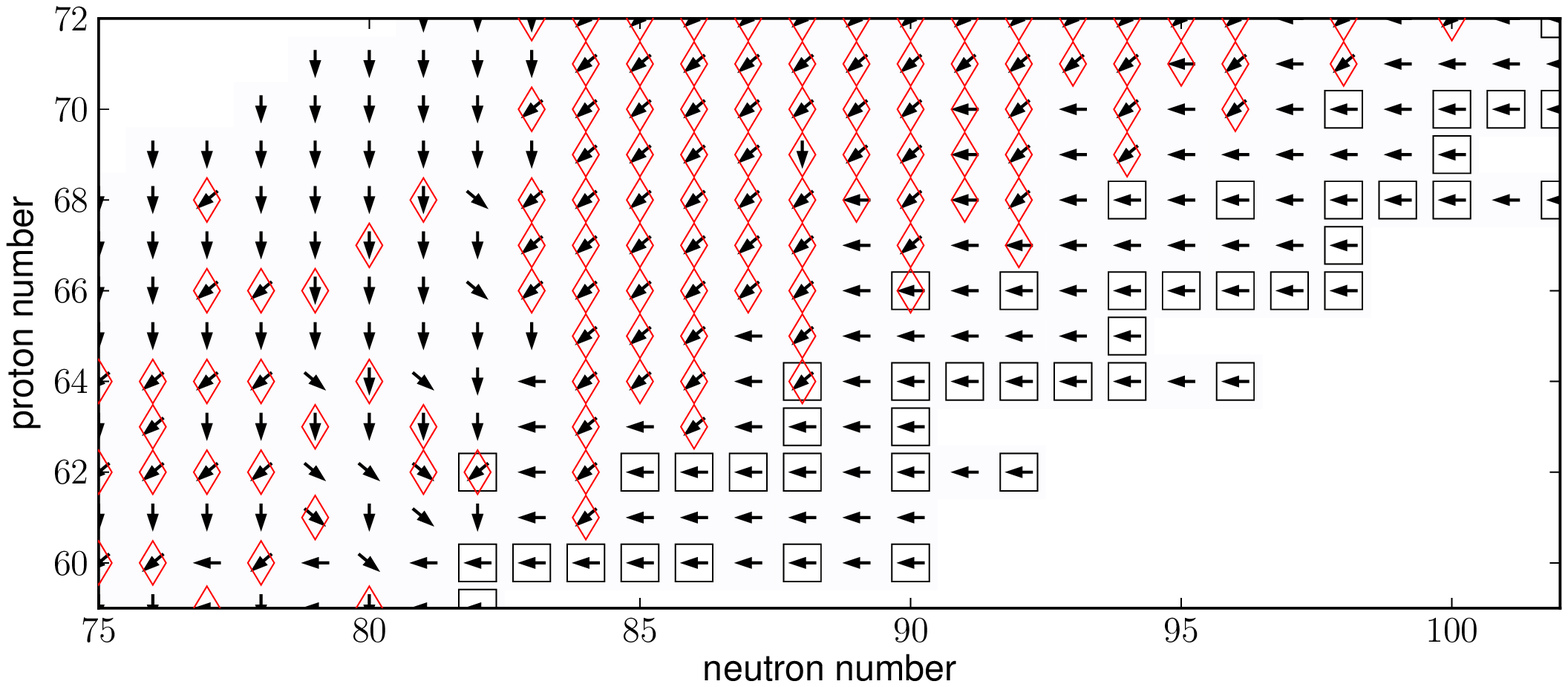}
\caption{Reactivity field plots for $59 \leq Z \leq 72$ (bottom) and $71\leq Z\leq 83$ (top) at 2 GK. The arrows give the dominant destruction reaction for each nucleus. Competition points are marked by red diamonds. The first competition point in a sequence of ($\gamma$,n) reactions from stability is mainly important.}
\label{fig:deflecthi}
\end{figure}

Not all reactions are equally important in all sections of the nucleosynthesis network. Figures \ref{fig:peak1} and \ref{fig:peak2} show the zonal production factors for all p-nuclides as function of the peak temperature reached in the zone of the ccSN model by \cite{hashimoto}. In a $\gamma$-process,
light p-nuclei are predominantly produced at higher temperatures (allowing efficient photodisintegration of the nuclei around mass $A\approx 100$)
whereas the production maximum of the heavy species lies towards the lower end of the temperature range.
Neutron captures and especially
($\gamma$,n) reactions are important throughout the $\gamma$-process network as the photodisintegration of stable nuclides
commences with ($\gamma$,n) reactions until sufficiently proton-rich nuclei have been produced and ($\gamma$,p) or ($\gamma$,$\alpha$)
reactions become faster. The released neutrons can be captured again by other nuclei and push the reaction path back to stability in
the region of the light p-nuclides.

In each isotopic chain we commence with initial ($\gamma$,n) on stable
isotopes and move towards the proton-rich side. The proton-richer a nucleus, the slower the ($\gamma$,n) rate while
($\gamma$,p) and/or ($\gamma$,$\alpha$) rates increase. At a certain isotope within the isotopic chain, a charged particle emission
rate will become faster than ($\gamma$,n) and thus deflect the reaction sequence to lower charge number. Historically, this endpoint
of the ($\gamma$,n) chain has been called ``branching'', inspired by the branchings in the s-process path \cite{woohow,branchings}.
A more appropriate term would be ``deflection (point)'', though, because -- unlike in the s-process -- the reaction path does not split necessarily
into two branches. The relative changes of the photodisintegration rates from one isotope to the next are so large that at each
isotope only one of the emissions dominates. As an example, a comparison of photodisintegration rates within a chain of Mo isotopes is shown in \fref{fig:ratecomp}. It can be seen how rapidly ($\gamma$,n) rates are decreasing with decreasing neutron number, whereas ($\gamma$,p) is strongly increasing because it becomes easier to emit a proton than a neutron. The ($\gamma$,n) rates exhibit strong odd-even staggering. Since in a reaction sequence, the timescale of the whole sequence is determined by the slowest reaction link(s), it is apparent that the ($\gamma$,n) sequences will be mostly sensitive to a variation in the slow ($\gamma$,n) rates and these have to be studied preferentially.

There are very few cases of two (or even three) types of emissions being comparable at a single isotope or over a short series of isotopes and these strongly depend on the optical potentials used. In these cases, true branchings would appear. They are mostly found in the range of the heavier p-nuclides, where ($\gamma$,$\alpha$) competes with ($\gamma$,n) as explained below, because the ($\gamma$,$\alpha$) rates do not vary as systematically as the rates for neutron or proton emission.

Examination of the deflection points easily shows that at higher mass
($\gamma$,$\alpha$) deflections are encountered whereas at lower mass most deflections are caused by ($\gamma$,p) due to the
distribution of reaction Q-values and Coulomb barriers \cite{branchings}. This is a well-understood nuclear structure effect and it can be seen in figures \ref{fig:deflectlow} and \ref{fig:deflecthi} that $\alpha$-emissions compete with neutron emission above $N=82$. The rate field plots in the two figures give the dominant destruction reaction for each nuclide, derived from a comparison of the rates per nucleus. For photodisintegration reactions the \textit{rate per nucleus} is just the reactivity $R^*_\gamma$ as defined in \sref{sec:ratedefs}. It should be noted that these are not reaction flows as they neglect the abundances of the interacting nuclei. In the $\gamma$-process, however, this is a good approximation because, except for neutron captures mentioned below, there are no reactions counteracting the photodisintegration of a single nucleus. Therefore, the rate per nucleus is very suitable to also study the competition of different reaction channels. The possible competition points are marked by diamond shapes in figures \ref{fig:deflectlow} and \ref{fig:deflecthi}. They are defined by two or more reaction channels having comparable rates within the assumed uncertainties in the prediction. The plots are based on calculations with the SMARAGD code \cite{raureview,smaragd} using the standard optical potentials of \cite{jlmm} for neutrons and protons, and \cite{mcfadden} for $\alpha$-particles. The assumed uncertainties were factors of two and three (up and down) for ($\gamma$,n) and ($\gamma$,p), respectively, while one fifth of the predicted  value was assumed to be the lower limit for the range of possible ($\gamma$,$\alpha$) rates. Since the reaction flow is coming from stable isotopes, the first competition point in each ($\gamma$,n) sequence is of major importance. Only if it is shown to be dominated by ($\gamma$,n), the second competition in a chain will also be of interest, and so on.

The above considerations explain the results of earlier studies with systematic variations of the ($\gamma$,p) and ($\gamma$,$\alpha$) rates by constant factors
\cite{rapp} or using a selection of different optical potentials for the ($\gamma$,$\alpha$) rates \cite{branchings,rauNIC2010}. They also found
that uncertainties in the ($\gamma$,p) rates mainly affect the lower half, whereas those in the ($\gamma$,$\alpha$)
rates affect the upper half of the p-nuclide mass range.

The higher the plasma temperature the further into the proton-rich side ($\gamma$,n) reactions can act. At low
temperature, even the ($\gamma$,n) reaction sequence may not be able to move much beyond the stable isotopes because it becomes
too slow compared to the explosive timescale (and neutron capture may be faster).
Obviously, ($\gamma$,p) and ($\gamma$,$\alpha$) are even slower in those cases and the photodisintegrations are ``stuck'', just slightly reordering stable abundances. Because of the tight limits on permissible temperatures to successfully produce p-nuclei, the competition points will not shift by more than $1-2$ mass units, if at all. More detailed investigations of the temperature dependence of deflections, competitions, and branchings can be found in \cite{woohow,branchings,rauNIC2010}, where each later work supersedes the previous one with updated reaction rate calculations and further studies of the dependence on optical potentials.

The free neutrons released by ($\gamma$,n) affect the final abundances of light p-nuclei (up to Sn) in the $\gamma$-process in several ways. The major effect is that of destruction of light p-nuclei by neutron captures in zones with sufficiently high temperature to photodisintegrate heavier nuclei but not enough for the lighter species. No change in the abundances of light p-nuclei would then be expected when only considering photodisintegration but such zones show strong destruction of these nuclei, as can be seen in figures \ref{fig:peak1} and \ref{fig:peak2}. This is because the neutrons released in the destruction of the heavier species destroy the pre-existing p-nuclei (if non-zero initial metallicity was assumed). Since all ejected zones are added up, this may affect the total yield of light p-nuclei. Neutron captures also counteract ($\gamma$,n) at temperatures at which the light p-nuclei are destroyed \cite{arngorp}. This may prevent the ($\gamma$,n) flow to move far into the proton-rich side. It was pointed out in \cite{rapp} that (n,p) reactions can push the $\gamma$-process path back to stability as well. Obviously, the availability of neutrons depends on the assumed seed abundance distribution and especially on the abundances in the region of heavier nuclei, which are destroyed already at low temperature. Therefore this will require special attention in models assuming strongly enhanced seeds, such as in the single-degenerate SNIa model (\sref{sec:WD}).

Finally, neutron captures \textit{before} the onset of the $\gamma$-process can indirectly influence the p-production by modifying the seed abundances. In massive stars, the weak s-processing sensitively depends on the $^{22}$Ne($\alpha$,n) rate, acting as the dominant neutron source (see, e.g., \cite{rhhw02,rauproc} for the effect of a variation of this rate). For s-processing in AGB stars, this rate and $^{13}$C($\alpha$,n) are important. Since the seed enhancement in thermonuclear burning of SNIa is supposed to come from the matter accreted from a companion star (or s-processing during the accretion), these rates would also affect the seeds and $\gamma$-processing in these models were they simulated self-consistently.

\subsection{Nuclear physics uncertainties}
\label{sec:sensi}

Before the main nuclear uncertainties in the synthesis of p-nuclei can be summarized, it is necessary to define the sensitivity of reaction rates and cross sections. While the definition of an ``error bar'' for theory is complicated by fundamental differences to attaching an experimental error (see \cite{sensipaper} for details), properly defined sensitivities immediately allow to see the impact of various uncertainties in nuclear properties and input. This also implies that it is easy to see which properties are in need of a better description in order to better constrain the astrophysical rate.
In order to quantify the impact of a variation of a model quantity $q$ (directly taken from input or derived from it) on the final result $\Omega$ (which is either a cross section or a reactivity), the relative sensitivity $s\left(\Omega,q\right)$ is defined as \cite{sensipaper,raureview}
\begin{equation}
\label{eq:sensi}
s\left(\Omega,q\right)=\frac{v_\Omega-1}{v_q-1} \quad.
\end{equation}
It is a measure of a change by a factor of $v_\Omega=\Omega_\mathrm{new}/\Omega_\mathrm{old}$ in $\Omega$ as the result of a change in the quantity $q$ by the factor $v_q=q_\mathrm{new}/q_\mathrm{old}$, with $s=0$ when no change occurs and $s=1$ when the final result changes by the same factor as used in the variation of $q$, i.e., $s\left(\Omega,q\right)=1$ implies $v_\Omega=v_q$.
Further information is encoded in the sign of the sensitivity $s$. Since both $v_\Omega>0$ and $v_q>0$ for the quantities studied in this context, a positive sign implies that $\Omega$ changes in the same manner as $q$, i.e., $\Omega$ becomes larger when the value of the quantity $q$ is increased. The opposite is true for $s<0$, i.e., $\Omega$ decreases with an increase of $q$.
The varied quantities $q$ in reaction rate studies are neutron-, proton-, $\alpha$-, and $\gamma$-widths. Sometimes also the NLD is varied although it can be shown that it mainly affects the $\gamma$-width in astrophysical applications. This is due to the fact that the particle widths are dominated by transitions to low-lying levels whereas $\gamma$-transitions to a continuum of levels at higher excitation of the compound nucleus determine the $\gamma$-width \cite{raugamma}.

Extended tables of sensitivities for reactions on target nuclei between the driplines and with $10\leq Z\leq 83$ have been published in \cite{sensipaper}, for rates as well as cross sections. These tables are used to determine which nuclear properties have to be known to accurately determine a rate, once it has been identified as relevant for nucleosynthesis. It is important to note that, according to \eref{eq:revrate} and \eref{eq:revphoto}, the same sensitivities apply to forward and reverse stellar rates, e.g., for the stellar capture rates as well as the stellar photodisintegration rates.

Despite of the number of suggested sites, the nuclear physics underlying the p-production and its uncertainty is similar in all of them, except for
the contribution of rapid proton capture processes far from stability. It has been argued in \sref{sec:extinct} that the latter cannot contribute significantly to the p-nuclei from Mo upwards. Since proton captures are in equilibrium in those models, the main uncertainties lie in the reactions bridging the waiting points close to $N=Z$. As mentioned in sections \ref{sec:neutrino} and \ref{sec:importantreactions}, (n,p) reactions are of highest importance in the $\nu$p-process. Because of their large Q-value, the compound nucleus is formed at high excitation energy and the statistical model is expected to be applicable well (\sref{sec:compound}). Contrary to astrophysical neutron captures at stability, no isolated resonances are contributing and the rates are only sensitive to the neutron width whereas in neutron captures at stability also the $\gamma$-width is important. The calculation of the neutron width depends on the optical potential used and on the knowledge of low-lying levels in both target and final nucleus. In the unstable region, the lack of accurately determined low-lying levels introduces the largest uncertainty (see below, however, for considerations regarding the optical potential).

The nuclei involved in the $\gamma$-process
are the stable nuclides and moderately unstable, proton-rich nuclei. The masses have been measured and thus the reaction Q-values
are well known. Half-lives are known in principle but the electron captures and $\beta^+$-decays need to be modified in the stellar plasma
by the application of theoretical corrections for ionization and thermal excitation. This has not been sufficiently addressed so far, with a few exceptions (see, e.g., \cite{arngorp}).

What is the actual impact of uncertainties in the photodisintegration rates on the production of the p-nuclides? Although the $\gamma$-process is not an equilibrium process and a reaction network with a large number of individual reactions has to be employed,
it has become apparent in \sref{sec:importantreactions} that not all possible reactions in the network have to be known with high
accuracy. Rather, only the dominant reaction sequences have to be known accurately, and within such a sequence the slowest
reaction as it determines the amount of processing within the given short timescale of the explosive process. Charged particle rates are
important only at or close to the deflection points because they are many orders of magnitude smaller than the ($\gamma$,n) rates
on neutron-richer isotopes.

As a general observation it was found that the $\gamma$-width is not relevant in astrophysical charged particle captures on intermediate and heavy mass nuclei \cite{sensipaper}. This is due to the fact that the Coulomb barrier suppresses the proton- and $\alpha$-widths at low energy and makes them smaller than the $\gamma$-widths at astrophysical interaction energies. The cross sections and rates will always be most sensitive to the channel with the smallest width. It also follows from this that whenever an $\alpha$-particle is involved, the reaction will be mostly sensitive to its channel. The situation for neutron captures is more diverse. For most nuclei close to stability, the rates are strongly sensitive to the $\gamma$-width. In between magic neutron numbers along stability and especially in the region of deformed nuclei, however, they are also or even more sensitive to the neutron-width (see figures 14, 15 in \cite{sensipaper}). This can be explained by the fact that the compound nuclei in these regions have higher level densities and this leads to comparable sizes of the neutron- and $\gamma$-widths at astrophysical energies. A comparison of theory to experimental data along stability revealed that the uncertainties are within a factor of two, with an average deviation of better than 30\% \cite{RTK96,raureview}. Since the $\gamma$-process does not go far out into the unstable region, similar uncertainties are expected.

\begin{figure}
\includegraphics[angle=-90,width=\textwidth]{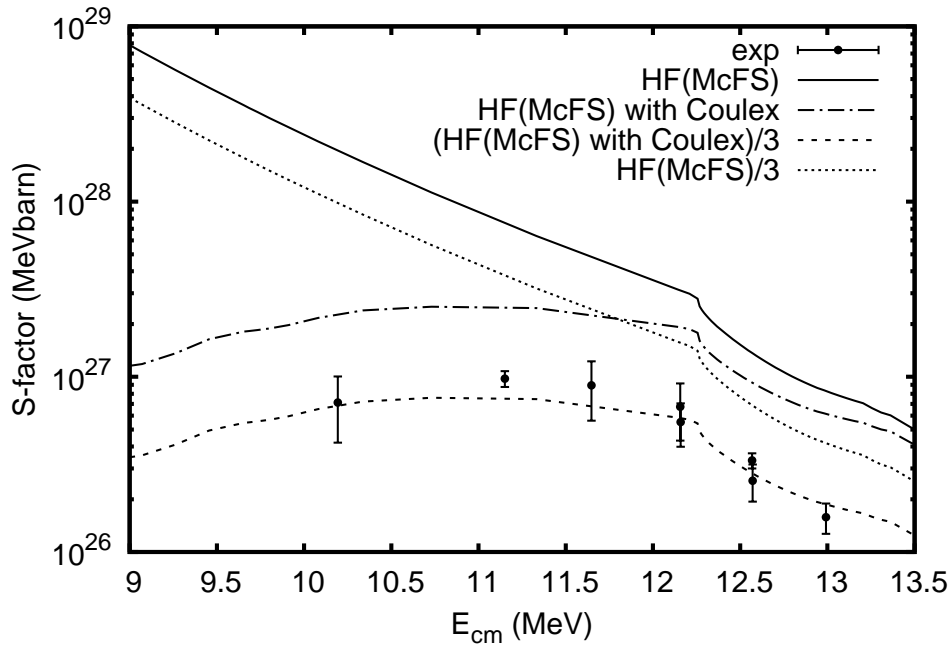}
\caption{\label{fig:sm144ag}Astrophysical S-factor for $^{144}$Sm($\alpha$,$\gamma$)$^{148}$Gd as function of c.m.\ energy. Data are from \cite{sm144}. The astrophysically relevant energy is about $8-9$ MeV. Shown are the standard prediction with the potential by \cite{mcfadden} (McFS, full line), the same but corrected for Coulomb excitation (dash-dotted line), and the Coulex corrected prediction with the $\alpha$-width divided by a constant factor of 3 (dashed line). Also shown is the standard prediction without correction but with the $\alpha$-width divided by 3 (dotted line). See text for an interpretation.}

\end{figure}

Charged particle captures and photodisintegrations are only sensitive to the optical potential at the astrophysically
relevant low energies. These optical potentials are usually derived from elastic scattering at higher energies and are thus
not well constrained around the Coulomb barrier. Especially the imaginary part should be energy-dependent (see \cite{raureview} for more details on optical potentials in astrophysical rate predictions and the influence of other nuclear properties). Particular problems persist with $\alpha$-captures at energies and in the mass region relevant for the $\gamma$-process. Comparisons of theoretical predictions with the few available data at low energy (see \sref{sec:expstatus}) revealed a mixed pattern of good reproduction and some cases of maximally $2-3$ times overprediction of the ($\alpha$,$\gamma$) cross sections when using the ``standard'' potential of \cite{mcfadden} (see, e.g., \cite{tm169,yalcin,gyurky06,gyurky10,sau12} and references therein; see also \sref{sec:charged}). So far, the
only known example of a larger deviation was found in $^{144}$Sm($\alpha$,$\gamma$)$^{148}$Gd (determining the $^{144}$Sm/$^{146}$Sm production ratio in the $\gamma$-process, see \sref{sec:extinct}) where the measured cross section is lower than the standard prediction by more than
an order of magnitude at astrophysical energies \cite{sm144}, inexplicable by global potentials and also not reproduced using a potential independently derived from elastic $\alpha$-scattering at higher energy \cite{moh97}.

As pointed out in \sref{sec:direct} and by \cite{raucoulexproc,raucoulexlett}, the inclusion of Coulomb excitation may alleviate the putative problem in the prediction of the ($\alpha$,$\gamma$) laboratory cross sections and provide a more consistent picture. The $^{144}$Sm($\alpha$,$\gamma$)$^{148}$Gd case is shown in \fref{fig:sm144ag}. The prediction using the standard $\alpha$+nucleus potential by \cite{mcfadden} is a factor of about 4 higher than the data at the upper end of the measured energy range. It has a completely different energy dependence, though, and yields a value higher by almost two orders of magnitude than the extrapolation of the data to the astrophysically relevant energy of 9 MeV. Applying \eref{eq:coulextrans} to correct the prediction of the laboratory value for the fact that part of the $\alpha$-flux is going into the direct inelastic channel, which is not included in the optical potential, leads to calculated cross sections which reproduce the energy dependence of the data but are too high by a factor of 3. The optical potential should be corrected only to account for this factor, i.e., the $\alpha$-width should be divided by this factor as shown in the figure. This corrected $\alpha$-potential -- but without further consideration of the Coulomb excitation (as explained in \sref{sec:direct}) -- has to be employed to calculate the stellar $\alpha$-capture reactivity from which the actually relevant ($\gamma$,$\alpha$) rate is derived. The S-factor is lower than the standard value by about a factor of two at 9 MeV. Elastic $\alpha$-scattering experiments at low energy, if feasible, may help to constrain the $\alpha$-width renormalization, see \sref{sec:alphascatt}, because the S-factor and cross sections at the upper end of the measured energy range are not only sensitive to the $\alpha$-width but also to the neutron and the $\gamma$-width. Other reactions may not even need changes to the optical potential and the apparent discrepancies may be explained by the laboratory Coulomb excitation alone \cite{raucoulexlett,raucoulexproc}. The uncertainties involved are the requirement to know precise B(E2) transition strengths (which is challenging for odd nuclei with non-zero g.s.\ spin) and the need to accurately calculate Coulomb barrier penetrabilities at very low $\alpha$-energies.

While the standard $\alpha$+nucleus potential most widely used in astrophysical applications is a purely phenomenological one, the optical potentials for protons and neutrons used in the prediction of astrophysical rates and also the interpretation of nuclear data are based on a more microscopic treatment,
a Br\"uckner-Hartree-Fock calculation with the Local Density Approximation including nuclear matter density distributions, which are
in turn derived from microscopic calculations \cite{raureview,jlm,jlmm,bauge1,bauge2}. It has been pointed out, however, that the isovector part of the potential is not well constrained by scattering experiments \cite{vector}. So far, this has not been found to give a noticeable effect in neutron captures at stability but may introduce an additional uncertainty both at the far neutron- and the proton-rich side, whereever the neutron-width is dominating the rate.
Also in comparison to measured low-energy (p,$\gamma$) cross sections, calculations using these potentials often have been in very good agreement with the data, sometimes being off by a maximal factor of two (see, e.g., \cite{raureview,pdpg,cdpg} and references therein). In general, the uncertainty in the astrophysical rates caused by the nucleon optical potentials seems to be much lower than for the $\alpha$-capture rates. Recent (p,$\gamma$) and (p,n) data of higher precision close to the astrophysically relevant energy window, however, revealed a possible need for modification of the imaginary part of the nucleon potentials \cite{raureview,cdpg,npa4proc,kisssupp,tomsupp}.
A consistently increased strength of the imaginary part at low energies improves the reproduction of the experimental data for a number of
reactions.

\subsection{Challenges and opportunities in the experimental determination of astrophysical rates}\label{sec:expgeneral}

\subsubsection{Status}
\label{sec:expstatus}
\begin{figure}
\includegraphics[angle=-90,width=0.8\textwidth]{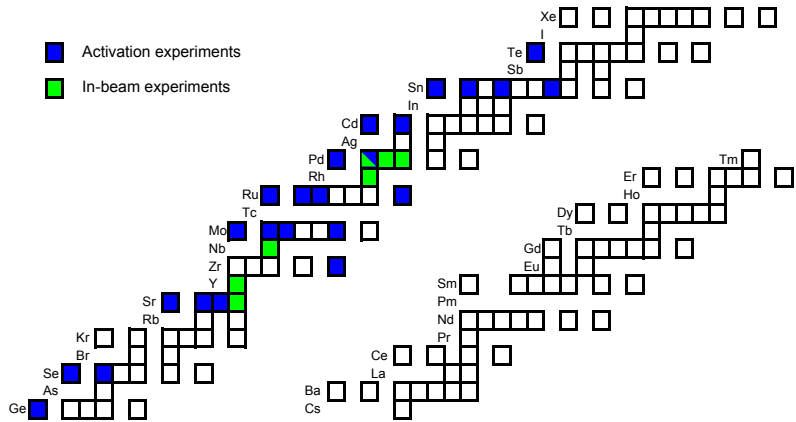}
\caption{Isotopes on which (p,$\gamma$) cross sections relevant for the $\gamma$-process have been measured. The upper part of the p-isotope mass region is not shown since there are no data available there. The measured cross section data can be found in  \cite{kiss07,gyu03,gyu01,gal03,tsa04,chl99,har01,sau97,bor98,spy07,pdpg,ozk02,cdpg,gur09,sau12,pre12}.}
\label{fig:pg_measured}
\end{figure}

\begin{figure}
\includegraphics[angle=-90,width=0.8\textwidth]{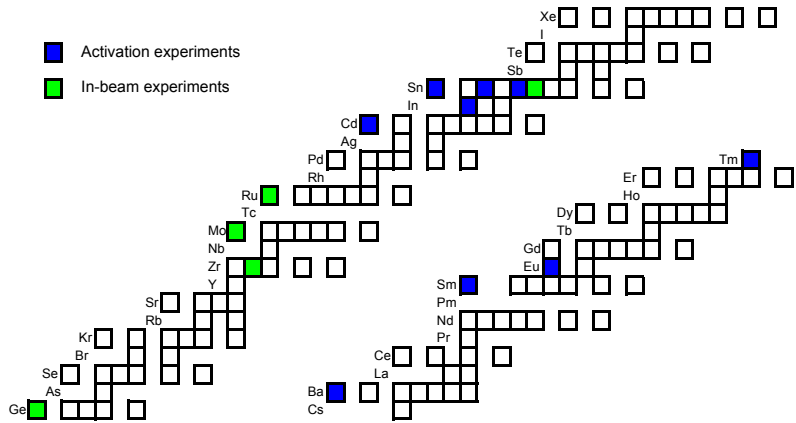}
\caption{Isotopes on which ($\alpha,\gamma$) cross sections relevant for the $\gamma$-process have been measured. The upper part of the p-isotope mass region is not shown since there are no data available  there with the exception of the $^{197}$Au($\alpha,\gamma$)$^{201}$Tl \cite{fil11}. The measured cross section data can be found in  \cite{ful96,har05,sau11,rap02,ozk07,rap08,has10,gyurky06,yalcin,cat08,gyurky10,sm144,hal12,tm169,kis12}.}
\label{fig:ag_measured}
\end{figure}

It has become apparent in the previous sections that several hundreds of reactions contribute to the synthesis of the comparatively few p-nuclides. The vast majority of the astrophysical reaction rates has only been predicted theoretically. Experimental data close to astrophysically relevant energies are very scarce, especially for charged-particle reactions. Figures \ref{fig:pg_measured} and \ref{fig:ag_measured} show those isotopes for which (p,$\gamma$) and ($\alpha,\gamma$), respectively, cross section measurements relevant for the $\gamma$-process are available. Only those isotopes are included where the motivation of the experiments was the origin of the p-nuclides. As can be seen, currently proton capture measurements are available for about 30 isotopes along the line of stability. The measurements are concentrated mainly in the lower mass region of the p-isotopes which is in line with the fact that ($\gamma$,p) reactions play the more important role in the lower mass range of a $\gamma$-process network (see \sref{sec:importantreactions}). Data for ($\alpha,\gamma$) reactions are even more scarce, leaving the theoretical reaction rate calculations largely untested. Moreover, the important higher mass region is almost completely unexplored, especially close to astrophysical energies.

Figures \ref{fig:pg_measured} and \ref{fig:ag_measured} show only those isotopes where radiative capture cross sections have been measured. On the other hand, particle emitting reaction cross sections, such as (p,n), ($\alpha$,n) or ($\alpha$,p), have also been measured along with the (p,$\gamma$) or ($\alpha,\gamma$) reactions, or on their own right. These reactions can provide valuable additional information to constrain the theoretical description of astrophysically relevant nuclear properties as explained below.

The situation is somewhat better regarding neutron-induced reactions along stability as the low-energy cross sections for neutron captures have been determined for s-process studies. The KADoNiS database \cite{kadonis0,kadonis,kadonisonline} centrally compiles data and provides recommended values for Maxwellian Averaged Cross Sections (MACS) at 30 keV and reactivities up to 100 keV. Unfortunately, this is only of limited use for the $\gamma$-process as peak temperatures of $2-3$ GK correspond to Maxwellian energies of $kT=170-260$ keV but require measured neutron capture cross sections up to 1400 keV, which is beyond the measured range for most nuclei. The standard way to deal with this is to renormalize Hauser-Feshbach calculations to the measured value at one energy (usually at 30 keV, but 100 keV have been discussed as possible alternative) in order to get an energy dependence for extrapolation to higher (and lower) energies. This can be problematic when the Hauser-Feshbach model is not applicable at the renormalization energy. Moreover, it has been pointed out recently that the renormalization procedure applied so far has not properly included the stellar effect of thermal population of excited states (see below) \cite{renormletter}. The proper renormalization is discussed in \sref{sec:expchallenges}.

Summarizing the available data for studying explosive burning it becomes apparent that, with few exceptions, almost no measurements in the relevant energy region are available. Although, as discussed in \sref{sec:expchallenges}, high-temperature environments severely limit the possibility to directly determine astrophysical reactivities, there is an urgent need to experimentally cover the relevant energy range to obtain data for the improvement of the theoretical predictions. Given the shortcomings and diversity of processes suggested for the production of p-nuclei (\sref{sec:sites}), a reliable database and reliable predictions are needed as a firm basis for future investigations. The accurate knowledge of reaction rates or, at least, their realistic uncertainties may allow to rule out certain astrophysical models on the grounds of nuclear physics considerations.

\subsubsection{Challenges and opportunities in the experimental determination of astrophysical rates}
\label{sec:expchallenges}

There are several challenges hampering the direct determination of stellar reactivities by experiments. Although such challenges would appear in many astrophysically motivated studies, they are more pronounced in high temperature environments and for intermediate and heavy nuclei. Reactions involved in the production of p-nuclei thus pose special problems, not or on a much smaller level encountered in the experimental study of reactions with light nuclei and in hydrostatic burning at lower temperature.
The main challenges include (i) reactions on unstable nuclei, (ii) tiny cross sections at astrophysically relevant energies, (iii) different sensitivities of the cross sections inside and outside of the astrophysical energy window, (iv) large differences between the stellar cross section and the laboratory cross section. In this section, we briefly address each of these points and its implications for experiments.

\paragraph{Small cross sections and unstable target nuclei}
The relevant energies for calculation of the astrophysical reactivity have been summarized in \sref{sec:energywindows} and can be found in \cite{energywindows}. In spite of high plasma temperatures, the interaction energies are small by nuclear physics standards. Although the energy windows for reactions involving charged particles in the entrance or exit channel are shifted to higher energy with increasing charge, the cross sections are also strongly suppressed by the Coulomb barrier. This results in low counting rates in laboratory experiments and requires efficient detectors as well as excellent background suppression. Although there has been recent experimental progress (\sref{sec:methods}), the astrophysically relevant energy window has only been reached in a few cases due to the tiny charged-particle cross sections, and a full coverage of such a window for $\gamma$-process conditions has not been achieved yet. Obviously, this does not apply to neutron captures but high accuracy measurements covering the full $\gamma$-process energy window are also scarce. Low nuclear absorption cross sections are also an obstacle for scattering experiments with charged particles. Such experiments are required to improve the optical potentials which are not well constrained at astrophysical energies (see sections \ref{sec:sensi} and \ref{sec:alphascatt}). At such energies, however, the obtained cross sections are almost indistinguishable from pure Coulomb scattering (Rutherford scattering). The situation is worsened by the fact that the majority of reactions for p-nucleosynthesis proceed on unstable nuclei. Although future facilities will allow the production of such nuclei it remains to be seen what types of reactions can be studied at which energies (see also \sref{sec:esr}). Since reactions with short-lived nuclei must be studied in inverse kinematics, neutron-induced reactions cannot be addressed because no neutron target exists.

\begin{figure}
\includegraphics[angle=-90,width=\columnwidth]{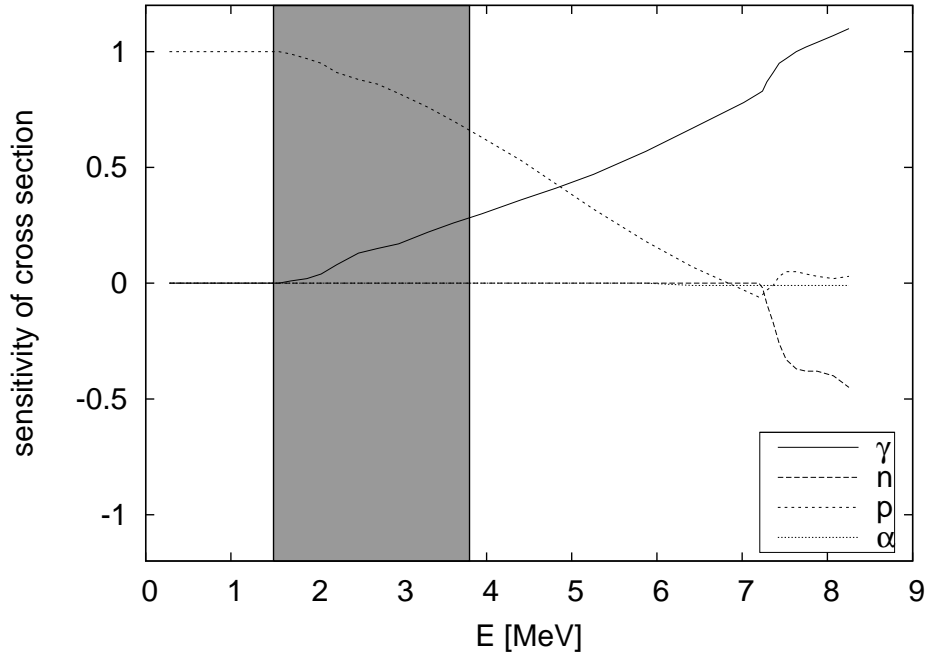}
\caption{Sensitivities $s$ of the laboratory cross section of $^{96}$Ru(p,$\gamma$)$^{97}$Rh to variations of nucleon-, $\alpha$-, and $\gamma$-widths, plotted as functions of c.m.\ energy \cite{sensipaper}. The shaded region is the astrophysically relevant energy range for $2\leq T \leq 3$ GK \cite{energywindows}. The sensitivity is defined in \eref{eq:sensi}.\label{fig:ruplot}}
\end{figure}

\paragraph{Sensitivities}
\label{sec:expsensi}
The sensitivity of a cross section or rate to a variation of nuclear properties is defined in \eref{eq:sensi}. It is very important to measure in the correct energy range or at least close to it. The reason is that cross sections above the astrophysical energies may exhibit a completely different sensitivity to nuclear properties which complicates extrapolation. For the same reason, comparisons with theoretical predictions outside the astrophysical energy range are only helpful when the measured cross sections show the same sensitivities as the astrophysical reactivity. Otherwise neither agreement nor disagreement between theory and data allows to draw any conclusion on the quality of the prediction of the rate.
\Fref{fig:ruplot} shows a striking example of how different the sensitivities can be in the astrophysical energy range and above it. If a measurement of $^{96}$Ru(p,$\gamma$)$^{97}$Rh showed discrepancies to predictions above, say, $5-6$ MeV it would be hard to disentangle the uncertainties from different sources. Moreover, such a discrepancy would imply nothing concerning the reliability of the prediction at astrophysical energies, as the cross section is almost exclusively sensitive to the proton width there, whereas the proton width does not play a role at higher energy. If good agreement was found, on the other hand, between experiment and theory at higher energy, this does not constrain the uncertainty of the reaction rate.
Such sensitivity considerations and plots have become an essential tool in the planning and interpretation of experiments. Further examples regarding the application of sensitivity plots to the astrophysical interpretation of experimental data are found in \cite{gyurky06,cdpg,kiss07,gyurky10,raureview,sauer11,hal12,rautm169,sau12} and references therein.

Also the impact of uncertainties in the (input) quantities on the final rate or cross section can be studied using the relative sensitivities defined in \eref{eq:sensi}. When an uncertainty factor $U_q$ is attached to a quantity $q$, it will appear as an uncertainty factor $^\Omega U=\left|s\left(\Omega,q\right) \right| U_q$ in the final result.

Laboratory cross sections $\sigma_0$ as defined in \eref{eq:labcs} exhibit different sensitivities in forward and reverse reactions, whereas it follows from the reciprocity relations \eref{eq:revrate}, \eref{eq:revphoto} that the sensitivities of the stellar reaction rates are the same for forward and reverse reaction. It is also to be noted that even in the astrophysical energy range laboratory cross sections do not necessarily show the same sensitivity as the stellar rate. This depends on the ground state contribution $X_0$ to the stellar rate (see below) and/or the sensitivities of the transitions from excited states in the target nucleus. Usually the cross section in the reaction direction with larger $X_0$ will behave more similar to the stellar rate. See \cite{sensipaper} for a more detailed discussion.

\Sref{sec:direct} pointed out a further complication, namely the change in reaction mechanism or appearance of further reaction mechanisms at low energy. This applies both to reaction and scattering experiments, as the absorptive part of an optical potential derived from scattering includes all inelastic channels but cannot distinguish between them. For the stellar rate and the application of the reciprocity relations, however, different mechanisms may have to be accounted for differently.

\paragraph{Stellar effects}
\label{sec:stellareffects}
At the conditions relevant to p-nucleosynthesis, all constituents of the stellar plasma, including nuclei, are in thermal equilibrium. This implies that a fraction of the nuclei will be present in an excited state. Thus, reactions not only proceed on target nuclei in the g.s.\ (as in the laboratory) but also on nuclei in excited states. This is automatically accounted for when using the stellar cross section $\sigma^*$ as defined in \eref{eq:effcs}. The relative contribution of transitions from g.s.\ and excited states to the stellar cross section, and thus to the stellar rate, depend on spin and excitation energy of the available levels, and on the plasma temperature. Again, nuclei with intermediate mass and heavy ones behave differently from light species. They show pronounced contributions of excited states already for neutron captures at s-process temperatures \cite{xfactor,renormletter}. Since the weights \eref{eq:weights} of the excited state contributions also include a dependence on the energy $E$, which is confined to the astrophysically relevant energy window, even more levels at higher excitation energy can contribute for charged particle reactions than for neutron captures at the same temperature. This is because the energy window is shifted to higher energies by the cross section dependence on the Coulomb barrier. And since the nuclear burning processes synthesizing p-nuclei occur at considerably higher temperature than the s-process, strong contributions of transitions from excited states are expected.

\begin{figure}
\includegraphics[angle=-90,width=\columnwidth]{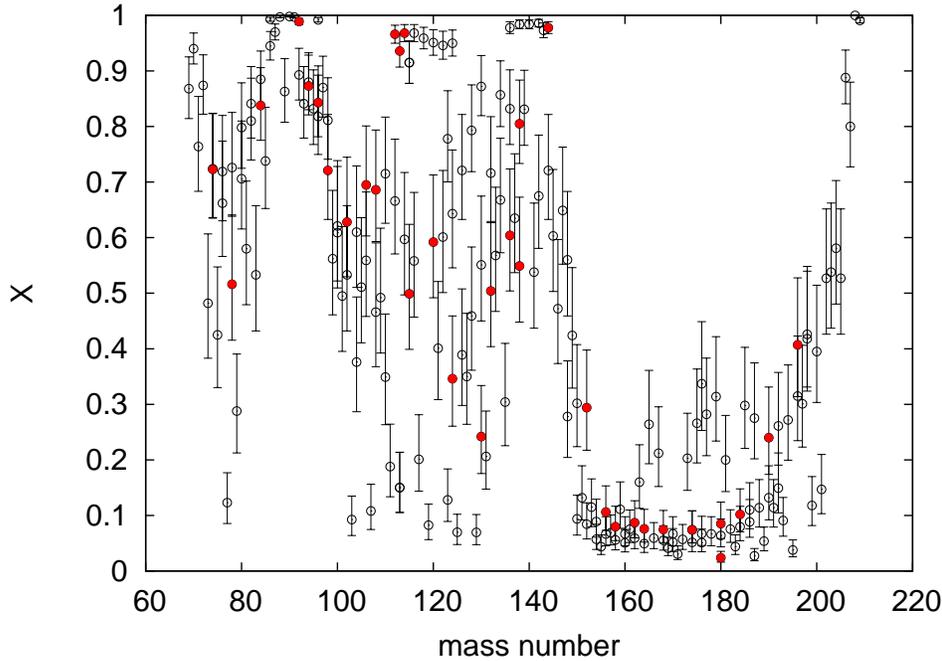}
\caption{Ground-state contributions $X_0$ (and isomeric state contribution $X_2$ for $^{180m}$Ta) to the stellar neutron capture rate at 2.5 GK for natural isotopes and their uncertainties, taken from \cite{sensipaper}. The filled symbols indicate p-isotopes.\label{fig:xplotng}}
\end{figure}

\begin{figure}
\includegraphics[angle=-90,width=\columnwidth]{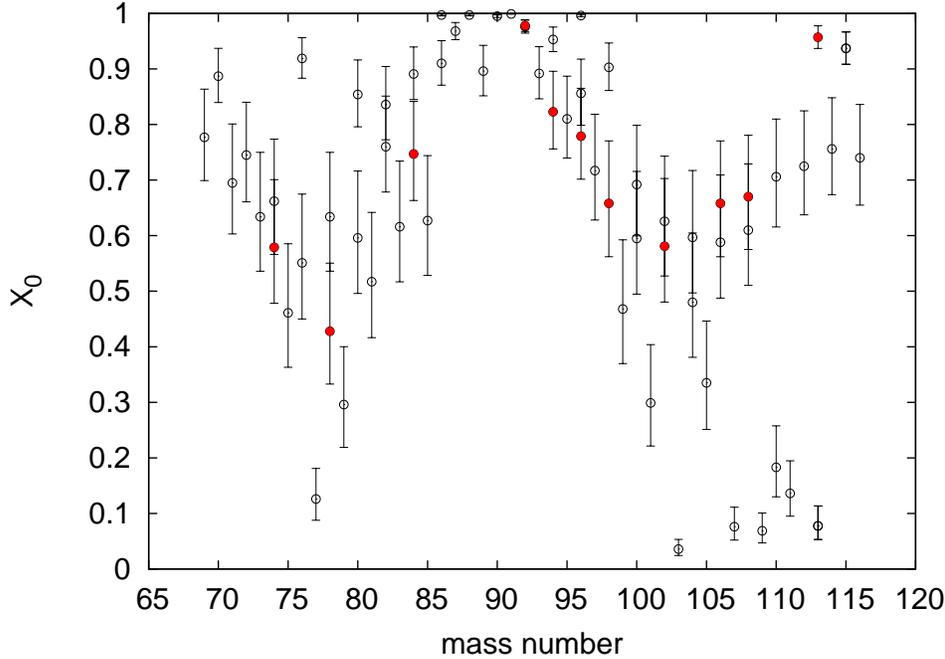}
\caption{Ground-state contributions to the stellar proton capture rate at 3 GK for natural isotopes and their uncertainties, taken from \cite{sensipaper}. The filled symbols indicate p-isotopes.\label{fig:xplotpg}}
\end{figure}

\begin{figure}
\includegraphics[angle=-90,width=\columnwidth]{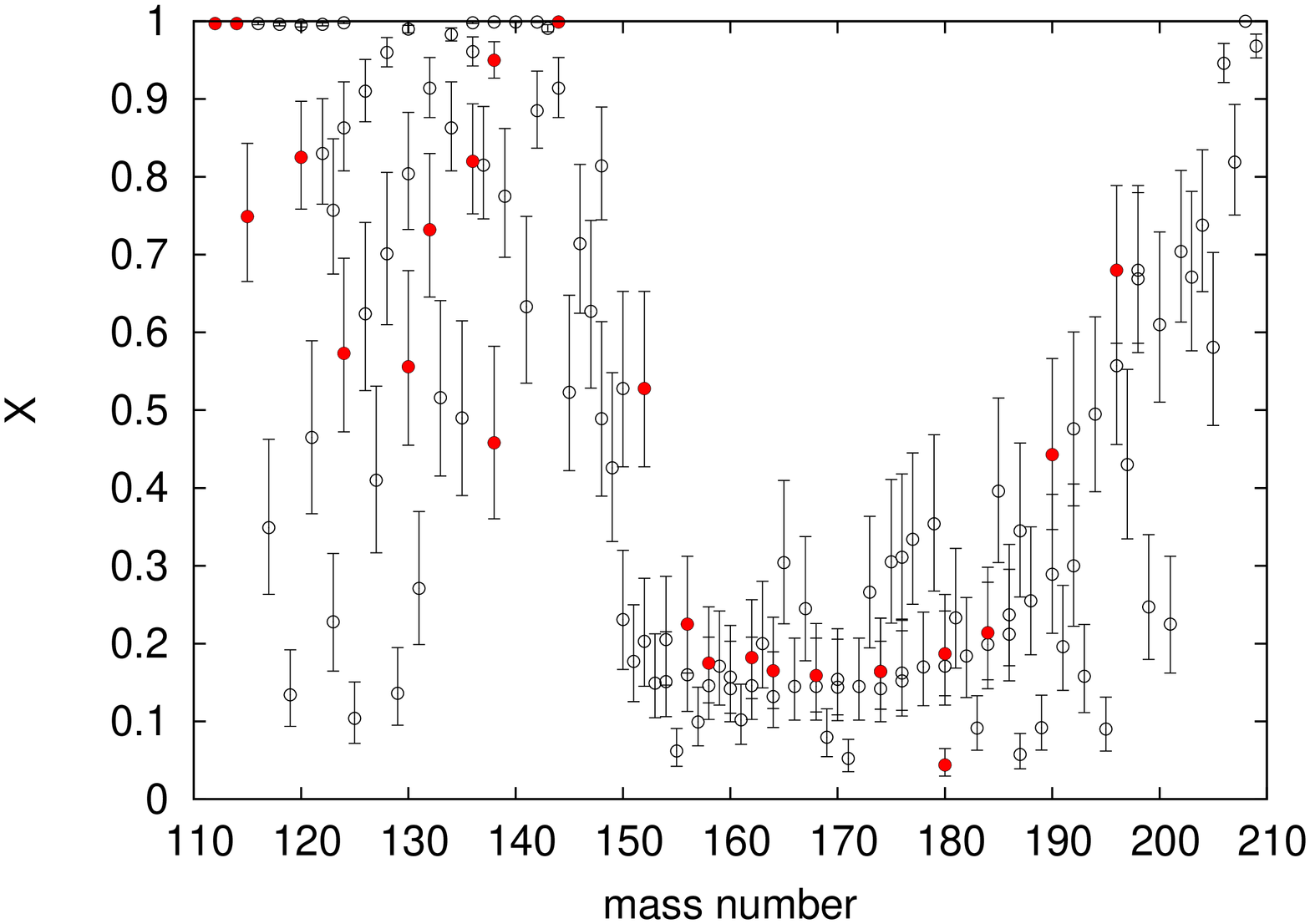}
\caption{Ground-state contributions (and isomeric state contribution $X_2$ for $^{180m}$Ta) to the stellar $\alpha$-capture rate at 2 GK for natural isotopes and their uncertainties, taken from \cite{sensipaper}. The filled symbols indicate p-isotopes.\label{fig:xplotag}}
\end{figure}

Even when measurements are possible directly at astrophysical energies, only the g.s.\ cross section $\sigma_0$ (or the cross section $\sigma_i$ of a long-lived isomeric state), as defined in \eref{eq:labcs}, can be determined in the laboratory. Therefore it is necessary to know the fraction of the stellar rate it actually constrains.
The relative contribution $X_i$ of a specific level $i$ to the total stellar rate $r^*$ is given by \cite{renormletter}
\begin{equation}
\label{eq:xfactor}
X_i(T)=\frac{2J_i+1}{2J_0+1}e^{-E_i/(kT)}\frac{\int\sigma_i(E)\Phi(E,T)dE}{\int\sigma^\mathrm{eff}(E)\Phi(E,T)dE} \quad,
\end{equation}
where $\sigma_i=\sum_j \sigma^{i \rightarrow j}$, as before, and $\sigma^\mathrm{eff}$ is the effective cross section as given in \eref{eq:effcs}. For the ground state, this simplifies to \cite{xfactor}
\begin{equation}
\label{eq:gsxfactor}
X_0(T)=\frac{\int\sigma_0(E)\Phi(E,T)dE}{\int\sigma^\mathrm{eff}(E)\Phi(E,T)dE}=\frac{R_0}{\int\sigma^\mathrm{eff}(E)\Phi(E,T)dE} \quad.
\end{equation}
It is very important to note that this is different from the simple ratio $R_0/R^*$ of g.s.\ and stellar reactivity, respectively, which has been called \textit{stellar enhancement factor} in the past.

The relative contribution $X_i$ has several convenient properties. It only assumes values in the range $0\leq X_i\leq 1$. The value of $X_0$ decreases monotonically with increasing plasma temperature $T$. It was shown in \cite{xfactor} that the magnitude of the uncertainty scales inversely proportionally with the value of $X_0$ (or generally $X_i$), i.e., $X_0=1$ has zero uncertainty as long as $G_0$ is known (this is the case close to stability),
and that the uncertainty factor $U_X\geq 1$ of $X_0$ is given by $\max(u_X,1/u_X)$, where
\begin{equation}
u_X=\overline{u}\left(1-X_0\right) + X_0 \quad,
\end{equation}
and $\overline{u}$ is an averaged uncertainty factor in the predicted ratios of the $R_i$. These ratios are believed to be predicted with better accuracy than the rates themselves and so it can be assumed $\overline{u}\leq U_\mathrm{th}$, with $U_\mathrm{th}$ being the uncertainty factor of the theoretical prediction.
In any case, the uncertainties are sufficiently small to preserve the magnitude
of $X_0$, i.e., small $X_0$ remain small within errors and large $X_0$ remain large, as can also be seen in figures \ref{fig:xplotng}--\ref{fig:xplotag}.

Complete tables of g.s.\ contributions $X_0$ for reactions on target nuclei between the driplines from $10\leq Z\leq 83$ are given in \cite{sensipaper}. Figures \ref{fig:xplotng}--\ref{fig:xplotag} show g.s.\ contributions (and isomeric state contributions $X_2$ for $^{180m}$Ta) for (n,$\gamma$), (p,$\gamma$), and ($\alpha$,$\gamma$) reactions, respectively, on natural isotopes at $\gamma$-process temperatures, including their uncertainties. It immediately catches the eye that excited state contributions are non-negligible for the majority of cases and that g.s.\ contributions are especially small in the region of the deformed rare-earth nuclei. This is because they have an inherently larger NLD. Near shell closures, on the other hand, NLDs are lower and the g.s.\ contributions larger. An additional effect acting for some of the ($\alpha$,$\gamma$) cases is the Coloumb suppression effect of the excited state contributions, which is explained further down below.

Laboratory measurements can only determine the stellar reactivity when the contribution $X_i$ of the target level (in most cases this is the g.s.) is close to unity. The stellar rate has to be derived by combining experimental data with theory when $X_i<1$.
Strictly speaking, the experimental cross section can only replace one of the contributions to the stellar rate while the others remain unconstrained by the data, unless further knowledge is present. Knowing $X_i$ and the theory values for $R_i$ and $R^*$, the proper inclusion of a new experimentally derived reactivity $R_i^\mathrm{exp}$ into a new stellar rate is performed by modifying the \textit{theoretical} stellar reactivity $R^*$ to yield the new stellar reactivity \cite{renormletter}
\begin{equation}
R^*_\mathrm{new}= f^* R^* \quad,
\end{equation}
with the renormalization factor
\begin{equation}
f^*=1+X_i\left(\frac{R_i^\mathrm{exp}}{R_i}-1\right) \label{eq:renormstellar}
\end{equation}
containing the experimental result.
Note that the renormalization factor is, of course, temperature-dependent.

Also the uncertainty in the experimental cross sections (the ``error bar'') can be included. Since the ultimate goal of a measurement is to reduce the uncertainty inherent in a purely theoretical prediction, it is of particular interest to know the final uncertainty of the new stellar reactivity $R^*_\mathrm{new}$. It is evident that the new uncertainty will only be dominated by the experimental one when $X_i$ is large. Using an uncertainty factor $U_{\mathrm{exp}}\geq 1$ implies that the ``true'' value of $R_i^\mathrm{exp}$ is in the range $R_i^\mathrm{exp}/U_{\mathrm{exp}}\leq R_i^{\mathrm{true}}\leq R_i^\mathrm{exp}U_{\mathrm{exp}}$, and analogous for the theoretical uncertainty factor $U_{\mathrm{th}}=U^*$ of the stellar reactivity $R^*$. For example, an uncertainty of 20\% translates into $U_{\mathrm{exp}}=1.2$. Experimental and theoretical uncertainty are then properly combined to the new uncertainty factor
\begin{equation}
U^*_\mathrm{new}=U_{\mathrm{exp}}+(U^*-U_{\mathrm{exp}})(1-X_i) \label{eq:uncertainty}
\end{equation}
for $R^*_\mathrm{new}$. Here, $U_{\mathrm{exp}}\leq U^*$ is assumed because otherwise the measurement would not provide an improvement.
Obviously, also the uncertainty factor is temperature-dependent because at least $X_i$ depends on the plasma temperature.
It is further possible to consider the uncertainty of $X_i$ in $U^*_\mathrm{new}$. Its impact, however, is small with respect to the other experimental and theoretical uncertainties. The fundamental differences between error determinations in experiment and theory are discussed in \cite{sensipaper} and appropriate choices are suggested in \sref{sec:sensi} and in \cite{renormletter}.

It should be kept in mind that the corrections for thermally excited nuclei have to be predicted even if a full database of
experimental (ground state) rates were available.
Therefore, designing an experiment it should be taken care to measure a reaction with the largest possible $X_i$. This also implies that the reaction should be measured in the direction of largest $X_i$. Since the astrophysically relevant energies of the reverse reaction $B(b,a)A$ are related to the ones of the forward reaction by $E_\mathrm{rev}=E+Q$ (see \sref{sec:compound}), it is obvious that transitions from excited states contribute more to the stellar rate in the direction of negative Q-value. This is due to the fact that the weights given in \eref{eq:weights} depend on $E$ and $E_\mathrm{rev}$, respectively, and decline more slowly with increasing excitation energy $E_i$ when $E$ is larger. This is especially pronounced in photodisintegrations because of the large $\left| Q \right|$. \Tref{tab:xphoto} gives examples for g.s.\ contributions to ($\gamma$,n) rates of intermediate and heavy target nuclides at typical $\gamma$-process temperatures. These numbers have to be compared to the ones for the neutron captures shown in \fref{fig:xplotng}. It is obvious that they are tiny in comparison. (This is also the reason why we do not show full plots similar to the ones for the capture reactions. All g.s.\ contributions and further plots can be found in \cite{sensipaper}.)

As argued above, the smallest stellar correction usually is found in the direction of positive Q-value. \Fref{fig:xplotag} and the tables in \cite{sensipaper}, however, contain a few exceptions to the Q-value rule. For example, ($\alpha$,$\gamma$) reactions with negative
Q-values appear in the $\gamma$-process network and show smaller corrections than their photodisintegration counterparts. Among them is, for example, the important case of $^{144}$Sm($\alpha$,$\gamma$)$^{148}$Gd with $X_0\approx 1$. These cases can be explained by a \textit{Coulomb suppression effect}, first pointed out in \cite{kisssupp} and studied in detail in \cite{tomsupp}. Since the relative interaction energies $E_i^\mathrm{rel}=E-E_i$ of transitions commencing on excited levels decrease with increasing $E_i$, cross sections $\sigma_i$ strongly depending on $E_i^\mathrm{rel}$ will be suppressed for higher lying levels. For large Coulomb barriers this can result in a much faster suppression of excited state contributions to the stellar rate than expected from the weights $W_i$. If there is a much higher Coulomb barrier in the entrance channel of a reaction with negative Q-value than in its exit channel (e.g., when there is no barrier present as for ($\alpha$,$\gamma$) or (p,n) reactions), this can yield a larger $X_0$ in the direction of negative Q-value than in the other direction. On this grounds it can be shown that it is generally much more advantageous to measure in capture direction for any type of capture reaction, regardless of Q-value, because the $X_0$ will be larger than for the photodisintegration and thus closer to the stellar value. This stellar value can then conveniently be converted to a stellar photodisintegration rate by applying \eref{eq:revphoto}.

\begin{table}
\caption{\label{tab:xphoto}Ground state contributions $X_0$ for selected ($\gamma$,n) reactions at 2.5 GK.}
\begin{indented}
\item[]\begin{tabular}{clclcl}
\br
Target & $X_0$ & Target & $X_0$ & Target & $X_0$ \\
\mr
$^{86}$Sr & 0.00059 &  $^{186}$W & 0.00049 &   $^{198}$Pt & 0.0018\\
$^{90}$Zr & 0.00034 &  $^{185}$Re & 0.00021 &   $^{197}$Au & 0.00035\\
$^{96}$Zr & 0.0061 &   $^{187}$Re & 0.00024 &  $^{196}$Hg & 0.00043\\
$^{94}$Mo & 0.0043 &     $^{186}$Os & 0.00016 &   $^{198}$Hg & 0.00084\\
$^{142}$Nd & 0.0028 &    $^{190}$Pt & 0.000069 &  $^{204}$Hg & 0.0088\\
$^{155}$Gd & 0.0012 &    $^{192}$Pt & 0.00011 &  $^{204}$Pb & 0.0059\\
\br
\end{tabular}
\end{indented}
\end{table}

\paragraph{How experiments can help}
\label{sec:help}
Due to the high plasma temperatures encountered in explosive nucleosynthesis and the large number of possible transitions between levels in target, compound, and final nucleus an experimental determination of the stellar rate is impossible for the majority of intermediate and heavy nuclei. The few exceptions can be selected by searching for reactions with large g.s.\ contribution $X_0$. A further constraint is the fact that most of the reactions involve unstable nuclei. Nevertheless, measurements can provide important information on specific transitions and parts of the stellar rate which can be compared to predictions of theoretical models. A good example are the past ($\alpha$,$\gamma$) and (p,$\gamma$) measurements at low energies which have shown deficiencies in the description of low-energy charged-particle widths (see \sref{sec:sensi}). Another example would be studies of single transitions to better constrain excited state contributions.
Essential for the development of reaction models, which can also be extended into the region of unstable nuclei, is to have systematic measurements across larger mass ranges. Such systematics are lacking even at stability for astrophysically important reactions and transitions at astrophysical energies.

The following \sref{sec:methods} provides an extensive overview of present and future experimental approaches to improve reaction rates for studying the synthesis of p-nuclides.

\section{Experimental approaches}\label{sec:methods}

\subsection{$\gamma$-induced measurements}\label{sec:gamma}

As explained in \sref{sec:stellareffects} and shown in \tref{tab:xphoto}, the experimental study of $\gamma$-induced reactions has only limited relevance for the direct astrophysical application. From a laboratory $\gamma$-induced cross section measurement with the target nucleus being in its g.s., no direct information can be inferred for the stellar reaction rate.
Nevertheless, as pointed out in \sref{sec:help}, the experimental study of $\gamma$-induced reactions can provide useful information for certain nuclear properties relevant to heavy element nucleosynthesis. The experimentally determined ($\gamma$,$x$) cross section (where $x$ can be a neutron, a proton, or an $\alpha$-particle)
probes particle transitions into the final nucleus initiated by a well-defined $\gamma$-transition from the g.s.\ of the target nucleus. Theoretical predictions for the relevant transitions can then be tested by comparing them with the measured photodisintegration cross section.

Again, the sensitivity of the reaction (see sections \ref{sec:sensi} and \ref{sec:expsensi}) has to be carefully checked to see what can be extracted from such a measurement. The sensitivities depend on the initial $\gamma$-energy and the particle separation energies. The latter define the relative interaction energies $E_i^\mathrm{rel}$ (i.e., the energies of the emitted particles) and thus the particle widths. As mentioned in \sref{sec:sensi}, the channel with the smallest width determines the sensitivity of the cross section. For laboratory ($\gamma$,n) reactions, this is always the $\gamma$-width. A comparison of predicted laboratory ($\gamma$,n) cross sections and measured ones, however, does not test a quantity that is of relevance in the astrophysical application for two reasons. Firstly, astrophysical charged-particle captures and photodisintegrations depend solely on the charged-particle widths as can be verified by inspection of the figures and tables given in \cite{sensipaper}. Secondly, although astrophysical neutron captures and ($\gamma$,n) do partially depend on the $\gamma$-widths, they are sensitive to a different part of the $\gamma$-strength function than can be tested in a laboratory experiment. The relevant $\gamma$-transitions are those with $2-4$ MeV downwards from states a few tens to a few hundreds of keV above the neutron threshold \cite{raugamma}. According to the reciprocity relation \eref{eq:revphoto}, this applies to stellar capture rates as well as photodisintegration rates. Assuming the validity of the Brink-Axel hypothesis \cite{brink,axel} and using low $\gamma$-energies on the nuclear g.s.\ does not help because the strength function cannot be probed below the particle emission threshold. A more promising approach is to study partial particle emission cross sections, such as ($\gamma$,n$_0$), ($\gamma$,n$_1$), and so on. Their ratio depends, apart from spin selection rules, on the particle emitting transitions and can thus be used to test the prediction of the ratios of g.s.\ particle transitions to transitions on excited states. This is then relevant to astrophysical applications, to study the interaction potentials and the stellar excitation effects (see, e.g., \sref{sec:stellareffects}). On-line detection of the outgoing particle (see below) is required for this.

The same can be done for charged particle emission. In addition, depending on the nucleus and the $\gamma$-energy used, photodisintegration with charged particle emission is sensitive to the particle channel and can thus be used to test, e.g., optical potentials. Comparing $\alpha$-emission data to predictions would be of particular interest regarding the action of the Coulomb excitation effect described in \sref{sec:direct}. It would imply different transmission coefficients for $\alpha$-capture and $\alpha$-emission and should not show up in the photodisintegration results. Partial cross sections are helpful also in this case because they allow testing at different relative interaction energies.

In the remainder of this section the available experimental techniques and those $\gamma$-induced reaction cross section measurements are reviewed for which the $\gamma$-process was specified as motivation. Experimental data on $\gamma$-induced reactions are mainly available around the giant dipole resonance, i.e., at much higher energy than important for astrophysics. Owing to the small cross section and other technical difficulties (see below), only very few measured cross sections are available at lower energies.

Different experimental approaches can be used to measure low-energy $\gamma$-induced reaction cross sections. The common requirement of these techniques is a high $\gamma$-flux in order to measure low cross sections. If the aim is to measure excitation functions (i.e., the cross section as a function of energy), a monoenergetic $\gamma$-beam is needed. Quasi-monoenergetic $\gamma$-rays can be obtained with the laser Compton scattering technique. In this method $\gamma$-rays are produced by head-on collision of laser photons with relativistic electrons. This method has been successfully applied for several reactions, e.g., at the National Institute of Advanced Industrial Science and Technology (AIST), Tsukuba, Japan \cite{uts03,shi05,uts08}. There are proposals for performing $\gamma$-process studies also at the High Intensity $\gamma$-Ray Source (HI$\gamma$S) at Duke University, USA.

Another method for $\gamma$-induced reaction studies is the application of Bremsstrahlung radiation. If a high-energy electron beam hits a radiator target, continuous energy Bremsstrahlung radiation is produced with an endpoint energy equal to the energy of the electron beam. Since the $\gamma$-energy spectrum is continuous, no direct measurement of the excitation function is possible. By the superposition of several Bremsstrahlung spectra of different endpoint energies, however, the high-energy part of a thermal photon flux can be approximated well \cite{moh00}. Therefore, the reaction rate of a photodisintegration reaction can be measured instead of the cross section itself. Since the target is always in its g.s., only the g.s.\ contribution to the reaction rate can, of course, be measured. Several $\gamma$-induced reactions have been studied with this method at the S-DALINAC facility in Darmstadt, Germany \cite{moh00,vog01,son04,has08,has09}, and at the ELBE facility in Dresden-Rossendorf, Germany \cite{nai08b,nai08c,erh10} (see \tref{tab:gammaexp}).

The Bremsstrahlung radiation method can be improved towards a quasi-monochromatic $\gamma$-beam by applying the so-called tagger technique. In this method, the electron beam hits a thin radiator target where it can be guaranteed that only one Bremsstrahlung photon is produced by one electron. If the remaining energy of the electron is then measured, the $\gamma$-energy can be inferred and the required $\gamma$-energy with relatively low energy spread can be selected. Such a method has been recently developed at the S-DALINAC facility \cite{sch10}.

Intense $\gamma$-beams created by ultra-high intensity lasers may open a new horzion in the study of $\gamma$-induced reactions in the near future. One of the four pillars of the European ELI (Extreme Light Infrastructure) facility to be built in Magurele, Romania, will be devoted to laser-based nuclear physics (ELI-NP). The white book of the ELI-NP project \cite{ELINPwhitebook} includes the study of the $\gamma$-process as one of the objectives of the facility. The very brilliant, intense $\gamma$-beam of up to 19 MeV, 0.1\% bandwidth, and 10$^{13}$ $\gamma$/s intensity is hoped to enable the measurement of ($\gamma$,$\alpha$) and ($\gamma$,p) reaction cross sections on many isotopes.

Two distinct methods can be applied to determine the number of reactions taking place during the $\gamma$-irradiation. The first method is based on the on-line detection of the outgoing particle. This method has been successfully applied only for ($\gamma$,n) reactions because charged particle emitting reactions have typically lower cross section and the detection of the resulting low yields in a high $\gamma$-flux environment requires special experimental technique.
The other method is photoactivation, where the cross section is determined from the off-line measurement of the induced activity of the irradiated target. This method is, of course, only applicable when the product nucleus is radioactive, but owing to its technical advantages the majority of the $\gamma$-process related photodisintegration measurements have been carried out with this technique. Further details of the in-beam and activation methods will be discussed in connection with the charged particle induced reaction cross section measurements in \sref{sec:charged}.

All the above mentioned experimental techniques can only be applied to stable target nuclei. For the $\gamma$-process, however, photon-induced reactions on proton-rich unstable isotopes are also important. These reactions can in principle be studied in inverse kinematics by the Coulomb Dissociation method. Coulomb Dissociation is a well known technique, and is, e.g., being studied at the GSI Helmholtz Center for Heavy Ion Research in Darmstadt, Germany, with the LAND (Large Array Neutron Detector) setup. Results for $^{92,93,100}$Mo($\gamma,n$) have been published recently, the analysis for $^{94}$Mo($\gamma,n$) is ongoing \cite{gsi10,ershova,goebel}.

\begin{table}
\caption{\label{tab:gammaexp} Experimentally studied $\gamma$-induced reactions relevant for the $\gamma$-process. Abbreviations: act.\,=\,activation, br.\,=\,Bremsstrahlung, CD \,=\, Coulomb dissociation, LCS\,=\,laser Compton scattering, n.c.\,=\,neutron counting}
\footnotesize\rm
\begin{tabular}{ccccc}
\hline
Target & Reaction & Method & Facility & Reference \\
\hline
$^{91,92,94}$Zr 			& ($\gamma$,n) & LCS, n.c. & AIST & \cite{uts08} \\
$^{92,100}$Mo, $^{144}$Sm & ($\gamma$,n),\,($\gamma$,p),\,($\gamma,\alpha$) & br., act. & ELBE & \cite{nai08b,nai08c,erh10} \\
$^{92,93,94,100}$Mo 			& ($\gamma$,n) & CD, n.c. & GSI-LAND & \cite{gsi10,ershova,goebel} \\
$^{148,150}$Nd,\,$^{154}$Sm,\,$^{154,160}$Gd & ($\gamma$,n) & br., act. & S--DALINAC & \cite{has08} \\
$^{181}$Ta 						& ($\gamma$,n) & LCS, n.c. & AIST & \cite{uts03} \\
$^{186}$W, $^{187}$Re, $^{188}$Os & ($\gamma$,n) & LCS, n.c. & AIST & \cite{shi05} \\
$^{192}$Os, $^{191,193}$Ir & ($\gamma$,n) & br., act. & S--DALINAC & \cite{has09} \\
$^{190,192,198}$Pt 		& ($\gamma$,n) & br., act. & S--DALINAC & \cite{moh00,vog01}\\
$^{196,198,204}$Hg, $^{204}$Pb & ($\gamma$,n) & br., act. & S--DALINAC & \cite{son04} \\
\hline
\end{tabular}
\end{table}

\Tref{tab:gammaexp} lists some of the $\gamma$-induced reactions studied experimentally in recent years. It is not an exhaustive list since only those measurements are listed where the $\gamma$-process was mentioned as a motivation of the work. As one can see, with the exception of the $^{92}$Mo and $^{144}$Sm isotopes, only ($\gamma$,n) reactions have been studied. The reason for this is the typically very low cross section of charged-particle emitting reactions at astrophysically relevant energies. The fast development of the different experimental methods, however, will most likely allow for the extensive study of ($\gamma$,$\alpha$) and ($\gamma$,p) reactions in the near future.

The photodisintegration cross sections are usually compared with the predictions of statistical model calculations. Calculations using different input parameters such as $\gamma$-ray strength functions are also considered and compared to the data in several works. It can be stated that in general the model calculations are able to reproduce the measured data within about a factor of two. For further details, the reader is referred to the original publications, for example the detailed overview on the direct determination of photodisintegration cross sections related to the $\gamma$-process in \cite{gammarev} and references therein.

\subsection{Charged-particle induced measurements}
\label{sec:charged}

Owing to the huge effect of the stellar enhancement factor in the case of $\gamma$-induced reactions, it is preferable to study experimentally the inverse capture reactions (see \sref{sec:stellareffects}). As shown in \sref{sec:sensi}, nuclear physics uncertainties influence strongly the result of a $\gamma$-process network calculation, therefore the cross section measurement of capture reactions in the relevant mass and energy range is of high importance. The case of neutron-induced reactions is discussed in \sref{sec:neutron}. In this section the cross section measurements of charged particle induced reactions are reviewed.

The fundamental difficulty of charged-particle induced reaction studies is the low values of the cross sections. \Tref{tab:chargedcrosssec} shows the case of two p-isotopes, the lowest mass $^{74}$Se and the highest mass $^{196}$Hg. The table shows the location of the relevant energy window for proton- and $\alpha$-capture reactions at some relevant temperatures for these two isotopes as well as the range of cross section within the energy window obtained with the NON-SMOKER code \cite{nonsmoker}. As one can see, the cross sections range from the millibarn region (in the case of proton capture on light nuclei) down to the 10$^{-18}$ barn region (for $\alpha$-capture on heavy isotopes). Unfortunately only the upper part of this cross section range is measurable, thus the experiments need to be carried out at higher energies and the results must be extrapolated to astrophysical energies. This is done based on theoretical cross section curves, so the involvement of some theory is inevitable. To minimize its effect, however, the experiments should be carried out at energies as low as possible and hence low cross sections need to be measured.

Due to the low cross sections encountered, experimental data on proton and $\alpha$-capture reactions are scarce at low energies in the mass region relevant to the $\gamma$-process. Experimental data started to accumulate only in the last 15 years. Still, the number of studied reactions remains relatively small (see figures \ref{fig:pg_measured} and \ref{fig:ag_measured} in \sref{sec:expstatus}) compared to the huge number of reactions involved in a $\gamma$-process network. There are two different methods for cross section measurements: the in-beam $\gamma$-detection technique and activation. In the following some features of these two methods are discussed.

\begin{table}
\caption{\label{tab:chargedcrosssec} Location of the Gamow window and the relevant cross sections of proton and $\alpha$-capture reactions on two p-isotopes}
\footnotesize\rm
\begin{tabular}{cccc}
\hline
reaction & temperature & Gamow window & cross section \\
				& 10$^9$\,K & MeV & barn \\
\hline
$^{74}$Se(p,$\gamma$)$^{75}$Br & 3.0 & 1.39\,--\,3.07 & 2$\cdot$10$^{-6}$\,--\,1$\cdot$10$^{-3}$\\
$^{74}$Se($\alpha,\gamma$)$^{78}$Kr & 3.0 & 4.65\,--\,7.15 & 5$\cdot$10$^{-8}$\,--\,3$\cdot$10$^{-4}$\\
$^{196}$Hg(p,$\gamma$)$^{197}$Tl & 2.0 & 2.56\,--\,4.64 & 1$\cdot$10$^{-10}$\,--\,6$\cdot$10$^{-6}$\\
$^{196}$Hg($\alpha,\gamma$)$^{200}$Pb & 2.0 & 6.93\,--\,10.03 & 1$\cdot$10$^{-18}$\,--\,1$\cdot$10$^{-10}$\\
\hline
\end{tabular}
\end{table}

\subsubsection{In-beam $\gamma$-detection technique}

The natural way of measuring a capture cross section is the detection of the prompt $\gamma$-radiation. In the relevant mass and energy range capture reactions mainly proceed through the formation of a compound nucleus which then decays to its ground state by the emission of $\gamma$-radiation. The excitation energy of the formed compound nucleus is typically above 10 MeV in $\gamma$-process related experiments and thus the level density is very high. Therefore, the particle capture can populate many nuclear levels which results in a complicated $\gamma$-decay scheme involving many primary and secondary transitions. In order to determine the total capture cross sections, practically all these transitions need to be detected. Any unnoticed transitions can result in an underestimation of the cross section (unless one dominant transition exists through which all de-excitation $\gamma$-cascades have to pass). This means that the in-beam $\gamma$-detection technique is very sensitive to laboratory background, and even more to beam-induced background. Weaker transitions in the investigated reaction can be buried under the peaks and Compton continuum of higher cross section reactions on target impurities, therefore an effective reduction of the background is crucial. Moreover, transitions during the decay of the compound nucleus can proceed between states with various and often unknown spin and parity. Therefore, the angular distribution of the measured $\gamma$-radiations is usually not isotropic. The measurement of these angular distributions for all the studied transitions is thus also necessary in order to obtain total, angle-integrated cross sections.

Despite the experimental obstacles, some cross sections relevant for the $\gamma$-process have successfully been measured. These measurements are dominated by proton capture reactions since in the case of ($\alpha,\gamma$) reactions the beam-induced background compared to the signal from the studied reaction is usually much stronger. First experiments with this technique have been carried out using a few HPGe detectors mounted on a turntable allowing for angular distribution measurements. Proton capture cross sections on $^{93}$Nb \cite{har01}, $^{88}$Sr \cite{gal03} and $^{74}$Ge \cite{sau12} isotopes have been measured with this technique in Athens, Greece and Stuttgart, Germany.

The detection efficiency can be increased and the measurement of angular distributions sped up by using a detector array consisting of many single $\gamma$-detectors arranged in a spherical geometry around the target. Additionally, such a configuration allows a substantial reduction of the background by requiring coincidence conditions between the single detectors in the array. Such a method has been developed recently at the University of Cologne, Germany \cite{has10,sau11}

Most of the disadvantages of the in-beam technique can be avoided by using a 4$\pi$ summing crystal for $\gamma$-detection \cite{spy07}. If the target is completely surrounded by, e.g., a large scintillator detector with a time resolution longer than the typical time interval between the successive $\gamma$-emissions during the compound nucleus decay, then one single $\gamma$-peak for all capture events will appear in the spectrum. The energy of this so-called sum-peak is the sum of the reaction Q-value and the center-of-mass energy. There is, of course, no need for angular distribution measurements in this case. Care must be taken, however, to remove some possible sources of uncertainty. The energy resolution of a scintillator is poor compared to a HPGe detector which results in a relatively wide sum-peak. The sum-peak may also contain unwanted events from reactions on target impurities, if those have similar Q-values than the reaction to be studied. This would lead to an overestimated cross section. The condition that all capture events generate a signal in the sum-peak is only valid if the detector has a 100\,\% efficiency for all $\gamma$-energies. Although the efficiency of a big summing crystal can be fairly large, it is never 100\%. Thus, the sum-peak efficiency must be determined experimentally which requires the knowledge of $\gamma$-ray multiplicities. Several reactions have been studied with this method in Athens, Greece \cite{tsa04}, in Bochum, Germany \cite{pdpg}, and in Bucharest, Romania \cite{pre12}. The analysis of most of the Bochum measurements is still in progress \cite{har05}.

\subsubsection{Activation method}

The overwhelming majority of $\gamma$-process related charged-particle capture cross-sections has been measured with the activation technique. In this method, the total number of reactions having taken place is determined through the number of product nuclei, instead of detecting the prompt $\gamma$-radiation following the capture process. This is feasible when the final nucleus is radioactive, decays with a convenient half-life, and the decay can be measured through the detection of a suitable high-intensity radiation.

When a target with areal number density $\mathcal{T}$ is bombarded by a proton or $\alpha$-beam with $\phi$ projectiles per second with energy $E$ for an irradiation time $t_\mathrm{irr}$, then the number of produced isotopes $N$ at the end of the irradiation is given by

\begin{equation}
\label{eq:acivation}
N = \sigma(E)  \mathcal{T}  \phi  \frac{1-e^{-\lambda  t_\mathrm{irr}}}{\lambda},
\end{equation}
where $\sigma(E)$ is the reaction cross section and $\lambda$ is the decay constant of the produced radioactive isotope. If the beam intensity is not constant in time, the formula becomes:

\begin{equation}
\label{eq:acivation2}
	N =\sigma(E)  \sum^{n}_{i=1} \mathcal{T}  \phi_i  \frac{1-e^{-\lambda  t_i}}{\lambda}  e^{-\lambda \tau_i}, \hspace{1cm} n t_i = t_\mathrm{irr}.
\end{equation}
Here the irradiation time is segmented into $n$ shorter intervals $t_i$. During the shorter intervals the beam intensity $\phi_i$ is considered to be constant. Supposing the produced isotope emits a $\gamma$-radiation with relative intensity $\eta$, then  the number of gammas detected during a subsequent counting interval of $t_c$ is
\begin{equation}
\label{eq:counting}
	n_\gamma=N  e^{-\lambda t_w}  (1-e^{-\lambda t_c}) \varepsilon_\gamma  \eta ,
\end{equation}
where $\varepsilon_\gamma$ is the detection efficiency for the studied $\gamma$-line and $t_w$ is the time elapsed between the end of the irradiation and the start of counting. The cross section can be deduced from the above equations when the other quantities are known. If the final nucleus has long-lived isomeric state(s), the above formulae become more complicated, see, e.g., \cite{sau97}.

The applicability of the activation method is, of course, limited to reactions leading to radioactive isotopes and no information about the details of the $\gamma$-transitions during the compound nucleus decay can be obtained. These limitations are, however, compensated by the relative easiness of the activation experiments compared to in-beam measurements. The total cross section is naturally obtained without any problems of possibly missed $\gamma$-transitions. No angular distribution effects have to be taken into account. The most important advantage is the typically much lower background. If target impurities -- on which long-lived radioactive isotopes would be produced -- are avoided, the background can essentially be reduced to the laboratory background which can be effectively shielded. Some beam-induced radioactivities of the target often cannot be completely avoided but the background level is always much lower than for in-beam experiments.

Owing to the lower background, it is possible to study more than one reaction in a single activation since different isotopes are characterized by different decay signatures. Using a target with natural isotopic composition target, the capture cross section of several isotopes of the same element can be measured simultaneously (see, e.g., \cite{gyu01}). Also, besides the radiative capture, some other reaction channels of the same isotope can be measured at the same time, such as ($\alpha$,n) and ($\alpha$,p) reactions along with ($\alpha$,$\gamma$) (see, e.g., \cite{gyurky06}).

In the mass range relevant for the $\gamma$-process the created radioisotopes are almost always $\beta$-radioactive and the decay is often followed by $\gamma$-radiation. Therefore, the majority of the activation measurements has been based on $\gamma$-detection. One exception was the study of the $^{144}$Sm($\alpha,\gamma$)$^{148}$Gd reaction, where the produced $^{148}$Gd is $\alpha$-radioactive and the cross section was measured via $\alpha$-detection. In some cases the $\beta$-decay is not followed by $\gamma$-radiation or its intensity is very low. If the decay, however, proceeds through electron capture, the detection of the emitted characteristic X-ray radiation can be used to determine the cross section. This method has been used for the first time in the case of the $^{169}$Tm($\alpha,\gamma$)$^{173}$Lu reaction \cite{tm169}.

It is worth noting that an activation experiment requires the knowledge of the decay parameters of the produced isotope, like the relative $\gamma$-intensities or the decay half-lives. Dedicated half-life measurements of several isotopes have been carried out recently to aid the activation experiments \cite{ful03,gyu05,gyu09,far11}.

\subsubsection{Indirect measurements with (d,p) or (d,n) reactions}
\label{sec:expdp}
There is considerable experience in performing indirect measurements of reaction cross sections with (d,p) and (d,n) reactions. With light target nuclei and at sufficiently high energy these can probe states and transitions also important in, say, capture reactions. The prerequisite for this is that direct reactions dominate and can be described, e.g., by the Distorted Wave Born Approximation (DWBA) \cite{sat83}. Given the success in the realm of light nuclei, such reactions at high energies -- several tens to hundreds of MeV, for instance -- have also been suggested to be used in studying reactions at higher mass for explosive thermonuclear burning, including those far off stability. This method has already been applied to study properties of nuclei at neutron shell closures in the r-process path \cite{kate,notkate}.

The benefits of this method for the $\gamma$- and $\nu$p-process, however, are more limited because the NLD at the compound formation energy is much higher in this case and the direct reaction mechanism does not contribute significantly at the astrophysically relevant energies. Therefore many more transitions are involved and the compound reaction mechanism dominates at astrophysical energies. Such measurements, however, could be used for spectroscopic studies of proton-rich, unstable nuclei at radioactive ion beam facilities. The extracted information on excited states and spectroscopic factors is useful in the calculation of the widths appearing in the treatment of compound reactions at low energy. Important is the determination of \textit{low-lying} states, as these are significant for the calculation of the particle widths and the stellar excitation effects (see sections \ref{sec:ratedefs}, \ref{sec:sensi}, \ref{sec:stellareffects} and \cite{raureview,sensipaper}). It remains to be seen whether the optical potentials required in the DWBA analysis of the experimental data can be predicted with sufficient reliability to extract the properties of the discrete states. (It should be noted that these potentials are different from the ones used in the statistical model at low energy.)

\subsubsection{Novel approaches}
\label{sec:novelexp}
One serious limitation of the activation method occurs when the half-life of the reaction product is too long and/or its decay is not followed by any easily measurable radiation. Then the methods as described above cannot be applied and novel techniques have to be developed. The number of produced isotopes can also be measured by Accelerator Mass Spectrometry (AMS). However, also in this case there is a restriction to reactions with stable target isotopes. To study reactions on unstable, proton-rich isotopes special experimental techniques are required. One of these new techniques is to use reactions in inverse kinematics in storage rings like, e.g., the Experimental Storage Ring (ESR) at the GSI Helmholtz Center for Heavy Ion Research in Darmstadt, Germany \cite{ESR}. Some details of these two approaches are discussed below.

\paragraph{Measurements using Accelerator Mass Spectrometry}
\label{sec:ams}
Accelerator Mass Spectrometry is an ultra-sensitive and ultra-selective analytical method for the detection of trace amounts (sub-ng range) of long-lived radioactive isotopes \cite{ams}. It is most commonly used for dating of archaeologic and geologic samples, e.g., with $^{14}$C (radiocarbon dating). The method relies on the ability of counting atoms rather than the respective decays and is thus superior to most other methods for the detection of long-lived isotopes ($t_{1/2}>10^{4}$ y) or for those isotopes which emit no or only weak $\gamma$-radiation. One of the major challenges for AMS measurements is the suppression and separation of (stable) isobaric interferences. For a more detailed description, see e.g.  \cite{paul}.

AMS is a relative method which measures the amount of a radionuclide versus the ion current of a stable isotope, ideally of the same element. It is up to now the most sensitive detection method and can reach isotope ratios down to $10^{-16}$. However, it requires standards with well-known isotope ratios (of the same order as the measured samples). The production of these calibration standards via nuclear reactions is not straightforward. Preparations for non-standard AMS isotopes, like the ones discussed in this section, involves additional complications.

A few years ago AMS has been successfully combined with astrophysical activation measurements, mainly for the determination of (n,$\gamma$) cross sections for s-process nucleosynthesis (see, e.g., \cite{ni62,rugel,ca40}), and for the $^{25}$Mg(p,$\gamma$)$^{26g}$Al and $^{40}$Ca$(\alpha,\gamma)$$^{44}$Ti cross sections \cite{al26,ti44}. Also first attempts to use AMS to determine the photodissociation cross section of the $^{64}$Ni($\gamma$,n)$^{63}$Ni reaction have been made \cite{ni64}. An overview of cross section measurements for nuclear astrophysics performed with AMS is given in  \cite{wallner}.

Searching the chart of nuclides on the proton-rich side of the valley of stability for radioactive isotopes with half-lives larger than 250 days and masses in the range $68\leq A\leq 209$ provides 32 matches. Discarding isotopes for which a measurement of the $\alpha$- or $\gamma$-activity is possible with normal efforts and those posing severe problems for the AMS technique, leaves $^{68}$Ge, $^{93}$Mo, $^{146}$Sm, $^{179}$Ta, $^{194}$Hg, $^{202}$Pb, and $^{205}$Pb.

First tests for AMS have been already performed with $^{93}$Mo \cite{mo93}, $^{146}$Sm \cite{sm146}, and $^{202}$Pb \cite{pb202}.
The activity of $^{68}$Ge could probably be deduced from its short-lived daughter $^{68}$Ga. However, the small cross section of the $^{64}$Zn($\alpha$,$\gamma$)$^{68}$Ge reaction prevented measurements at astrophysically relevant energies ($E_\alpha<7.5$ MeV) up to now \cite{pori59,ruddy69,bas07}. Tests are presently carried out to determine this cross section with AMS at lower energies \cite{wallner12}.

Rare earth elements are known to poison the ionizer and lead to strongly declining beam currents.
Only $^{146}$Sm has the potential for further investigations, using special cathodes. The
reaction $^{142}$Nd($\alpha$,$\gamma$)$^{146}$Sm is planned to be measured in the near future \cite{paul12}.

The reaction $^{190}$Pt($\alpha$,$\gamma$)$^{194}$Hg is also in reach of AMS but applications in other fields are missing and the sample material is expensive.
The two long-lived isotopes $^{202}$Pb and $^{205}$Pb are produced in lead-cooled fast (Generation IV) reactors. However, lead sputters quickly in the ion sources, preventing the production of ion beams which are stable for long periods of time. The measurement of the $^{198}$Hg($\alpha$,$\gamma$)$^{202}$Pb cross section is under consideration \cite{wallner12}.

AMS could be an alternative for some cross section measurements in the $\gamma$-process region. However, the efforts needed to develop the method for non-standard isotopes are complicated and time-consuming. Unfortunately they can only be justified with additional requests from other research fields, e.g., geology, archeology, and fission or fusion technology.

\paragraph{Measurements using a Storage Ring}
\label{sec:esr}
All the charged particle capture experiments discussed so far are restricted to reactions on stable target isotopes. For explosive nucleosynthesis processes, however, reactions on unstable, proton-rich isotopes are also important. The deflection points in the $\gamma$-process path lie off stability (see \sref{sec:importantreactions}). Cross section measurements on radioactive isotopes require special experimental techniques. Radioactive beam facilities have been developed over the last years which are suitable for such measurements.

A pioneering experiment was carried out at the storage ring ESR in 2009 \cite{zho10}. Proton-induced reactions were measured in inverse kinematics involving fully stripped $^{96}$Ru$^{44+}$ ions which had been injected into the storage ring and slowed down. This stable beam experiment was an important step for future investigations of charged-particle reactions with radioactive beams in inverse kinematics.
First results at high energy were published recently \cite{zho10}, yielding an upper limit of 4 mb for $^{96}$Ru(p,$\gamma$)$^{96}$Rh, which is in good agreement with the predictions from the Hauser-Feshbach code NON-SMOKER \cite{nonsmoker}. This first test measurement was limited to energies above the astrophysical interesting energy range for technical reasons. For future experiments, improved detectors are developed to allow measurements at or close to the astrophysical energies.

\subsection{$\alpha$-elastic scattering}
\label{sec:alphascatt}

As it was emphasized in \sref{sec:charged} and also earlier in this paper, one of the most important input parameters of the statistical model calculation in the case of $\alpha$-induced reactions is the $\alpha$-nucleus optical potential. Therefore, the direct experimental investigation of this potential is of high importance. The optical potential can be studied by measuring elastic $\alpha$-scattering cross sections. The measurement of elastic scattering cross sections at several tens or hundreds of MeV is a well established experimental tool in nuclear physics and high-energy optical potentials for many stable nuclei are thus quite well known. At low, astrophysically relevant energies, however, elastic scattering is dominated by the Coulomb interaction (Rutherford cross section), making the experimental study of the nuclear part of the potential rather challenging.

In order to find substantial deviation from the Rutherford cross section, elastic scattering experiments for $\gamma$-process purposes are carried out at energies somewhat higher than astrophysically relevant. Angular distributions in a wide angular range need to be measured in order to see the typical oscillation pattern of the cross section as a function of scattering angle. The experiments must be carried out with high precision in order to measure the tiny deviations from the Rutherford cross section. The precise determination of the scattering angle is also of high importance since the Rutherford cross section, to which the scattering data are normalized, depends very sensitively on the angle.

\begin{table}
\centering
\caption{\label{tab:scattering} Some parameters of elastic $\alpha$-scattering experiments relevant for the $\gamma$-process. All measurements have been carried out at ATOMKI with the exception of the ones on the Te isotopes which have been studied at the University of Notre Dame.}
\begin{tabular}{rlll}
\hline
isotope(s) & $\alpha$-energies & angular range & reference \\
				& MeV & degree &  \\
\hline
$^{144}$Sm & 20 & 15\,--\,172 & \cite{moh97} \\
$^{92}$Mo & 13.8, 16.4, 19.5 & 20\,--\,170 & \cite{ful01} \\
$^{112,124}$Sn & 14.4, 19.5 & 20\,--\,170 & \cite{gal05} \\
$^{106}$Cd & 16.1, 17.7, 19.6 & 20\,--\,170 & \cite{kis06} \\
$^{89}$Y & 16.2, 19.5 & 20\,--\,170 & \cite{kis09} \\
$^{110,116}$Cd & 16.1 19.5 & 20\,--\,175 & \cite{kis11} \\
$^{120,124,126,128,130}$Te & 17, 19, 22, 24.5, 27 & 22\,--\,168 & \cite{pal09}\\
\hline
\end{tabular}
\end{table}

Several elastic $\alpha$-scattering experiments have been carried out at the Institute of Nuclear Research (ATOMKI) in Debrecen, Hungary. A similar research program has been initiated recently at the University of Notre Dame, USA. The studied isotopes, as well as some parameters of the experiments, are listed in \tref{tab:scattering}.

The measured angular distributions can be compared with predictions using different optical potential parameterizations. Global optical potentials (i.e., potentials that are designed for broad mass and energy regions) are preferred since they can be used in extended $\gamma$-process networks. Several global potentials are available in literature, for a list see, e.g., \cite{kis09}. It is found that different potentials often lead to largely different angular distributions and by comparison with the measured data the one with the best match can be selected. More sensitive analyses can be performed when the ratio of the measured cross sections of two isotopes studied at the same energy is calculated and compared to the corresponding ratio given by global potentials. Such a comparison has been carried out, e.g., for the $^{112,124}$Sn ($Z=50$) and the $^{106,110,116}$Cd ($Z=48$) isotopes, and the $^{89}$Y, $^{92}$Mo ($N=50$) isotones. None of the available global potentials seem to describe well the cross section ratios, clearly indicating the need for improved potential parameterizations at low energies.

In addition to the comparison with global potentials, the measured angular distributions can also be used to constrain some parameters relevant for the given isotope. By fitting the measured cross sections using various approaches, such as Woods-Saxon parameterizations or double folding potentials, local optical potential parameters can be obtained. By studying several isotopes, the evolution of the best fit potential parameters can also be investigated. If the angular distribution is measured in an almost complete angular range, total reaction cross sections can be easily obtained by simply calculating the missing flux from the elastic channel \cite{moh10}. The calculated total cross section can then be compared to experimental data, if they are available, or with model predictions \cite{gyu12}.

The possible appearance of additional reaction channels at low energy, such as the Coulomb excitation introduced in \sref{sec:direct} complicate the interpretation of the scattering data for $\alpha$-particles. If an optical potential is derived from scattering data at an energy where compound nucleus formation is the dominant reaction mechanism, its absorptive part will only account for this loss of $\alpha$-flux from the elastic channel. When the extrapolated potential is then used at lower energies at which low-energy Coulomb excitation (or any other additional mechanism) acts, it will underestimate the inelastic (reaction) cross section at these energies. Nevertheless, it may be correctly describing the compound formation probability and thus will be appropriate for calculating stellar ($\gamma$,$\alpha$) rates, as explained in \sref{sec:direct}. If an optical potential, on the other hand, is derived from scattering data for energies at which the additional mechanism is non-negligible, its absorptive part will include this additional mechanism. Unfortunately, this does not help by itself in the application to the stellar rate, as the compound formation cross section is not constrained separately. This would only be possible by using additional theory, i.e., by calculating the expected cross section for the additional mechanism (e.g., Coulomb excitation) and then adjusting the absorption in the optical potential in such a way that it yields a flux into inelastic channels that is the original one subtracted by the one going into the direct channel. Most likely, such a modification will introduce additional parameters.

Regardless of the possible complications at low energies, precise low-energy elastic $\alpha$-scattering experiments can provide useful information to better understand the optical potential behavior at astrophysical energies and further experiments are needed.

\subsection{Neutron-induced measurements}
\label{sec:neutron}

\subsubsection{Neutron captures}
\label{sec:expng}

Neutron captures in astrophysics have been comprehensively studied along stability for s-process nucleosynthesis. A series of publications has compiled the data and provided recommended values \cite{kadonis,bao87,bao00}. These focussed, however, on the energy range relevant for the s-process which is at lower energies than required in the $\gamma$-process. For many of the quoted reactions cross sections are not even available across the s-process energy range, as only MACS at 25 keV were studied. Quasi-stellar neutron spectra at $kT=25$ keV (which is close to the dominant s-process temperature) were produced. When an energy-dependence was required, renormalized theoretical cross sections were used. As mentioned in \sref{sec:expstatus}, it was shown only recently how to correctly account for stellar effects in the renormalization. The newly renormalized reactivities as function of $kT$, obtained by application of \eref{eq:renormstellar} are given in \cite{renormletter}.

As has been pointed out in \sref{sec:expstatus}, almost no data is available for $\gamma$-process energies. This includes cross sections from major libraries (e.g., ENDF/B \cite{endfb71}, JEFF \cite{jeff31}, JENDL \cite{jendl40}) which are based on theoretical values. A cross-comparison is made difficult by the fact that different, partly undocumented calculations were adopted. Even within a given library, different reaction codes were used. Renormalized theory values could in principle also be used to cover the energy range up to several hundreds of keV, as required for the $\gamma$-process. This has two disadvantages, however. Firstly, it is not always clear whether the statistical model, used for cross section predictions, is applicable at low energy, where data is available. If not, the renormalization should rather be performed at a higher energy. Secondly, according to \eref{eq:uncertainty} and realizing that g.s.\ contributions \eref{eq:gsxfactor} are small at $\gamma$-process temperatures (see, e.g., \fref{fig:xplotng}), it turns out that reactivities at higher temperature are not constrained strongly by the experimental data. This may also discourage direct measurements at $\gamma$-process energies but nevertheless can such data be used to test theoretical reaction models for g.s.\ reactivities, as it is done for reactions with charged particles.

The only way to obtain laboratory reactivities for higher temperatures is to provide cross section measurements covering the astrophysical energy region. Similar as for the study of charged-particle reactions described in \sref{sec:charged}, activation of target material can be used or in-beam measurements at time-of-flight facilities can be performed. Also, a combination of AMS (\sref{sec:ams}) and activation can be applied. When using activation techniques it is crucial not to apply Maxwellian neutron spectra. Such neutron spectra are only useful when the g.s.\ contribution \eref{eq:gsxfactor} is large and the stellar rate can be measured directly. This is not the case for neutron captures in the $\gamma$-process.

A number of neutron time-of-flight facilities have contributed to astrophysical measurements in the past, among them the Oak Ridge Electron Linear Accelerator (ORELA), USA; a similar facility, GELINA, at the IRMM in Geel, Belgium; the DANCE setup at the Los Alamos Neutron Science Center (LANSCE).
Also dedicated to astrophysical measurements, the nTOF facility at CERN has been very productive in recent years. The facility uses high energy protons impinging on a lead spallation target to produce a pulsed neutron beam. A large neutron energy range and a high instantaneous neutron flux combined with high resolution due to the long neutron flightpath are among the key characteristics of the facility. Since 2010, the experimental area has been modified to allow the extension of the physics program to include neutron-induced reactions on radioactive isotopes \cite{gunsing}. This facility would also be well suited to provide the data required for calculating laboratory reactivities for neutron captures in the $\gamma$-process.

A new facility with focus on nuclear astrophysics, the Frankfurt Neutron Source (FRANZ), is under construction at the University of Frankfurt \cite{franz}. It will provide the highest neutron flux in the keV region worldwide and therefore again be best suited for neutron capture in the s-process.

The study of neutron captures on unstable nuclei at radioactive ion beam facilities using inverse kinematics is not possible due to the unavailability of a neutron target. See \sref{sec:expdp} for a discussion of the application of (d,p) reactions instead.

\subsubsection{Inelastic neutron scattering}

It has been pointed out in \sref{sec:gamma} that it is useful to study transitions to excited states in the final nucleus through particle emission. This way, transitions appearing due to the thermal excitation of states in the stellar plasma can be investigated. Inelastic neutron scattering, i.e., (n,n'), provides another approach to achieve this.

For example, inelastic neutron scattering has been used to probe neutron transitions in $^{187}$Os. Measurements with astrophysical motivation were performed, e.g., at ORELA \cite{macklin} and at nTOF \cite{mosconi}. This nucleus is of interest because of its importance in the Re-Os cosmochronometer \cite{mosconi0}. Due to low-lying excited states, it has a non-negligible contribution of excited state transitions to the stellar rate already at s-process temperatures.

Using (n,n') at higher energies for the $\gamma$-process would be even more important (and an interesting complement to neutron capture measurements) because of the even larger stellar effects.

\subsubsection{Studying the optical potential via (n,$\alpha$) reactions}
\label{sec:nalpha}

An interesting alternative to studying low-energy $\alpha$-transitions is the use of (n,$\alpha$) reactions, as suggested by \cite{nalpha0,nalpha1}. There is a number of reasons for this. Except at threshold, the reaction cross sections are sensitive to the $\alpha$-width and therefore to the optical $\alpha$+nucleus potential \cite{sensipaper}. It is also important to note that, when using stable target nuclei, the Q-values of (n,$\alpha$) reactions on nuclei in the mass range relevant to the $\gamma$-process are such that the relative energies of the emitted $\alpha$-particles are in the astrophysically interesting energy range. It is further advantageous that the neutron energy can be varied to probe the energy dependence of the $\alpha$+nucleus optical potential. Measuring partial cross sections, i.e., (n,$\alpha_0$), (n,$\alpha_1$), (n,$\alpha_2$), \dots, also allows to probe this energy dependence and to test the prediction of stellar excitation effects.

The small cross sections expected for low neutron energies, from a few to a few hundred keV, may be problematic. Using predicted cross sections and scaling sample sizes of previous measurements, it was estimated that as many as 30 nuclides across a wide range of masses should be accessible to measurements \cite{nalpha0}.

So far, (n,$\alpha$) on $^{143}$Nd \cite{nalpha1} and on $^{147}$Sm \cite{nalpha0,nalpha1,koe04} have been measured at neutron energies below 1 MeV. Neutron energies of several MeV have been applied in the recent measurements of (n,$\alpha$) on $^{143}$Nd \cite{gle09}, $^{147}$Sm \cite{gle09}, and $^{149}$Sm \cite{gle10,zhanglett}.

Such experiments also provide an excellent possibility for complementary studies to clarify the Coulomb excitation effect suggested to explain the result of the $^{144}$Sm($\alpha$,$\gamma$)$^{148}$Gd measurement (see sections \ref{sec:direct}, \ref{sec:sensi}, and \fref{fig:sm144ag}). No additional reduction of the cross section should be observed in $\alpha$-emission. Although a measurement of $^{147}$Gd(n,$\alpha$)$^{144}$Sm is impossible, it is interesting that the measured (n,$\alpha$) cross sections of the Nd and Sm isotopes are consistently about a factor of $2-3$ below the predictions obtained with several codes. This is in line with the factor of three discrepancy remaining in the
$^{144}$Sm($\alpha$,$\gamma$) case \textit{after} correcting for Coulomb excitation and can be attributed to deficiences in the description of the optical potential alone.

The facilities listed in \sref{sec:expng}, also those focussing on low neutron energies, could also be used for such (n,$\alpha$) studies in the future.

\section{Conclusion}
\label{sec:conclusion}

We have come a long way since \cite{b2fh,agw} but the mystery of the origin of the p-nuclides is still with us. Modern models of low-mass stars show strong s-process contributions to several nuclei previously considered to be p-nuclei. Detailed models of nucleosynthesis in massive stars, coupling large reaction networks to the hydrodynamic evolution of the star, have confirmed the working of the $\gamma$-process, synthesizing p-nuclei through photodisintegration. Supplementing such photodisintegration with neutrino processes required in the production of $^{138}$La and $^{180}$Ta explains the bulk of the p-abundances. Deficiencies are found at higher mass, $150\leq A\leq 165$, and for light p-nuclei with $A<100$. While nuclear reactions are uncertain in the high mass part and further developments of theory and additional experimental data are needed there, the severe underproduction of the light p-nuclei is a long-standing problem and may point to a principle difficulty encountered when trying to obtain light p-nuclei from massive stars. Therefore a number of alternatives have been suggested and some have been studied in detail recently. Consistent hydrodynamical and nucleosynthetic treatment are still missing, however. Meteoritic specimens provide strong constraints for any new process under investigation and thus partially circumvent the problem that p-isotopic abundances cannot be determined from stellar spectra. They also allow to invoke GCE models providing further constraints.
Although there are considerable uncertainties in the astrophysical modeling of the sites possibly producing
p-nuclei, a sound base of nuclear reaction rates is essential for all such investigations. As long as an experimental determination of the rates around the deflection points is impossible, measurements of low energy cross sections of stable nuclides are essential to test and improve the theoretical calculations. This has been underlined by the recent results regarding optical potentials for the interaction of charged nuclei. Further measurements (at even lower energy) are highly desireable but have to be designed carefully, taking into account the astrophysically relevant energy ranges, sensitivities of stellar rates to nuclear input, and the principally possible contribution of the laboratory cross section to the stellar rate. Combining continuing nuclear physics efforts, both in experiment and theory, with
improved, self-consistent hydrodynamic simulations of the possible production sites will gradually improve
our understanding of p-nucleosynthesis.

\ack

We thank T. Davinson, M. Paul, G. Rugel, K. Sonnabend, A. Wallner, and R. Gallino for useful discussions, and A. Heger for providing stellar models considering the latest solar abundance determinations. TR is supported by the Hungarian Academy of Sciences, the MASCHE collaboration in the ESF EUROCORES programme EuroGENESIS, the THEXO collaboration within the 7th Framework Programme of the EU, the European Research Council, and the Swiss NSF. This work was supported in part by NASA (NNX09AG59G), NSF EAR Petrology, and Geochemistry (EAR-1144429), and by a Packard Fellowship to ND.
ID is supported by the German Helmholtz association with a Helmholtz Young Investigator project VH-NG-627.
CF acknowledges support from the DOE Topical Collaboration "Neutrinos and Nucleosynthesis in Hot and Dense Matter" under contract DE-FG02-10ER41677. GG is supported by the ERC StG 203175, and through grants OTKA K101328 and NN83261 (EuroGENESIS).

\end{document}